\documentclass[12pt]{report}
\usepackage[utf8]{inputenc}
\usepackage{braket}
\usepackage{graphicx}
\usepackage{amsmath}
\usepackage[hypertexnames=false]{hyperref}
\usepackage{amsfonts}
\usepackage{slashed}
\usepackage{float}
\usepackage{simpler-wick}
\usepackage{geometry}
\usepackage{tikz-feynman}
\usepackage{amssymb, setspace}
\usepackage{cite}
\usepackage{amsmath, amssymb, graphics, setspace}

\newcommand{\mathsym}[1]{{}}
\newcommand{\unicode}[1]{{}}

\newgeometry{tmargin=2cm, bmargin=2cm, lmargin=1.5cm, rmargin=1.5cm}

\begin{document}
\newgeometry{tmargin=4cm, bmargin=3cm, lmargin=3cm, rmargin=3cm}
\newtheorem{theorem}{Theorem}

\pagenumbering{gobble}
\begin{center}
    {\huge University of Warsaw}\\
    {\LARGE Faculty of Physics}\\[1.5 cm]
    {\Large Oskar Grocholski}\\
    {\large Record book number: 382618} \\[2 cm]
    {\Huge\textbf{Factorization in hard exclusive}}\\[0.3cm]
    {\Huge\textbf{processes.}}\\[0.1cm]
    {\Huge\textbf{Computation of one-loop corrections to the diphoton
	photoproduction on proton.}}\\[1 cm]
    {\Large Master’s thesis}\\ 
	{\Large in the field of Physics}\\[2.5 cm]
\end{center}

\begin{flushright}
%    \begin{tabular}{l}
%    {\large Thesis written}\\
%    {\large under the supervision of}\\[0.1cm]
    {\large The thesis was written under the supervision of}\\[0.1cm]
    {\large Dr hab. Jakub Wagner}\\[0.1 cm]
    {\large National Centre for Nuclear Research}\\
    {\large and Prof. Piotr H. Chankowski}\\
    {\large University of Warsaw, Faculty of Physics}\\
    {\large Institute of Theoretical Physics}\\
    {\large Chair of Theory of Hadrons and Leptons}\\[1.5 cm]
%    \end{tabular}
\end{flushright}

\begin{center}
    Warsaw, July 2021
\end{center}

\newpage

\begin{center}
    {\large \textbf{Summary}}\\
\end{center} 
Generalized Parton Distributions (GPDs) carry information on the internal structure of 
hadrons such as the angular momentum of quarks and gluons, or their spacelike distribution. 
They can be experimentally studied in exclusive experiments with hadrons, 
i.e. processes in which all initial and final states are measured.
The tool that creates the necessary bridge between the theoretical predictions and the 
experiments is the collinear factorization. It allows disentangling perturbatively computable parts of 
the amplitude, which describes interactions of quark and gluons with the external particles, from the 
non-perturbative quantities, which are identified as GPDs.

In this work, I extend the theoretical analysis of the process photoproduction of photon pairs on
a proton to the next-to-leading order in perturbative
Quantum Chromodynamics within the framework of collinear factorization.
I give the proof that all collinear divergences which arise in one-loop
computations cancel at the level of the amplitude. 
This result enlarges the family of reactions, which can be studied using the
collinear factorization by processes of the type $2 \rightarrow 3$, which
have not been previously studied within this theoretical framework
beyond the leading QCD order.
Furthermore, I compute the full form of the amplitude of the discussed process
at the one-loop order. That improves the accuracy of theoretical predictions for this experiment, which may be used for planning
future experiments in JLAB or EIC.\\[1 cm]
    
\begin{center}
    %{\large \textbf{Area of study}}\\[0.2cm] 
    %13.2 Physics\\[1 cm]
    %{\large \textbf{Theme classification}}\\[0.2cm]
    %Hadron physics \\[1 cm]
    {\large \textbf{The title of the thesis in Polish}}\\[0.2  cm] 
    {\Large Faktoryzacja w twardych procesach ekskluzywnych.
     Poprawki jednop\c etlowe do fotoprodukcji par foton\'ow na protonie.}
\end{center}
\newpage
\newgeometry{tmargin=2cm, bmargin=3cm, lmargin=2cm, rmargin=2cm}
\pagenumbering{arabic}
\tableofcontents

\newpage
\chapter{Introduction}
As far as we know, hadrons are the smallest composite objects in the
Universe. The first indication that nucleons, unlike electrons, are
not pointlike particles came from the measurement of the magnetic moment
of the proton performed by Stern and collaborators \cite{Stern} already
in 1933. More than twenty years later, analysis of elastic scattering
of several hundreds MeV electrons off nuclei allowed R. Hofstadter 
to determine the electric charge distributions in protons and neutrons
\cite{Hof} (the charge r.m.s. radius of the proton is now estimated to
be approximately $0.87\times10^{-15}$~m \cite{proton-radius}). Finally, at
the end of the sixtieth of the XX century, the experiments, in some respect
analogous to the historic one of Rutherford which revealed the existence
of atomic nuclei, with deeply inelastic scattering (the so-called DIS)
of leptons on nucleons performed at Stanford Linear Accelerator Center
(SLAC) \cite{DIS} inspired Feynman to formulate the parton model
\cite{feynman,partons-altarelli-parisi,Wallon,Muta} in which hadrons
probed by highly virtual photons can be treated as composed of almost
noninteracting pointlike constituents. These, called by Feynman partons,
were subsequently identified with quarks, first proposed by Zweig and
Gell-Mann \cite{Gell-Mann,Zweig1}, and spin 1 bosons (gluons)
which are now known to be the elementary constituents of hadrons.

The proper theory of interactions of quarks and gluons is a non-abelian
quantum gauge field theory known as Quantum Chromodynamics (QCD). The
effective coupling constant of this theory becomes large at long distances
(or, equivalenty, at small transfers of momenta) \cite{PolGrossWil}.
Therefore, applications of the perturbative expansion to QCD are limited
to only special classes of high energy processes which are characterized
by the existence of at least one ``hard scale'' $Q$, e.g. of a large
(compared to a typical hadronic mass scale set by $\Lambda_{\rm QCD}$ - the
intrinsic energy scale of QCD) four-momentum transfer $-q^2=Q^2$
($\sqrt{|q^2|}\gg\Lambda_{\rm QCD}$). This makes the study of hadrons, 
their reactions and structure,
a great challenge: rather advanced quantum field theory tools
are required in order to extract the relevant information from QCD.
Already to obtain precise theory predictions for scaling violation in DIS
experiments (i.e. to go beyond the simple Feynman parton picture) one needs
to employ the so-called operator product expansion. It allows in this
case to rigorously split the computed {\it cross section} of the inclusive
process into the part (Wilson coefficients) which is dominated by the
aforementioned hard scale $Q$ and can, owing to this, be computed
perturbatively, and the other part which can only be phenomenologically
parametrized in terms of the parton distribution functions
(PDFs) $f_p(x)$ introduced originally at the intuitive level by Feynman in
his (so-called naive) parton model and which are given the interpretation
\cite{partons-altarelli-parisi,Wallon,Muta,Factorization-Collin-book}
of the probability densities that inside the hadron a parton $p$ of a given
type carries a fraction $x$ of the total hadron four-momentum.
Measurements which confirmed the scaling violation, i.e. the predicted
by QCD changes of the distributions $f_p(x)$ with the changes of the hard
scale $Q$, were important tests of QCD which together with the observation
by the DORIS and Petra experiments in Hamburg of the
three-jet events in $e^-e^+$ collissions essentially established it as a
correct theory of strong interactions.

Amplitudes of more complicated processes, in particular of exclusive
processes in which characteristics of not only the initial but also of
the final state hadron(s) are specified (and measured experimentally),
usually cannot be analysed as rigorously as DIS. Nevertheless, one commonly
believes that {\it amplitudes} of the studied process which are
characterized by a hard scale can still be split into a
perturbatively computable (within QCD) hard part (dominated by the
scale $Q$) and the rest which is nonperturbative and must be treated 
phenomenologically, i.e. parametrized in agreement with the known
symmetries of the underlying theory (QCD), determined experimentally
in one group of experiments and then used to predict results of another
group of experiments. For example, analysis of the data accumulated over
years in DIS-type experiments carried out with the help of the HERA accelerator
 (experiments ZEUS and H1) allowed to obtain parametrizations of
PDFs of the parton model and are now routinely being used to predict
rates of processes studied experimentally in the proton-proton inelastic
collisions in LHC.

The principle on which the possibility of splitting amplitudes of exclusive hadronic processes into perturbative
and nonperturbative parts relies is the so-called {\it factorization} which
operates at the level of amplitudes. It is not as rigorous as the OPE,
the use of which is limited to only a rather narrow class of directly
experimentally measurable processes, but can, at least in the case of
some special processes characterized by a hard scale $Q$, be proven with
a fair degree of rigour (enough
to satisfy most physicists but few pure mathematicians - as my and prof.
Chankowski's hero, A.B. Pippard, would say). Demonstrating
factorization requires, however, a careful analysis of Feynman diagrams
and their singularities in various kinematical regimes, both ultraviolet
(UV), infrared and/or collinear, in order to be able to isolate them and
properly absorb into the nonperturbative part of the amplitude.
As far as exclusive processes with one and the same single hadron
(usually nucleon) in the initial and final state are concerned, the
nonperturbative parts of their amplitudes are parametrized in terms of the 
\textit{Generalized Parton Distributions} (GPDs), which are also known
as Off-Forward Parton Distributions. GPDs which, similarly to PDFs,
describe properties of a given hadron and are, therefore, process-independent,
significantly generalize PDFs (which can be obtained
from the GPDs by taking their forward limit) and as
such contain much more detailed information on the hadron structure, e.g.
on angular momenta of partons of a given kind or on their spacelike
distribution inside the hadron \cite{Schweizer,Osborne,Ji-dvcs,Ji-spin},
 than do the PDFs. Still, the strategy of
their use is similar to that of the latter: they should be extracted
from the data on one class of processes and
applied to predict rates of other classes of processes.

Extraction of GPDs from experimental data is complicated, because cross
sections of processes sensitive to parton distributions are rather small
and because what actually can be extracted are only
convolutions of GPDs with some other, process-dependent, distributions
\cite{Wallon,Ji-dvcs,tcs1}. Hence, to reliably determine generalized
distributions, processes of different kinds must be considered in
conjunction. Moreover, since at energy scales of order few GeV
($|q|^2\sim10$~GeV$^2$) relevant for these processes
\cite{Ji-dvcs,tcs1,GK,cytuj-promotora1}, the strong coupling constant
$\alpha_s$ is
of order $\sim0.1$, higher order corrections must necessarily be taken
into account. One-loop results concerning electroproduction
of a real photon, of photoproduction of lepton pair or electroproduction
of a meson are already available - they have been given in the papers 
\cite{Factorization-Ji}, \cite{tcs-nlo} and \cite{Collins-mesons},
respectively. Recently, the process of photoproduction of two photons
on nucleon has been considered, but only at the leading order
\cite{cytuj-promotora1,cytuj-promotora11}.

The main goal of this work
%\footnote{\bf To sa w\l a\'snie sakramentalne
% s\l owa, kt\'ore MUSZA pa\'s\'c we wstepie ka\.zdej pracy magisterskiej!}
is to extend the analysis of factorization of the amplitude of diphoton
photoproduction on proton, i.e. of the process
$\gamma p\rightarrow \gamma^\prime\gamma^{\prime\prime} p^\prime$, given in
\cite{cytuj-promotora1,cytuj-promotora11}, to the first nontrivial order.
The work consists of two parts. In Chapter \ref{chapter-factorization}
I describe in some details the collinear factorization.
%, which allows to connect
%Generalized Parton Distributions to experimental data extracted from hard
%exclusive experiments, i.e. from processes of the (schematic)
%form $\mathrm{hadron} + A\rightarrow\mathrm{hadron}+B$, with definite
%states $A$ and $B$ which are identified experimentally, and there is at
%least one hard kinematical scale $Q$ such that $|Q^2|\gg m^2$, where $m$
%is the mass of the hadron. In such a case, it is possible
%to split the amplitude of the process into two parts. The first one, which
%can be represented by Feynman diagrams and computed perturbatively,
%accounts for the interaction of partons with other pointlike particles.
%The second one, which cannot be computed perturbatively, is given by a GPD.
%The possibility of spliting the amplitude relies on the so-called
%factorization principle which is relatively easy to demonstrate in
%the lowest order approximation in the case of processes characterized by a
%hard scale $Q$. 
In particular I explain how factorization should be handled in higher
orders, when the perturbatively computed part of the amplitude develops
in addition to UV also infrared divergences resulting from diagrams
containing loops with propagators of partons treated as massless quanta.
It will be seen that if factorization is to hold also in higher orders,
the infrared (collinear) divergences must cancel against the ones present in 
parton distributions defined as (Fourier transforms - see below - of) matrix
elements between hadron states of appropriately defined renormalized nonlocal
composite operators. Parton distributions defined in this way are
frequently called, somewhat misleadingly, renormalized
distributions.\footnote{Parton distributions defined in terms of
  unrenormalized operators are accordingly and also typically misleadingly
  %(the speciality of a multitude of idiots cultivating theoretical physics),
  called bare parton distributions. To be
  sure: whether defined in the leading order or in higher orders they are
  extracted from the same measured quantities. It it thus only their theoretical
  definition which is given either in terms of bare or in terms of renormalized
  operators; the GPDs are always the same quantities.}
If such a cancellation does indeed hold, the amplitude
can still be split (i.e. factorized) into the perturbative (hard)
part computable with the help of renormalized perturbative expansion and the
nonperturbative one parametrized now in terms of the renormalized GPDs.
Since this cancellation is not obvious (and there are known
situations, in which it does not obtain \cite{Factorization-not-working}),
it must be checked for each process independently. Proofs that 
factorization holds in the leading order in $|Q|^{-1}$ and up to an
arbitrary order in perturbative QCD can be carried out only for a limited
family of processes. Since renormalization of GPDs
(i.e. the construction of renormalized operators in terms of
which GPDs must be defined in higher orders) is more complicated than,
for example, renormalization of conventional $S$-matrix elements or
Green's functions of elementary or local composite operators (here the
renormalized operators must be defined in a gauge invariant way, i.e. must
be supplemented by the Wilson gauge link operator),  I provide in Section
\ref{section-renormalization} a detailed demonstration of the necessary
renormalization procedure at the one-loop level, and derive the resulting
evolution equations (analogous to the ones governing the scale dependence
of the ordinary PDFs)
satisfied by the resulting generalized parton densities.

The second part of the work is devoted to the analysis in the
next-to-leading order of photoproduction of photon pairs with large
invariant mass (which sets the hard scale $Q$) on proton. This
process is especially interesting from the point of view of the
theoretical analysis, because it is the simplest one of all
$2\rightarrow 3$ processes that can be studied within the framework
of collinear factorization discussed in the first part of the
work. The corresponding
amplitude is sensitive only to combinations
of partons distributions, which are odd with respect to the charge
conjugation. That provides the additional source of information on GPDs.
The leading-order analysis \cite{cytuj-promotora1} shows,
that this process can in principle be studied experimentally at the
JLab facilities which provide intense beams of quasi-real photons
and in the planned Electron-Ion Collider (EIC). Moreover, some
experimental data pertaining to this process probably already exists, as
exclusive experiments with protons and highly energetic quasi-real photons
have already been performed to measure another process, the Timelike
Compton Scattering. In Chapter \ref{chapter-diphoton} I present a detailed
computation of one loop corrections to the amplitude of the
$\gamma~\! p\rightarrow\gamma^\prime\gamma^{\prime\prime}p^\prime$ process.
I give a proof of the validity of the collinear factorization in it at
NLO, compute the finite part of the amplitude and the
differential cross section of the process. I also introduce some techniques which may, I believe, prove useful also in
theoretical studies of other processes. 
\vskip0.2cm
 
Before proceeding with the main part of the work, it is convenient to define
the notation used throughout it. I will introduce here also the definitions
of the Generalized Parton Distributions since, in my opinion, it is better
to do it \textit{before} explaining how they arise in actual computations
of amplitudes of specific physical processes. 

In defining GPDs it is customary to introduce the so-called Sudakov
decomposition of four-vectors $v^\mu$ defined in the following way:
\begin{equation}
v^\mu = v^+ n^\mu_+ + v^- n^\mu_- + v^\mu_\perp,
\end{equation}
where 
\begin{equation}
n^\mu_\pm\equiv\frac{1}{\sqrt{2}}[1,0,0,\pm 1]^\mu,
\end{equation}
and $v^\mu_\perp$ denotes the spatial part of the vector projected onto
the $xy$ plane. In the analogous way one introduces the lightcone
basis of the gamma matrices:
\begin{equation}
\gamma^\pm\equiv\frac{1}{\sqrt{2}}\big(\gamma^0\pm\gamma^3\big).
\end{equation}

Consider now two states $|N(\mathbf{p}_1,\sigma_1)\rangle$ and
$|N(\mathbf{p}_2,\sigma_2)\rangle$ of the same hadron $N$ with the
four-momenta $p_1^\mu=(E_1,\mathbf{p}_1)$ and
$p_2^\mu=(E_2,\mathbf{p}_2)$ ($E_i=\sqrt{m^2_N+\mathbf{p}_i^2}$, $i=1,2$,
where $m_N$ is the mass of the hadron), and $\sigma_i$ its spin
projection. The "$+$" components $p_i^+=(E_i+p_i^z)/\sqrt2$, $i=1,2$
of the four-momenta $p_1$ and $p_2$ will be parametrized with the help
of two parameters:  $\xi$ ($-1<\xi<1$) and $p^+\geq0$ defined by 
\begin{equation}
  p_1^+=(1+\xi)~\!p^+, \quad p_2^+=(1-\xi)~\!p^+.
\end{equation}
It follows that
$p^+=(p_1^++p_2^+)/2\geq0$ and $\xi=(p_1^+-p_2^+)/(p_1^++p_2^+)$. 
We assume that both hadrons move nearly with the speed of
light in the direction of the $z$-axis, so that $p^+\gg m_N$.

Generalized Parton Distribution can be defined by considering in
the limit $p^+\rightarrow\infty$ of the following Fourier transforms
of matrix elements of non-local composite operators:
\begin{eqnarray}
&&\int\!\frac{dz^-}{2\pi}~\!e^{ixp^+ z^-}\left.\!\!\langle N(\mathbf{p}_2, \sigma_2)|
\bar\psi_q(-z/2) [-z/2, z/2]\gamma^+\psi_q(z/2)|N(\mathbf{p}_1, \sigma_1)\rangle
\right|_{z^+,z_\perp=0},\label{eq-gpd-q1} \\
&&\int\!\frac{dz^-}{2\pi}~\!e^{ixp^+ z^-}\left.\!\!\langle N(\mathbf{p}_2, \sigma_2)|
\bar\psi_q(-z/2)[-z/2, z/2]\gamma^+\gamma_5~\!\psi_q(z/2)|N(\mathbf{p}_1, \sigma_1)\rangle
\right|_{z^+,z_\perp =0},\phantom{aa}\label{eq-gpd-q2}\\
&&\int\!\frac{dz^-}{2\pi}~\!e^{ixp^+ z^-}\left.\!\!\langle N(\mathbf{p}_2, \sigma_2)|
G^{+\mu}(-z/2)[-z/2, z/2]G^{\phantom{a}+}_\mu(z/2)|N(\mathbf{p}_1, \sigma_1)\rangle
\right|_{z^+,z_\perp =0}, \label{eq-gpd-g1}\\
&&\int\!\frac{dz^-}{2\pi}~\!e^{ixp^+ z^-}\left.\!\!\langle N(\mathbf{p}_2, \sigma_2)|
G^{+\mu}(-z/2)[-z/2, z/2]\tilde G^{\phantom{a}+}_\mu(z/2)|N(\mathbf{p}_1, \sigma_1)\rangle
\right|_{z^+,z_\perp =0}.\label{eq-gpd-g2}
\end{eqnarray}
$\psi_q$ denotes here the bare (i.e. the Heisenberg picture operator
the scale of which is fixed by the canonical kinetic term in the QCD
lagrangian density) quark field operator
of flavor $q$, $G^{\mu\nu}$ is the bare field strength tensor of the gluon field
and $\Tilde{G}^{\mu\nu}=\frac12\epsilon^{\mu\nu\kappa\lambda}G_{\kappa\lambda}$ the
tensor dual to it; $\epsilon^{\mu\nu\kappa\lambda}$ is the completely antisymmetric
tensor (I use $\epsilon^{0123}=+1$). The matrix $\gamma_5$ is defined as
$\gamma_5=\gamma^5=i\gamma^0\gamma^1\gamma^2\gamma^3$. Strictly speaking,
in the leading order the operator denoted in (\ref{eq-gpd-q1}-\ref{eq-gpd-g2})
$[-z/2, z/2]$, called the gauge link is replaced by the unit operator.
It must be included in higher orders (together with
appropriate operator counterterms which complete the
definitions of the renormalized nonlocal operators) and is
defined in the following way:
\begin{equation}
[z_1, z_2]\equiv P\exp\Big(ig\!\int^{z_1}_{z_2}\!d\xi_\mu\mathbb{A}^\mu(\xi)\Big).
\end{equation}
The symbol $P$ denotes here the path ordering and 
\begin{equation}
\mathbb{A}^\mu\equiv \sum_a t^a A^{\mu, a},
\end{equation}
where $t^a$ are the generators (taken in the fundamental representation)
of the $SU(3)$ gauge group; $a=1,\dots8$ is the adjoint color index.
Presence of this operator ensures gauge invariance of GPDs.
How it arises in actual calculations is explained in Section \ref{Subsection-gauge-link}.
In the literature, see e.g. \cite{string1,string2} products of two operators
separated by a light-like distance and the gauge link, as in the matrix elements
(\ref{eq-gpd-q1}-\ref{eq-gpd-g2}), are often called \textit{string operators}.

In most applications the hadron $N$ is a spin-$1/2$ particle
(proton or neutron). In such cases the most general forms of
the Fourier transforms \eqref{eq-gpd-q1}-\eqref{eq-gpd-g2} can be,
taking into account Lorentz invariance, parity, time reversal and charge
conjugation symmetries of the underlying theory, parametrized
in terms of the functions $H^{q,g}$, $\tilde H^{q,g}$, $E^{q,g}$ and
$\tilde E^{q,g}$ (their definitions given here agree with those used
in \cite{Diehl}) called \textit{generalized parton distributions}
($\Delta^\kappa\equiv(p_2-p_1)^\kappa$):
\begin{align}
&\eqref{eq-gpd-q1}=\frac{1}{p^+}\Big[H^q(x,\xi,t)~\!\bar u(\mathbf{p}_2,\sigma_2)\gamma^+
u(\mathbf{p}_1,\sigma_1)+E^q(x,\xi,t)~\!\bar u(\mathbf{p}_2,\sigma_2)~\!
\frac{i\sigma^{+\kappa}\Delta_\kappa}{2m}~\!u(\mathbf{p}_1, \sigma_1)\Big],
  \label{eq-gpd-q1GPD}\\
&\eqref{eq-gpd-q2}=\frac{1}{p^+}\Big[\Tilde{H}^q(x,\xi,t)~\!\bar u(\mathbf{p}_2,\sigma_2)
\gamma^+\gamma_5~\!u(\mathbf{p}_1,\sigma_1)+\Tilde{E}^q(x,\xi,t)
\bar u(\mathbf{p}_2,\sigma_2)~\!\frac{\gamma_5\Delta^+}{2m}~\!
u(\mathbf{p}_1,\sigma_1)\Big], \label{eq-gpd-q2GPD}\\
&\eqref{eq-gpd-g1}=\frac{1}{2}\Big[H^g(x,\xi,t)~\!\bar u(\mathbf{p}_2,\sigma_2)
\gamma^+ u(\mathbf{p}_1,\sigma_1)+E^g (x,\xi,t)~\!\bar u(\mathbf{p}_2,\sigma_2)~\!
\frac{i\sigma^{+\kappa}\Delta_\kappa}{2m}~\! u(\mathbf{p}_1,\sigma_1)\Big],
  \label{eq-gpd-g1GPD}\\
&\eqref{eq-gpd-g2}=\frac{i}{2}\Big[\Tilde{H}^q(x,\xi,t)~\!\bar u(\mathbf{p}_2,\sigma_2)
\gamma^+\gamma_5~\! u(\mathbf{p}_1,\sigma_1) +\Tilde{E}^q (x,\xi,t)~\!
\bar u(\mathbf{p}_2,\sigma_2)~\!\frac{\gamma_5\Delta^+}{2m}~\!u(\mathbf{p}_1,\sigma_1)\Big].
  \label{eq-gpd-g2GPD}
\end{align}
Owing to the Lorentz invariance, the distributions $H$, $\Tilde{H}$, $E$ and
$\Tilde{E}$ depend on only 3 variables: $x$, $\xi$ and $t=\Delta^2$;
moreover, as can be shown \cite{Time-ordering},  $-1<x<1$.

As already said, beyond the leading order GPDs (called then
\textit{renormalized} Generalized Parton Distributions) must be defined in
the analogous way but as matrix elements of renormalized string operators.
The form of these renormalized operators is established in Section \ref{section-string-renormalization}.
Because of cancellation of divergences, which is an essential element of
factorization, all observables must be then expressed in terms of these
renormalized GPDs.
How the matrix elements (\ref{eq-gpd-q1}-\ref{eq-gpd-g2}), and
therefore the generalized parton distributions, appear in actual computation
of amplitudes of physical processes, and how the generalized parton distributions
vary with changes of the hard scale $Q$ (i.e. what
renormalization group they satisfy)
will be shown in Chapter \ref{chapter-factorization}.

\chapter{The QCD factorization in hard exclusive processes}
\label{chapter-factorization}

\section{Leading order factorization \\  and generalized parton distributions}
\label{section-fact-lo}

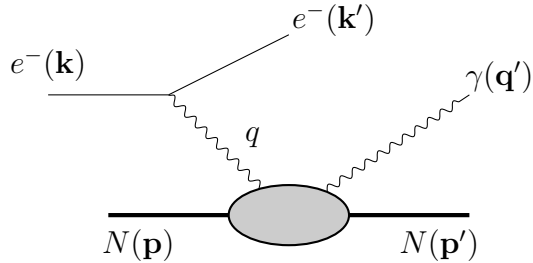
\begin{figure}[H]
\centering
\begin{tikzpicture}[scale = 0.8]
	\draw [ultra thick] (1,0) -- (3,0);
	\node at (1.5, -0.5) {$N(\mathbf{p})$};
	\draw [ultra thick] (5,0) -- (7,0);
	\node at (6.5, -0.5) {$N(\mathbf{p}^\prime)$};
	\draw (0,2) -- (2,2);
	\node at (0, 2.5) {$e^-(\mathbf{k})$};
	\node at (3.4,1.3) {$q$};
	\draw (2,2) -- (4,3);
	\node at (4.75, 3.25) {$e^-(\mathbf{k}^\prime)$};
	\draw [decorate,
  decoration={snake, segment length=2mm, amplitude=0.5mm,post length=1mm}] (2,2) -- (4,0);
	\draw [decorate,
  decoration={snake, segment length=2mm, amplitude=0.5mm,post length=1mm}] (4,0) -- (7,2);
	\node at (7.5,2.3) {$\gamma(\mathbf{q}^\prime)$};
	\filldraw[color=black, fill=gray!40, thick] (4,0) ellipse (1 and 0.5);
\end{tikzpicture}
\caption{DVCS as a subprocess of the photon electroproduction on nucleon.
  $k$ and $k^\prime$ are the four-momenta of the initial and final electron.
  The hard scale is in this case set by the virtuality of the
  exchanged photon $-q^2=-(k-k^\prime)^2$. The gray blob is the part of the
  amplitude which cannot be computed perturbatively. Another subprocess
  contributing to the same photon electroproduction on nucleon
  is the Bethe-Heitler subprocess in which
  nucleon absorbs only the virtual photon and the final state photon is
  emitted from one of the electron lines and, therefore, the ``hadronic part''
  of the amplitude is parametrized in term of well-known elastic
  nucleon form-factors \cite{Hof}.}
\label{fig-eN-egammaN}
\end{figure}

One of the simplest processes allowing to extract GPDs from the experimental
data is the so-called Deeply Virtual Compton Scattering (DVCS) \cite{Ji-dvcs}
in which a nucleon $N$ of mass $m_N$ absorbs a highly virtual photon carrying
the four-momentum $q$ ($-q^2\equiv Q^2\gg m^2_N$) and subsequently emits a
real photon:
$\gamma^\ast(q)N(\mathbf{p})\rightarrow\gamma(\mathbf{q}^\prime)N(\mathbf{p}^\prime)$.
This process can be experimentally studied as a sub-process shown
in Fig. \ref{fig-eN-egammaN} of the photon electroproduction that
is, in the reaction $e^-N\rightarrow e^-\gamma N$.

Another process sensitive to the generalized parton distributions
\cite{tcs1} is the Timelike Compton Scattering (TCS),
$\gamma(\mathbf{q})N(\mathbf{p})\rightarrow\gamma^\ast(q^\prime)N(\mathbf{p}^\prime)$,
which
can be regarded as a ``reversed'' DVCS: the nucleon absorbs a real photon and
emits a virtual photon, which subsequently produces a lepton pair. Both DVCS
and TCS are the limiting cases of a more general process called Double Deeply
Virtual Compton Scattering (DDVCS), which can be represented as
$\gamma^\ast(q) N(\mathbf{p})\rightarrow\gamma^\ast(q^\prime)N(\mathbf{p}^\prime)$,
in which at least one photon is far off-shell, so that $|q^2|+|q^{\prime2}|\gg m^2_N$.
An analysis of DDVCS at the next-to-leading order, together with the proof of
factorization of its amplitude
at the 1-loop level, has been presented in \cite{cytuj-promotora2}.

Before considering the process
$\gamma(q) N(\mathbf{p})\rightarrow\gamma(\mathbf{q}^\prime)
\gamma(\mathbf{q}^{\prime\prime})N(\mathbf{p}^\prime)$ which is the main goal of
this work, let us discuss factorization of the photon electroproduction
amplitude starting from the basic principles of quantum field theory.
In view of the fact that hadrons are not ``quanta'' of elementary
QCD field operators, the $S$-matrix element
\begin{eqnarray}
  \langle\left(N(\mathbf{p}^\prime,\sigma^\prime) e^-(\mathbf{k}^\prime)
  \gamma(\mathbf{q}^\prime)\right)_{\rm out}|\left(N(\mathbf{p},\sigma)e^-(\mathbf{k})
  \right)_{\rm in}\rangle~\!,\label{eq-S-mat-el}
\end{eqnarray}
(the subscripts ``in''
and ``out'' denote the in and out asymptotic eigenstates of the QCD+QED
Hamiltonian) corresponding to the experimentally realizable process
$N(\mathbf{p})e^-(\mathbf{k})\rightarrow N(\mathbf{p}^\prime)e^-(\mathbf{k}^\prime)
\gamma(\mathbf{q}^\prime)$
%$$S=\braket{\big(\big)_-|\big( N(p,\sigma_1) e^-(k) \big)_+}$$
must be extracted from the (connected part of the) vacuum Green's function
\begin{eqnarray}
  \langle\Omega_{\rm out}|T\!\left[O(y^\prime)O^\dagger(y)\psi_{(e)}(w^\prime)
    \bar\psi_{(e)}(w)A^\mu(x^\prime)\right]\!|\Omega_{\rm in}\rangle~\!,\nonumber
\end{eqnarray}
where $O(y^\prime)$ is an appropriately constructed composite Heisenberg picture
operator ($\psi_{(e)}$ and $A^\mu$ are the electron and photon Heisenberg
picture elementary field operators)  using the Lehmann-Symanzik-Zimmermann
reduction formula \cite{Chank-LSZ}. If $N$ stands for proton, as
$O$ one can take e.g. the composite operator
transforming as the spin $1/2$ representation of the Lorentz group
constructed out of three quark (Heisenberg picture) field operators
$\varepsilon_{ijk}(C^{-1})^{\beta^\prime\beta}
\psi^{(u)i}_{\beta^\prime}\psi^{(d)j}_\beta\psi^{(u)k}_\alpha$,
where $C^{-1}$ is the charge conjugation matrix (which makes a Lorentz
singlet out of two quark fields) and $\varepsilon_{ijk}$ makes the operator
a colour singlet. Because the quantum numbers match, the 
operator $O(y^\prime)$ certainly has a nonzero matrix element
between the vacuum state $|\Omega\rangle$ and the (in or out)
state $|N(\mathbf{p},\sigma)\rangle$ (an eigenstate of the complete Hamiltonian
of QCD+QED) which by symmetry principles must take the form
\begin{equation}
  \langle\Omega_{\rm out}|O^H_\alpha(y^\prime)|(N(\mathbf{p}^\prime,\sigma^\prime)_{\rm out}\rangle
   =\mathcal{Z}_O^{1/2} e^{-ip^\prime\cdot y^\prime} u_\alpha(\mathbf{p}^\prime,\sigma^\prime),
\end{equation}
in which $u_\alpha(\mathbf{p}^\prime, \sigma^\prime)$ is the ordinary spinor
and $\mathcal{Z}_O$ is some non-zero constant (which is the property of the
operator $O$). However
since the electromagnetic interaction is weak, it is sufficient to
consider the amplitude of the process
$N(\mathbf{p})e^-(\mathbf{k})\rightarrow N(\mathbf{p}^\prime)e^-(\mathbf{k}^\prime)
\gamma(\mathbf{q}^\prime)$
in the lowest possible order in the coupling $e$.  This allows to write it
as a sum of two terms: one the main part of which is the amplitude of the DVCS
subprocess
and another one corresponding to the Bethe-Heitler subprocess. The latter
is well-known (it can be written in terms of the elastic nucleon form-factors)
and will not be discussed here. Thus in this approximation\footnote{As usually
we consider only connected parts of $S$-matrix elements and of the
  corresponding Green's functions; this will not be indicated explicitly.}
\begin{eqnarray}
  \langle\left(N(\mathbf{p}^\prime,\sigma^\prime) e^-(\mathbf{k}^\prime)
  \gamma(\mathbf{q}^\prime)\right)_{\rm out}|\left(N(\mathbf{p},\sigma)e^-(\mathbf{k})
  \right)_{\rm in}\rangle=(2\pi)^4\delta^4(q^\prime+k^\prime+p^\prime-k-p)
  \phantom{aaaaaaaaaaaaaaaa}\nonumber\\
  \times\left\{\epsilon^*_\nu(\mathbf{q}^\prime)~\!(-ie)~\!
  W^{\mu\nu}(\mathbf{p}^\prime,\mathbf{p};q^\prime,q)~\!(-ie)~\!{-i\over q^2}
 ~\!ie~\! [\bar{u}(\mathbf{k}^\prime,s^\prime) \gamma_\mu u(\mathbf{k},s)]
   -i{\cal A}_{\rm BeHe}\right\},\nonumber
\end{eqnarray}
where ${\cal A}_{\rm BeHe}$ stands for the invariant amplitude of the Bethe-Heitler
subprocess, $\epsilon^\ast_\nu(\mathbf{q}^\prime)$ is the polarization vector
of the outgoing photon, $-i/q^2$ is the virtual photon propagator
(marked $q$ in Fig. \ref{fig-eN-egammaN}) and the tensor
$W^{\mu\nu}(\mathbf{p}^\prime,\mathbf{p};q^\prime,q)$ is defined by
the double Fourier transform 
\begin{eqnarray}
  (2\pi)^4\delta^{(4)}(q^\prime+p^\prime-q-p)~\!
  W^{\mu\nu}(\mathbf{p}^\prime,\mathbf{p};q^\prime,q)
  =\int\!d^4x^\prime d^4x~\!e^{iq^\prime\cdot x^\prime}e^{-iq\cdot x}~\!
  W^{\mu\nu}(\mathbf{p}^\prime,\mathbf{p};x^\prime,x)~\!,
  \label{eq-W-munu-Mom}
\end{eqnarray}
of the matrix element
\begin{eqnarray}
    W^{\mu\nu}(\mathbf{p}^\prime,\mathbf{p};x^\prime,x)=
    \langle(N(\mathbf{p}^\prime,\sigma^\prime))_{\rm out}|
    T J^\mu_H (x) J^\nu_H (x^\prime) |(N(\mathbf{p},\sigma))_{\rm in}\rangle~\!,
    \label{eq-W-munu-Pos}
\end{eqnarray}
of the chronological product of two (Heisenberg picture) operators $J_H^\mu$
of the hadronic electromagnetic current 
\begin{equation}
J_H^\mu(x)=\sum_f e_f \bar{\psi}^H_{(f)}(x)\gamma^\mu\psi^H_{(f)}(x)~\!,
\end{equation}
where the sum is over all light quark flavours $f$ and $e_f$ are the
quark electric charges (in units of $e>0$). The matrix element
(\ref{eq-W-munu-Pos}) can be extracted using the LSZ prescription
\cite{Chank-LSZ} from the appropriate vacuum Green's function
similarly as the $S$-matrix element (\ref{eq-S-mat-el}), e.g. as
\begin{eqnarray}
W^{\mu\nu}(\mathbf{p}^\prime,\mathbf{p};x^\prime,x)=
  \left(i\mathcal{Z}_O^{-1/2}\right)^2\!\!
  \int\!d^4y^\prime d^4y~\!e^{ip^\prime\cdot y^\prime}e^{-ip\cdot y}~\!
  \bar{u}(\mathbf{p}^\prime,\sigma^\prime)~\!
  \phantom{aaaaaaaaaaaaaaaaaaaaaaaaaaaaa}\label{eq-LSZ-OOJJ}\\
  (i\overset{\rightarrow}{\slashed{\partial}}_{y^\prime}-m_N)
  \langle\Omega_{\rm out}|T J^\mu_H (x) J^\nu_H (x^\prime)O_H(y^\prime) \bar{O}_H(y)|
  \Omega_{\rm in}\rangle
     (i\overset{\leftarrow}{\slashed{\partial}}_y
    + m_N)u(\mathbf{p},\sigma)~\!.\nonumber
    \end{eqnarray}
The application of the Gell-Mann - Low formula\footnote{It is interesting
  to note that the Gell-Mann - Low formula can be directly applied here
  owing to nonzero quark masses which are sources of {\it explicit}
  chiral symmetry breaking which removes degeneracy of a continuum
  of QCD vacua which, if quark masses were vanishing, would exist making
  impossible adiabatic reaching the full QCD vacuum from the free one.}
(see e.g. \cite{Chank-LSZ}) allows in principle to compute {\it vacuum} matrix
elements like the one in (\ref{eq-LSZ-OOJJ})
perturbatively expressing them through the chronological products of
the interaction picture operators according to the standard rule
\begin{eqnarray}
\langle\Omega_{\rm out}|T J^\mu_H(x) J^\nu_H(x') O_H(y^\prime) \bar{O}_H(y)
|\Omega_{\rm in}\rangle\nonumber\phantom{aaaaaaaaaaaaaaaaaaaaaaaaaaaaaaaa}\\
= \langle\Omega_0|T J^\mu_I(x) J^\nu_I(x')  O_I(y^\prime)\bar{O}_I(y)
  \exp\!\left(-i\!\!\int\!d^4w~\!\mathcal{H}^{\rm int}_I(w)\!\right)\!|\Omega_0\rangle~\!,
  \label{eq-Gell-Mann-Low}
\end{eqnarray}
in which
$\mathcal{H}^{\rm int}_I(w)$ is the interaction hamiltonian density of QCD
expressed in terms of the interaction picture quark and gluon field
operators and the operators $O_I$ and $J^\mu_I$ in this picture are obtained
by simply replacing in their Heisenberg counterparts the elementary
Heisenberg picture operators by the interaction picture ones. The Green's
function can be therefore computed perturbatively by expanding the exponens and
using the Wick theorem. Because of the large value of the QCD coupling $g_s$,
it is clear, however, that the expansion obtained in this way
cannot be truncated to its few
first terms only and for this reason the entire prescription for computing
$W^{\mu\nu}(\mathbf{p}^\prime,\mathbf{p};q^\prime,q)$ remains rather formal.
To make some progress, one has therefore to reorganize the formal
perturbative expansion obtained from (\ref{eq-Gell-Mann-Low}) in order to
pick up this part of it which through which the large momentum transfer
$q$ flows and which, owing to this, can be reliably computed evaluating
only a few subdiagrams (as a small value of the QCD coupling can be
used for this purpose), and the rest which can be identified as giving
the expansion of matrix elements analogous to the left
hand side of (\ref{eq-Gell-Mann-Low}), but with the electromagnetic
currents replaced by some other operators. In other words one wants
to represent the right hand side of the formula (\ref{eq-W-munu-Mom})
defining the tensor $W^{\mu\nu}(\mathbf{p}^\prime,\mathbf{p};q^\prime,q)$
by the sum of convolutions of  ``hard''
and ``nonperturbative'' parts of the schematic form (possible contractions
of spinor and/or Lorentz indices other than ${\mu\nu}$ between the
amplitudes $\mathcal{T}^{\mu\nu}_{W}$ and $W$ are not displayed)
\begin{eqnarray}
  (2\pi)^4\delta^{(4)}(q^\prime+p^\prime-q-p)~\!
  W^{\mu\nu}(\mathbf{p}^\prime,\mathbf{p};q^\prime,q)
  \phantom{aaaaaaaaaaaaaaaaaaaaaaaaaaaaaaaaaaaaaaaaaa}\nonumber\\
  =\sum_{W}\!\int\!{d^4k_1\over(2\pi)^4}\dots{d^4k_n\over(2\pi)^4}~\!
  (2\pi)^4\delta^{(4)}(q^\prime-q-k_n-...-k_1)~\!
  \mathcal{T}^{\mu\nu}_{W}(q^\prime,q,k_n,\dots,k_1)\nonumber\\
  \times(2\pi)^4\delta^{(4)}(p^\prime+k_1+...+k_n-p)~\!
  W(k_n,\dots,k_1;\mathbf{p}^\prime,\mathbf{p})~\!,\phantom{aaaaa}
  \label{eqn:SumOfConvolutions}
  \end{eqnarray}
where the expressions in the second line are Fourier transforms of amplitudes
defined similarly as (\ref{eq-W-munu-Pos}) but 
with  the product of two electromagnetic currents replaced 
by various products of different numbers of other operators.
One such amplitude (in fact the most important one in the leading
order) is
\begin{eqnarray}
  W_{(f)\beta\alpha}(\mathbf{p}^\prime,\mathbf{p};z)=
  \langle(N(\mathbf{p}^\prime,\sigma^\prime))_{\rm out}|
  T\bar\psi^H_{(f)\beta}(-z/2)~\!\psi^H_{(f)\alpha}(z/2)|(N(\mathbf{p},\sigma))_{\rm in}
  \rangle~\!.\label{eq-W-barpsipsi}
\end{eqnarray}
It too, similarly as the amplitude (\ref{eq-LSZ-OOJJ}),
can formally be represented by a perturbative expansion by applying
the Gell-Mann - Low theorem to the appropriate vacuum matrix element.
The contribution of (\ref{eq-W-barpsipsi})
to $W^{\mu\nu}(\mathbf{p}^\prime,\mathbf{p};q^\prime,q)$
can be written in the form
\begin{eqnarray}
  W^{\mu\nu}(\mathbf{p}^\prime,\mathbf{p};q^\prime,q)
  =\int\!{d^4k\over(2\pi)^4}~\!\mathcal{T}^{\mu\nu}_{\beta\alpha}(q^\prime,q,k)
  \int\!d^4z~\!e^{ik\cdot z/2 + i(k+q-q^\prime)\cdot z/2}~\!  
  W_{(f)\beta\alpha}(\mathbf{p}^\prime,\mathbf{p};z),\label{eq-T-W-pre-factorized}
\end{eqnarray}
To see this, let's write explicitly the corresponding term in the
sum (\ref{eqn:SumOfConvolutions}) in the form (see Fig. \ref{fig-dwa-jajeczka}):
%%%%%%%%%%%%%%%%%%%%%%%%%%%%%%%%%%%%%%%%%%%%%%%%%%%%%%%%%%%%%
%% FIG 2.2: DWA JAJECZKA
%%%%%%%%%%%%%%%%%%%%%%%%%%%%%%%%%%%%%%%%%%%%%%%%%%%%%%%%%%%%%
\begin{figure}[H]
\centering
\begin{tikzpicture}
\draw (-1, 0.5) -- (-1, 2.5);
\draw (1, 0.5) -- (1, 2.5);
\draw [double, thick] (-1.7, 3.2) -- (-1, 2.5);
\draw [double, thick] (1.7, 3.2) -- (1, 2.5);
\draw [->, >=stealth] (-1.7, 3.5) -- (-1.2, 3);
\draw [<-, >=stealth] (1.7, 3.5) -- (1.2, 3);
\node at (-1, 3.5) {$q$};
\node at (1, 3.5) {$q'$};
\draw [very thick] (-2, 0.5) -- (2, 0.5);
\node at (-3, 0.5) {$N(\mathbf{p}, \sigma_1)$};
\node at (3, 0.5) {$N(\mathbf{p}', \sigma_2)$};
\draw  [->, >=stealth] (-1.3, 1) -- (-1.3,2);
\node at (-1.8, 1.5) {$k_1$};
\draw  [->, >=stealth] (1.3, 2) -- (1.3,1);
\node at (1.8, 1.5) {$k_2$};
\filldraw[color=black, fill=gray!40, thick] (0,0.5) ellipse (1.5 and 0.5);
\filldraw[color=black, fill=gray!40, thick] (0,2.5) ellipse (1.5 and 0.5);
\node at (0, 0.5) {\tiny{Nonperturbative part}};
\node at (0, 2.5) {\tiny{Hard part}};
\end{tikzpicture}
\caption{Graphical representation of the
  contribution \eqref{eq-T-W-pre-factorized} to the amplitude
  $W^{\mu\nu}(\mathbf{p}^\prime,\mathbf{p};q^\prime,q)$:
  a quark of flavour $f$ and four-momentum $k_1$ emerges
  from the nonperturbative part, interacts absorbing a large momentum
  transfer $q$ (the ``hard'' part of the amplitude) and reimmerses
  in the nonperturbative part with four-momentum $k_2=k_1+q-q^\prime$.}
\label{fig-dwa-jajeczka}
\end{figure}
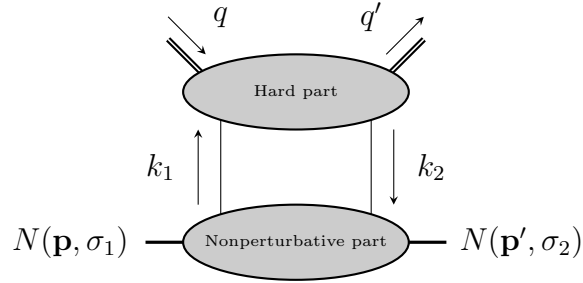
\begin{eqnarray}
  \int\!{d^4k_2\over(2\pi)^4}{d^4k_1\over(2\pi)^4}~\!
  (2\pi)^4\delta^{(4)}(q^\prime-q-k_2-k_1)~\!
  \mathcal{T}^{\mu\nu}_{\beta\alpha}(q^\prime,q,k)
  \phantom{aaaaaaaaaaaaaaaaaaaaaaaaa}\nonumber\\
  \int\!d^4z_2~\!d^4z_1~\!e^{ik_2\cdot z_2}~\!e^{ik_1\cdot z_1}~\!
  \langle(N(\mathbf{p}^\prime,\sigma^\prime))_{\rm out}|
  T\bar\psi^H_{(f)\beta}(z_2)~\!\psi^H_{(f)\alpha}(z_1)|(N(\mathbf{p},\sigma))_{\rm in}
  \rangle~\!.\nonumber
\end{eqnarray}
Making the change of variables: $z_2={1\over2}(w-z)$, $z_1={1\over2}(w+z)$
(the Jacobian is $2^{-4}$) and using the Poincar\'e covariance of
the matrix element
\begin{eqnarray}
  \langle(N(\mathbf{p}^\prime,\sigma^\prime))_{\rm out}|
  T\psi^H_{(f)\beta}(z_2)~\!\bar\psi^H_{(f)\alpha}(z_1)|(N(\mathbf{p},\sigma))_{\rm in}
  \rangle\phantom{aaaaaaaaaaaaaaaaaaaaaa}\nonumber\\
  =e^{i(p^\prime-p)\cdot w/2}~\!
  \langle(N(\mathbf{p}^\prime,\sigma^\prime))_{\rm out}|
  T\bar\psi^H_{(f)\beta}(-z/2)~\!\psi^H_{(f)\alpha}(z/2)|(N(\mathbf{p},\sigma))_{\rm in}
  \rangle~\!,\nonumber
\end{eqnarray}
    enables one to take explicitly the integral over $w$ which gives
    $(2\pi)^4\delta^{(4)}(q^\prime+p^\prime-q-p)$ (times $2^4$ which cancels the
    Jacobian) and comparing the result with the definition (\ref{eq-W-munu-Mom})
    one arrives at the contribution (\ref{eq-T-W-pre-factorized})
    to the tensor $W^{\mu\nu}(\mathbf{p}^\prime,\mathbf{p};q^\prime,q)$.
\vskip0.3cm

\begin{figure}[H]
\centering
	\begin{tikzpicture}[scale = 0.7]
		\draw [ultra thick] (-2, 0) -- (2,0);
		\draw [double, thick] (-1.2, 1.2) -- (-1,1);
	\draw [double, thick] (1.2,1.2) -- (1,1);
	\draw [->, >=stealth] (-1.8, 1.8) -- (-1.3, 1.3);
	\node at (-1.2, 1.7) {$q$};
	\node at (1.2, 1.7) {$q'$};
	\draw [->, >=stealth] (1.3, 1.3) -- (1.8, 1.8);
	%\node at (-1.5, 1.6) {$J^\mu(x_1)$};
		%\node at (1.5, 1.6) {$J^\nu(x_2)$};
	\draw (-1,0) -- (-1,1) -- (1,1) -- (1,0);
	\filldraw[color=black, fill=gray!40, thick] (0,0) ellipse (1.5 and 0.5);
	\node at (0, -1.5) {$(a)$};
	\end{tikzpicture}
\qquad
	\begin{tikzpicture}[scale = 0.7]
		\draw [ultra thick] (-2, 0) -- (2,0);
		\draw [double, thick] (-1.2, 1.2) -- (-1,1);
	%\node at (1.5, 1.6) {$J^\mu(x_1)$};
		%\node at (-1.5, 1.6) {$J^\nu(x_2)$};
	\draw [double, thick] (1.2,1.2) -- (1,1);
	\node at (1.2, 1.7) {$q$};
	\node at (-1.2, 1.7) {$q'$};
	\draw [<-, >=stealth] (-1.8, 1.8) -- (-1.3, 1.3);
	\draw [<-, >=stealth] (1.3, 1.3) -- (1.8, 1.8);
	\draw (-1,0) -- (-1,1) -- (1,1) -- (1,0);
	\filldraw[color=black, fill=gray!40, thick] (0,0) ellipse (1.5 and 0.5);
	\node at (0, -1.5) {$(b)$};
	\end{tikzpicture}
\qquad
	\begin{tikzpicture}[scale = 0.7]
		\draw [ultra thick] (-2, 0) -- (2,0);
		\draw [double, thick] (-1.2, 1.2) -- (-1,1);
	\draw [double, thick] (1.2,1.2) -- (1,1);
	%\node at (-1.5, 1.6) {$J^\mu(x_1)$};
	%	\node at (1.5, 1.6) {$J^\nu(x_2)$};
	\draw (-1,0) -- (-1,1) -- (0,0) -- (1,1) -- (1,0);
	\node at (-1.2, 1.7) {$q$};
	\node at (1.2, 1.7) {$q'$};
	\draw [->, >=stealth] (-1.8, 1.8) -- (-1.3, 1.3);
	\draw [->, >=stealth] (1.3, 1.3) -- (1.8, 1.8);
	\filldraw[color=black, fill=gray!40, thick] (0,0) ellipse (1.5 and 0.5);
	\node at (0, -1.5) {$(c)$};
	\end{tikzpicture}
        \caption{Contributions to the DVCS amplitude (to the tensor
          $W^{\mu\nu}(\mathbf{p}^\prime,\mathbf{p};q^\prime,q)$) 
          of the nonperturbative amplitude (\ref{eq-W-barpsipsi})
          combined with the lowest order contributions 
          (diagrams a and b) to the associated perturbative
          amplitude $\mathcal{T}^{\mu\nu}_{\beta\alpha}(q^\prime,q,k)$.        
%          An example of discussed contribution to DVCS with different
%          number of normally ordered quark fields inside the amplitude.
%          The double lines denote insertions of hadronic current operator.
%         The first two graphs, $(a)$ and $(b)$, correspond to the situation,
%          when two quark fields are contracted, and two normally ordered
%          quark fields are contracted with operators $O(y_1)$ and $O(y_2)$
          %          (that is represented by the gray blob).
          Diagram c is the possible contribution of the nonperturbative
          amplitude with four quark fields. Such contributions
%          corresponds to a possible term in the perturbative expansion,
%          in which there are 4 normally ordered quark fields -- such terms
          can be shown to be of a lower order in the hard scale \cite{Muta},
          and are, hence, neglected.}
\label{fig-contr}
\end{figure}
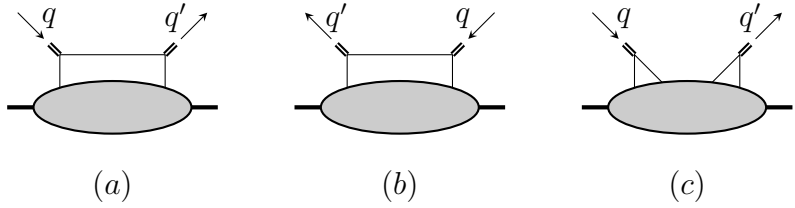

The contribution of the amplitude (\ref{eq-W-barpsipsi}) to the tensor
$W^{\mu\nu}(\mathbf{p}^\prime,\mathbf{p};q^\prime,q)$, taking into account
only the zeroth order (in $g_s$) contributions to
$\mathcal{T}^{\mu\nu}_{\beta\alpha}(q^\prime,q,k)$, is shown in
Figs. \ref{fig-contr}a and \ref{fig-contr}b.
Fig. \ref{fig-contr}c shows one of possible contributions
to $W^{\mu\nu}(\mathbf{p}^\prime,\mathbf{p};q^\prime,q)$ of the amplitude
analogous to (\ref{eq-W-barpsipsi}) but with four quark fields under
the chronological product (and with the lowest order
contribution to the associated amplitude 
$\mathcal{T}^{\mu\nu}_{\bar\alpha\alpha\bar\beta\beta}(k_1,k_2,q,q^\prime)$).
This contribution, which can be interpreted as absorption and emission
by nucleon of  two different quarks, is
subleading in the limit of high $|q^2|$ \cite{Muta}, and will
therefore be neglected in our analysis. 

%%%%%%%%%%%%%%%%%%%%%%%%%%%%%%%%%%%%%%%%%%%%%%%%%%%%%%%%%%%%%
%% FIG 2.3: TRZY BABLE
%%%%%%%%%%%%%%%%%%%%%%%%%%%%%%%%%%%%%%%%%%%%%%%%%%%%%%%%%%%%%

The perturbative amplitudes $\mathcal{T}^{\mu\nu}_W$ can be computed
using the standard Feynman rules and are simply equivalent to
connected Green's functions with amputated external lines. The lowest
order contribution to ${\cal T}^{\mu\nu}_{(f)\beta\alpha}(q^\prime,q,k)$
therefore is (cf. Figs \ref{fig-contr}a and \ref{fig-contr}b):
\begin{eqnarray}
  {\cal T}^{\mu\nu}_{(f)\beta\alpha}(q^\prime,q,k)=
  e_f^2\left[\gamma^\nu~\!{i(\not\!k~\!+\not\!q+m_f)\over
      (k+q)^2-m^2_f+i0}~\!\gamma^\mu
    +\gamma^\mu~\!{i(\not\!k~\!-\not\!q^\prime+m_f)\over
      (k-q^\prime)^2-m^2_f+i0}~\!\gamma^\nu\right]_{\beta\alpha}~\!.
  \label{eq-T-quark}
\end{eqnarray}
It can be viewed as the amplitude of the elastic scattering
of the (highly) virtual photon of four-momentum $q$ on an 
off-shell quark of flavour $f$ and four-momentum $k$.
\vskip0.2cm

In the limit $-q^2\rightarrow\infty$ (physically realized when $-q^2\gg m_N$)
it is convenient to go to the infinite momentum frame, in which the $"+"$
components of the four-momenta $p$, $p^\prime$ and $q$, as well as the $"-"$
component of $q$ are much larger than $m_N$. Recall, that in this frame
the momenta of hadrons are  parametrized in the following way:
\begin{equation}
  p^\mu \approx (1+\xi)~\!p^+ n^\mu_+~\!,
  \qquad p^{\prime\mu} \approx (1-\xi)~\! p^+ n^\mu_+~\!.
\end{equation}
At this point one invokes the basic assumption of the parton model,
namely that in the hard amplitude, that is in the lowest order in
the formula \eqref{eq-T-quark}, the four-momentum of the incoming
quark can be taken in the form $k^\mu = k^+ n_+^\mu$ (see: \cite{Wallon},
page 112, Fig. 5.2). Heuristically one can argue that this quark,
being a constituent of a hadron with large "$+$" component of
its (on-shell) four-momentum $p^\mu$, should also have a large "$+$"
component of the four-momentum $k^\mu$, while all other components
of $k^\mu$ should be much smaller
compared to the components of the $q$ and $q^\prime$ four-momenta
(note, that $q-q^\prime \approx 2\xi p^+ n_+$).
It should, however, be clearly said, that this is a model assumption
which cannot be justified from the first principles. Using this
approximation, the integration over $dk^- d^2 k_\perp$ in the formula 
(\ref{eq-T-W-pre-factorized}) can be explicitly performed
giving the Dirac delta which allows to perform the integration over
$dz^+ d^2 z_\perp$. 
Parametrizing the four-momentum of the quark 'entering' the hard
part using the variable $x$ as $k^+=(x+\xi)p^+$, one
can represent the contribution (\ref{eq-T-W-pre-factorized})
to $W^{\mu\nu}(\mathbf{p}^\prime,\mathbf{p},q^\prime,q)$ in the form
\begin{eqnarray}
  W^{\mu\nu}(\mathbf{p}^\prime,\mathbf{p},q^\prime,q)
  =p^+\!\!\int\!dx~\!\mathcal{T}^{\mu\nu}_{\beta\alpha}(q^\prime,q,k=(x+\xi)p^+n_+)
  \!\int\!{dz^-\over2\pi} ~\!e^{ixp^+ z^-}~\!
  W_{\beta\alpha}(\mathbf{p}^\prime,\mathbf{p},z=z^- n_-)~\!.\phantom{aa}
  \label{eqn:WTcontraction}
  \end{eqnarray}
In order to perform the contraction of the spinor indices in
(\ref{eqn:WTcontraction}) one can write the product $\bar{\psi}_\beta\psi_\alpha$
of field operators in \eqref{eq-W-barpsipsi} using
%Since it is inconvenient to work with expressions with two free spinor
%indices, as in Eqs \eqref{eq-W-barpsipsi} and \eqref{eq-T-quark}, one
%decomposes the product of quark fields from the formula
%\eqref{eq-W-barpsipsi} in the basis of gamma matrices. To this end, we apply
the Fierz decomposition (see for example \cite{Wallon}, p. 33):
\begin{equation}
\bar{\psi}_\beta\psi_\alpha= \frac14 \big(1\big)_{\alpha\beta} \bar{\psi}\psi
+\frac14\big(\gamma_\mu\big)_{\alpha\beta}  \bar{\psi}\gamma^\mu \psi
+\frac14\big(\gamma_5 \gamma_\mu\big)_{\alpha\beta}\bar{\psi}\gamma^\mu\gamma_5\psi
+\frac14\big(\gamma_5\big)_{\alpha\beta}\bar{\psi}\gamma_5\psi
+\frac18\big(\sigma^{\mu\nu}\big)_{\alpha\beta}\bar{\psi}\sigma^{\mu\nu}\psi,
\end{equation}
where $1$ denotes the unit matrix and
$\sigma^{\mu\nu}=\frac{i}{2}\big(\gamma^\mu\gamma^\nu-\gamma^\nu\gamma^\mu\big)$.
If the quark mass $m_f$ in the numerator of \eqref{eq-T-quark} is neglected,
only two structures, the one with $\gamma^\lambda$ and that
with $\gamma^\lambda \gamma_5$, survive after taking the trace.
%A priori there are $16$ possible contributions from all gamma matrices.
%However, contraction of indices $\alpha\beta$ yields  a trace of these
%matrices and the structure from the hard part in Eq. \eqref{eq-T-quark}
%${\cal T}^{\nu\mu}_{(f)\beta\alpha}(q^\prime,q,k)$. It turns out that, 
%if we assume that for massless quarks the only non-vanishing traces are
%those with $\gamma^\lambda$ and $\gamma^\lambda \gamma_5$.
Moreover, Lorentz invariance implies that the expressions
\begin{equation}\label{eq-quarks-preLSZ}
\begin{aligned}
&\!\int\!{dz^-\over2\pi} ~\!e^{ixp^+ z^-}
  \langle N(\mathbf{p}^\prime,\sigma^\prime)_{\rm out}|
  T\bar\psi^H_{(f)}(-z^-n_-/2)\gamma^\lambda ~\!\psi^H_{(f)}(z^-n_-/2)
  |N(\mathbf{p},\sigma)_{\rm in}\rangle~\!, \\
&\!\int\!{dz^-\over2\pi} ~\!e^{ixp^+ z^-}
  \langle N(\mathbf{p}^\prime,\sigma^\prime)_{\rm out}|
  T\bar\psi^H_{(f)}(-z^-n_-/2)\gamma^\lambda \gamma_5 ~\!
  \psi^H_{(f)}(z^-n_-/2)|N(\mathbf{p},\sigma)_{\rm in}\rangle~\!,
\end{aligned}
\end{equation}
can depend on the index $\lambda$ only through $p^\lambda$ and $p^{\prime\lambda}$.
%-- so that, in the Sudakov frame, the dependence on the "$+$" component
%is dominant, while terms corresponding to $\lambda =$ "$-$" or "$\perp$"
%can be neglected. The same argument can be applied for $\gamma^\lambda \gamma_5$.
%Hence, at the leading order 
Therefore, in the Sudakov frame in which the ``tranverse'' ("$\perp$") and "$-$"
components of $p^\lambda$ and $p'^\lambda$ can be neglected compared to the "$+$" one one can approximate (\ref{eqn:WTcontraction})
by
\begin{eqnarray}
W^{\mu\nu}(\mathbf{p}^\prime,\mathbf{p},q^\prime,q)\approx\sum_f\Bigg[
    {p^+\over4}\!\int\!d^4x~\!{\rm tr}\Big[\mathcal{T}^{\mu\nu}(q^\prime,q,(x+\xi)p^+n_+)
      ~\!\gamma^-\Big] \phantom{aaaaaaaaaaaaaaaaaaaaaaaaaaa}
    \label{eqn:FactorizedTwoquarkContrib}\\ 
    \times\!\int\!\frac{dz^-}{2\pi}~\!e^{ixp^+ z^-}\!
    \langle N(\mathbf{p}^\prime,\sigma^\prime)_{\rm out}|
    T\bar\psi^H_{(f)}(-z^-n_-/2)\gamma^+ ~\!\psi^H_{(f)}(z^-n_-/2)
    |N(\mathbf{p},\sigma)_{\rm in}\rangle \nonumber\\
    +{p^+\over4}\!\int\!d^4x~\!{\rm tr}\Big[\mathcal{T}^{\mu\nu}(q^\prime,q,(x+\xi)p^+n_+)
      ~\!\gamma^-\gamma_5 \Big]\phantom{aaaaaaaaaaaaaaaaaaaaaaaa} \nonumber\\
    \times\!\int\!\frac{dz^-}{2\pi}~\!e^{ixp^+ z^-}\!
    \langle N(\mathbf{p}^\prime,\sigma^\prime)_{\rm out}|
    T\bar\psi^H_{(f)}(-z^-n_-/2)\gamma^+\gamma_5 ~\!\psi^H_{(f)}(z^-n_-/2)
    |N(\mathbf{p},\sigma)_{\rm in}
  \rangle\Bigg].\nonumber
\end{eqnarray}
%\begin{equation}
%\begin{aligned}
%&W^{\nu\mu}(\mathbf{p}^\prime,\mathbf{p},q^\prime,q)\approx\sum_f\Bigg[ \\
%    &{p^+\over4}\!\int\!d^4x~\!{\rm tr}\Big[\mathcal{T}^{\nu\mu}(q^\prime,q,(x+\xi)p^+n_+)
%    \gamma^- \Big] \times \\ 
%    &\qquad\int\!\frac{dz^-}{2\pi}~\!e^{ixp^+ z^-}
%    \langle N(\mathbf{p}^\prime,\sigma^\prime)_{\rm out}|
%    T\bar\psi^H_{(f)}(-z^-n_-/2)\gamma^+ ~\!\psi^H_{(f)}(z^-n_-/2)
%    |N(\mathbf{p},\sigma)_{\rm in}  \rangle~\!  \\
% & +{p^+\over4}\!\int\!d^4x~\!{\rm tr}\Big[\mathcal{T}^{\nu\mu}(q^\prime,q,(x+\xi)p^+n_+)
%      \gamma^- \gamma_5 \Big] \times \\
%    & \qquad\int\!\frac{dz^-}{2\pi}~\!e^{ixp^+ z^-}
%    \langle N(\mathbf{p}^\prime,\sigma^\prime)_{\rm out}|
%    T\bar\psi^H_{(f)}(-z^-n_-/2)\gamma^+\gamma_5 ~\!\psi^H_{(f)}(z^-n_-/2)
%    |N(\mathbf{p},\sigma)_{\rm in}  \rangle~\! \Bigg].
%\end{aligned}
%\end{equation}
%Results of the work \cite{Time-ordering} show, that in the case of a
%product of two fields, using the normal ordering {\bf JAKI ZNOWU NORMAL
%  ORDERING HEISENBERGOWSKICH OPERATOR\'OW??? KTO TAK NAPISA\L !?},
%time ordering, or no ordering is equivalent.
The $1/4$ from the Fierz decomposition is conventionally absorbed in the hard part. As shown e.g. in \cite{Time-ordering}, in the kinematics of the Sudakov
frame  in the matrix elements (\ref {eq-quarks-preLSZ}) the symbols $T$ of
the chronological products can be removed.

%%%%%%%%%%%%%%%%%%%%%%%%%%%%%%%%%%%%%%%%%%%%%%%%%%%%%%%%%%%%%
%% FIG 2.4: GLUONS THROUGH THE QUARK LOOP
%%%%%%%%%%%%%%%%%%%%%%%%%%%%%%%%%%%%%%%%%%%%%%%%%%%%%%%%%%%%%
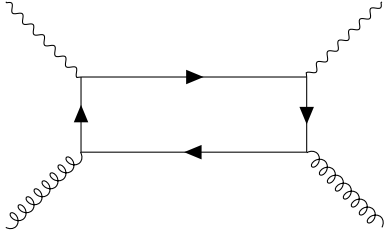
\begin{figure}[H] 
\centering
\begin{tikzpicture}
\begin{feynman}
\vertex (q1) at (0,0);
\vertex (k1) at (1, -1);
\vertex (k2) at (4, -1);
\vertex (k3) at (4, -2);
\vertex (k4) at (1, -2);
\vertex (q2) at (5, 0);
\vertex (g1) at (0, -3);
\vertex (g2) at (5, -3);

\diagram*{
	(k1) -- [fermion] (k2) -- [fermion] (k3) -- [fermion] (k4) -- [fermion] (k1);
	(q1) -- [photon] (k1);
	(q2) -- [photon] (k2);
	(g1) -- [gluon] (k4);
	(g2) -- [gluon] (k3);
};
\end{feynman}
\end{tikzpicture}
\caption{The lowest order (in $g_s$) diagram which
  introduces the dependence of $W^{\nu\mu}(\mathbf{p}^\prime,\mathbf{p},q^\prime,q)$
  on the gluon generalized distributions.}
\label{fig-gluon-loop}
\end{figure}

Higher order (in $g_s$) corrections to the hard amplitude (\ref{eq-T-quark})
will be discussed in section \ref{section-renormalization}.
Since gluons can interact with photons through quark loops, in the
sum (\ref{eqn:SumOfConvolutions}) one has to take into account also
other terms. In particular in order $g_s^2$ there will appear terms
corresponding to the emergence from the nonperturbative part of
two gluons. The order $g_s^2$ contribution to the corresponding
hard amplitude $\mathcal{T}^{\mu\nu}_{\phantom{aa}\rho\kappa}(q^\prime,q,k)$
is shown in Fig. \ref{fig-gluon-loop}. The additional Lorentz structure
of this amplitude (the indices $\rho\kappa$)  can be decomposed
into $g_{\rho\kappa}$ and $\epsilon_{\rho\kappa\sigma\lambda}n_+^\sigma n_-^\lambda$
and the analysis similar to the one discussed above
shows that the corresponding contribution to the
$W^{\mu\nu}(\mathbf{p}^\prime,\mathbf{p},q^\prime,q)$ amplitude takes the
form analogous to (\ref{eqn:FactorizedTwoquarkContrib}) but with the
nonperturbative part represented by the matrix elements
%Generalization of the hard amplitude to the higher QCD orders is
%straightforward, since any contractions, after representing propagators
%in the momentum picture and performing integrals in the position space
%will result in expressions of the form analogous to
%\eqref{eq-T-W-pre-factorized} -- a hard amplitude of scattering of a
%parton with momentum $k$ convoluted with the nonperturbative Green's
%function.. Note, that gluons can interact with colourless particles via
%quark loops (like the one in Fig. \ref{fig-gluon-loop} ), so at higher
%orders one also obtains terms with the hard amplitude corresponding to
%scattering of a gluon with momentum $k$ times the Green's function with
%the gluon fields instead of the quark ones. In this case, the gluon
%operators entering the nonperturbative matrix elements are the elements of
%gluon field strength tensor $G^{+\mu}(-z/2)G^{\nu+}(z/2)$. In the considered
%kinematics, they can be used instead of the product of gauge fields. Indices
%$\mu,\nu$ are contracted with the corresponding indices from the hard
%amplitude (i.e. the scattering amplitude of a gluon with amputated
%polarization vectors -- hence the free indices). Analogously to the Fierz
%decomposition, one can decompose the hard amplitude using tensors
%$g_{\mu\nu}$ and $\epsilon_{\mu\nu\alpha\beta}n_+^\alpha n_-^\beta$. The
%same manipulations as those for the quark fields discussed above allow
%to parametrize the resulting gluon contributions by the matrix elements
\begin{align}
  &\int\!\frac{dz^-}{2\pi}~\!e^{ixp^+ z^-}\!\bra{N(\mathbf{p}^\prime,\sigma^\prime)_{\mathrm{out}}}
  G^{+\lambda}\big(-z^- n_-/2\big) G^{\phantom{a}+}_\lambda \big(z^- n_-/2\big)
  \ket{N(\mathbf{p},\sigma)_{\mathrm{in}}},\\
  &\int\!\frac{dz^-}{2\pi}~\!e^{ixp^+ z^-}\!\bra{N(\mathbf{p}^\prime,\sigma^\prime)_{\mathrm{out}}}
  G^{+\lambda} \big(-z^- n_-/2\big)\Tilde{G}^{\phantom{a}+}_\lambda \big(z^- n_-/2\big)
  \ket{N(\mathbf{p},\sigma)_{\mathrm{in}}},
\end{align}
the gluon field strength
tensor $G^{\lambda\rho}$ and and its dual $\tilde G^{\lambda\rho}$ are
used instead of the gauge potentials -- for reference see Sec. 3.2.6 in \cite{Radyushkin}.  \\
\vskip0.2cm

\begin{figure}[H]
\centering
\begin{tikzpicture}
\draw [ultra thick] (-2, 0) -- (2,0);
\draw [decorate, decoration={snake, segment length=2mm, amplitude=0.5mm,post length=1mm}] (-2, 1) -- (0,0);
	\draw [decorate,
  decoration={snake, segment length=2mm, amplitude=0.5mm,post length=1mm}] (2,1) -- (0,0);
	\filldraw[color=black, fill=gray!40, thick] (0,0) ellipse (1 and 0.5);
	\node at (5, 0) {$= \quad \sum_i \int_{-1}^1 p^+ dx$};
\draw [ultra thick] (7, -1) -- (11,-1);
		\draw [decorate,
  decoration={snake, segment length=2mm, amplitude=0.5mm,post length=1mm}] (7, 2) -- (9,1);
	\draw [decorate,
  decoration={snake, segment length=2mm, amplitude=0.5mm,post length=1mm}] (11,2) -- (9,1);
	\draw (8.5, -1) -- (8.5, 1);
	\draw (9.5, -1) -- (9.5, 1);
	\draw [->, >=stealth] (8.25, -0.4) -- (8.25, 0.5);
	\draw [->, >=stealth] (9.75, 0.5) -- (9.75, -0.4);
	\node at (7.5, 0) {\tiny{$(x+\xi)p^+$}};
	\node at (10.5,0) {\tiny{$(x-\xi)p^+$}};
	\filldraw[color=black, fill=green!40, thick] (9,1) ellipse (1.25 and 0.5);
	\filldraw[color=black, fill=blue!30, thick] (9,-1) ellipse (1.25 and 0.5);
	\node at (9, -1) {\small{$\mathrm{GPD}^i(x,\xi,t)$}};
	\node at (9, 1) {\small{$\mathcal{T}_i$}};
\end{tikzpicture}
\caption{Graphical representation of the factorization of the  DVCS amplitude.
  $\int_{-1}^1 p^+ dx$ corresponds to integration over all possible values of
  the "$+$" components of momentum of the parton entering the hard part
  of the amplitude (the upper green blob).}
\label{fig-factorization-graphically}
\end{figure}
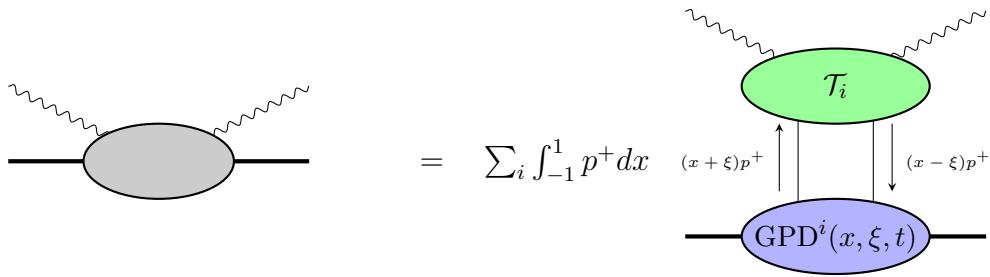

%Factorization formula, presented graphically in Fig.
%\ref{fig-factorization-graphically}, allows to treat independently
%contributions corresponding to the short- and long distance dynamics.
%Since it assumes, that momenta of all partons making up the hadron
%are in the same direction, it is referred as the collinear factorization.
%Assuming its validity, one can use it to write the amplitude of an
%exclusive process as
Summarizing, the factorization, sketched above on the example of the
DVCS subprocess, should in general allow to represent the amplitude
${\cal A}$ ($W^{\mu\nu}(\mathbf{p}^\prime,\mathbf{p},q^\prime,q)$
in the case of DVCS) of an exclusive process in the schematic  form
(depicted in Fig. \ref{fig-factorization-graphically})
\begin{equation}\label{eq-factorization-ch2}
\mathcal{A}=\sum_i\int_{-1}^1\!dx~\!\mathrm{GPD}^i(x,\xi, t)~\!\mathcal{T}_i(x,\xi,\dots),
\end{equation}
where the summation over $i$ runs over several kinds of quark and gluon
generalized distributions $\mathrm{GPD}^i(x,\xi, t)$
($H^{q,g}$, $\tilde H^{q,g}$, $E^{q,g}$ defined by the decompositions
(\ref{eq-gpd-q1GPD})-(\ref{eq-gpd-g2GPD}) in the case of DVCS), $t=(p^\prime-p)^2$, and $\mathcal{T}_i(x,\xi,\dots)$
are the corresponding hard parts computable perturbatively (the ellipses
stand for all kinematics other kinematical characteristics on which
the hard amplitude may depend).

%, $\mathcal{T}_i$ denotes the corresponding hard part (traced with
%the appropriate gamma matrix or contracted with a given second rank tensor),
%and "$\dots$" in the parenthesis stands for all kinematics parameters,
%on which depends the hard part of the amplitude for a given parton $i$.
%\begin{equation}\label{eq-factorization-ch2}
%\mathcal{A} = \sum_i \mathrm{GPD}^i (x,\xi, t) \mathcal{T}_i (x, \xi, \dots),
%\end{equation}
%where the index $i$ goes over all kinds of quark and gluon distributions,
%$\mathcal{T}_i$ denotes the corresponding hard part (traced with the
%appropriate gamma matrix or contracted with a given second rank tensor),
%and "$\dots$" in the parenthesis stands for all kinematics parameters, on
%which depends the hard part of the amplitude for a given parton $i$.

%%%%%%%%%%%%%%%%%%%%%%%%%%%%%%%%%%%%%%%%%%%%%%%%%%%%%%%%%%%%%
%% FIG 2.5: GENERAL FACTORIZATION
%%%%%%%%%%%%%%%%%%%%%%%%%%%%%%%%%%%%%%%%%%%%%%%%%%%%%%%%%%%%%

\section{Higher order QCD corrections}
\label{section-renormalization}

In this section we explain, how perturbative corrections in QCD affect
the theoretical predictions of amplitudes of exclusive processes. It
turns out, that whether the collinear factorization can be applied at
higher orders of perturbative calculation is a non-trivial question,
since validity of the factorization requires that infrared and collinear divergences from
bare parton distributions cancel with those present in the hard
amplitude. For example, see
\cite{Factorization-Ji, Factorization-proof, cytuj-promotora2}.
\begin{figure}[H]
\centering
	\begin{tikzpicture}[scale = 0.7]
	\begin{feynman}
		\vertex (ph1) at (0,0);
		\vertex (q1) at (1, -1);
		\vertex (q11) at (2, -1);
		\vertex (q21) at (3, -1);
		\vertex (q2) at (4, -1);
		\vertex (ph2) at (5, 0);
		\vertex (q-in) at (0, -3);
		\vertex (q-out) at (5, -3);

		\diagram *{
			(q-in) -- [fermion] (q1) -- [fermion] (q11) -- [fermion] (q21) -- [fermion] (q2) -- [fermion] (q-out);
			(ph1) -- [photon] (q1);
			(ph2) -- [photon] (q2);
			(q11) -- [gluon, half right] (q21);
		};
	\end{feynman}
	\node at (2.5, -3.5) {$(a)$};
	\end{tikzpicture}
\qquad
	\begin{tikzpicture}[scale = 0.7]
	\begin{feynman}
		\vertex (ph1) at (0,0);
		\vertex (q1) at (1, -1);
		\vertex (q11) at (2, -1);
		\vertex (q2) at (4, -1);
		\vertex (ph2) at (5, 0);
		\vertex (q-in) at (0, -3);
		\vertex (q-in1) at (0.5, -2);
		\vertex (q-out) at (5, -3);

		\diagram *{
			(q-in) -- [fermion] (q-in1) -- [fermion] (q1) -- [fermion] (q11) -- [fermion] (q2) -- [fermion] (q-out);
			(q-in1) -- [gluon, quarter right] (q11);
			(ph1) -- [photon] (q1);
			(ph2) -- [photon] (q2);
		};
	\end{feynman}
	\node at (2.5, -3.5) {$(b)$};
	\end{tikzpicture}
\qquad
	\begin{tikzpicture}[scale = 0.7]
	\begin{feynman}
		\vertex (ph1) at (0,0);
		\vertex (q1) at (1, -1);
		\vertex (q2) at (4, -1);
		\vertex (ph2) at (5, 0);
		\vertex (q-in) at (0, -3);
		\vertex (q-in1) at (0.5, -2);
		\vertex (q-out) at (5, -3);
		\vertex (q-out1) at (4.5, -2);

		\diagram *{
			(q-in) -- [fermion] (q-in1) -- [fermion] (q1) -- [fermion] (q2) -- [fermion] (q-out1) -- [fermion] (q-out);
			(ph1) -- [photon] (q1);
			(ph2) -- [photon] (q2);
			(q-in1) -- [gluon] (q-out1);
		};
	\end{feynman}
	\node at (2.5, -3.5) {$(c)$};
	\end{tikzpicture}
\caption{Examples of diagrams in the hard-part of the amplitude at the $\alpha_S$ order in QCD. There are two kinds of divergences. The first one -- ultraviolet divergence -- is present in diagrams $(a)$ and $(b)$. By power counting one sees that the diagram $(c)$ is UV safe. However, there are also so-called collinear divergences present in diagrams $(b)$ and $(c)$, which result from the integration region, where the momentum of the gluon is nearly collinear with the one of the on-shell quark. These divergences are discussed in more detail in Subsec. \ref{subsec-divergences}.}
\label{fig-dvcs-loops}
\end{figure}
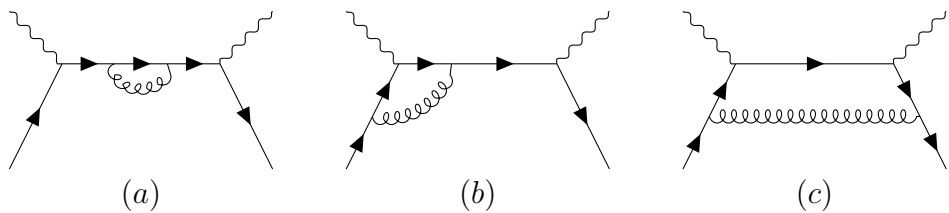
In this work we use the $\overline{MS}$ renormalization scheme together with the dimensional regularization, where the number of dimensions is $d=4-\varepsilon$.
In general, one should use the renormalized perturbation theory, so that the hard amplitude is UV-finite, however, at the one-loop level the renormalization constants cancel,
so that in calculations presented here, for our convenience, we can use the bare fields and unrenormalized hamiltonian (so that there is no need to include the diagrams with counter-terms). For reference, see the beginning of Part III in \cite{Factorization-Ji}. 
At higher orders, that argument does not necessarily work, so that one has to use the renormalized theory.
Computation of regularized loop corrections to the hard part is a well-known procedure, which, at this point, is not discussed. In Chapter \ref{chapter-diphoton} we present such calculations in the process of photoproduction of photon pairs. The NLO analysis of DDVCS can be found in \cite{cytuj-promotora2}. 

Apart from hard amplitudes, also bare parton distributions develop divergences at higher orders in strong coupling $\alpha_S$. To control them,
one needs to renormalize string operators used in Eqs. \eqref{eq-gpd-q1}-\eqref{eq-gpd-g2} by finding such combinations of these operators, that their matrix elements are finite. Subsequently, one defines renormalized GPDs denoted by $\mathrm{GPD}^i_R$ using Green's functions of the renormalized string operators. The relation between the bare parton distributions used in the formula \eqref{eq-factorization-nlo} and the renormalized ones reads:
\begin{equation}\label{eq-gpdB-gpdR}
 \mathrm{GPD}^i (x,\xi, t) = \mathrm{GPD}^i_R (x,\xi, t; \mu_F) + \frac{2}{\varepsilon} \Big( \frac{\mu_F^2 e^\gamma}{4\pi \mu_R^2 } \Big)^{-\varepsilon/2} \int dx' \: K^{ij}(x, x') \mathrm{GPD}^j_R(x',\xi,t;\mu_F) + \mathcal{O}(\alpha_S^2),
\end{equation}
 where $\gamma$ is the Euler constant, $\mu_R$ and $\mu_F$ denote the renormalization scale and factorization scale, respectively. Meaning of these scales is elaborated in Section \ref{section-string-renormalization}. The matrix $K^{ij}$ is called the evolution kernel, and there is summation over index $j$. In Section \ref{section-string-renormalization} it is explained, how it is derived by considering Green's functions of string operators defining GPDs.

Recall the factorization formula \eqref{eq-factorization-ch2}:
\begin{equation}\label{eq-factorization-nlo}
	\mathcal{A} = \sum_i \int_{-1}^1  \mathrm{GPD}^i (x,\xi, t) \mathcal{T}_i (x, \xi, \dots) dx.
\end{equation}
The divergent hard part can be organized in the following way (for simplicity we omit arguments other than $x$):
\begin{equation}\label{eq-div-part-def}
	\mathcal{T}^i (x) = \mathcal{C}^i_0 (x) + \bigg[ -\frac{2}{\varepsilon}\Big( \frac{|Q^2| e^\gamma}{4\pi \mu_R^2 } \Big)^{-\varepsilon/2} \mathcal{C}^i_{coll.}(x) + \mathcal{C}^i_1 (x) \bigg],
\end{equation}
where $\mathcal{C}^i_1$ and $ \mathcal{C}^i_{coll.}$ are of order $\alpha_S$.
Let us write the divergent part of the amplitude at the order $\alpha_S$:
\begin{equation}
\begin{aligned}
	\mathcal{A}_{div} = &-\frac{2}{\varepsilon} \sum_j \int_{-1}^1 dx \: \mathrm{GPD}^j_R (x) \mathcal{C}^j_{coll.}(x) + \\
	+&\frac{2}{\varepsilon} \sum_{i,j} \int_{-1}^1 dx \: dx' \: K^{ij}(x,x') \mathrm{GPD}^j_R (x') \mathcal{C}^i_0 (x').
\end{aligned}	
\end{equation}
The cancellation of divergences occurs if
\begin{equation}\label{eq-cancelation-of-div}
	\mathcal{C}^j_{coll.} (x) = \sum_i \int_{-1}^1 K^{ij} (y,x) \mathcal{C}^i_0 (y) dy.
\end{equation}
If that is the case, the the amplitude can be expressed using the finite quantities:
\begin{equation}\label{eq-final-factorization-nlo}
	\sum_i \int_{-1}^1  \mathrm{GPD}^i_R (x,\xi, t; \mu_F) \bigg[  \mathcal{C}^i_0 (x, \xi, \dots) +  \mathcal{C}^i_1(x,\xi,\dots) + \log\Big(\frac{|Q^2|}{\mu_F^2}\Big)  \mathcal{C}^i_{coll.}(x,\xi, \dots)  \bigg].
\end{equation}
Dependence on the scale $\mu_R^2$ vanished due to the simple relation 
$$\log \Big( \frac{|Q^2|}{\mu_R^2} \Big) - \log \Big( \frac{\mu_F^2}{ \mu_R^2} \Big) = \log \Big( \frac{|Q^2|}{\mu_F^2} \Big).$$
Generalization to higher QCD orders is straightforward, and can be described in the following points:
\begin{enumerate}
	\item Using the renormalized fields and couplings, compute all diagrams in the hard part of the amplitude within a given regularization scheme. Renormalization makes them only UV-finite, but the still possess infrared/collinear divergences.
	\item Using the same regularization, express the bare parton densities in terms of renormalized ones. In particular, the divergent part of it must be expressed as a product of an infinite constant and a term linear in renormalized GPDs.
	\item Verify cancellation of infrared/collinear divergences and write the amplitude in terms of finite quantities.
\end{enumerate}

\subsection{Gauge link}\label{Subsection-gauge-link}
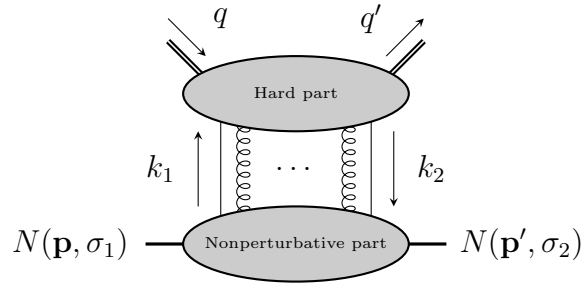
\begin{figure}[H]
\centering
\begin{tikzpicture}
\begin{feynman}
\vertex (g1up) at (-0.7, 2.5);
\vertex (g2up) at (0.7, 2.5);
\vertex (g1down) at (-0.7, 0.5);
\vertex (g2down) at (0.7, 0.5);
\diagram*{
	(g1up) -- [gluon] (g1down);
	(g2up) -- [gluon] (g2down);
};
\end{feynman}
\node at (0, 1.5) {$\dots$};
\draw (-1, 0.5) -- (-1, 2.5);
\draw (1, 0.5) -- (1, 2.5);
\draw [double, thick] (-1.7, 3.2) -- (-1, 2.5);
\draw [double, thick] (1.7, 3.2) -- (1, 2.5);
\draw [->, >=stealth] (-1.7, 3.5) -- (-1.2, 3);
\draw [<-, >=stealth] (1.7, 3.5) -- (1.2, 3);
\node at (-1, 3.5) {$q$};
\node at (1, 3.5) {$q'$};
\draw [very thick] (-2, 0.5) -- (2, 0.5);
\node at (-3, 0.5) {$N(\mathbf{p}, \sigma_1)$};
\node at (3, 0.5) {$N(\mathbf{p}', \sigma_2)$};
\draw  [->, >=stealth] (-1.3, 1) -- (-1.3,2);
\node at (-1.8, 1.5) {$k_1$};
\draw  [->, >=stealth] (1.3, 2) -- (1.3,1);
\node at (1.8, 1.5) {$k_2$};
\filldraw[color=black, fill=gray!40, thick] (0,0.5) ellipse (1.5 and 0.5);
\filldraw[color=black, fill=gray!40, thick] (0,2.5) ellipse (1.5 and 0.5);
\node at (0, 0.5) {\tiny{Nonperturbative part}};
\node at (0, 2.5) {\tiny{Hard part}};
\end{tikzpicture}
\caption{At higher orders one needs to include contributions, in which apart from two ``main'' partons (denoted by straight lines connecting the hard and nonperturbative parts) there is an arbitrary number of gluons leaving the hard part and entering the nonperturbative matrix element.}
\label{fig-dwa-jajeczka-z-gluonami}
\end{figure}

\begin{figure}[H]
\centering
\begin{tikzpicture}[scale = 0.6]
	\begin{feynman}
		\vertex (ph1) at (0,0);
		\vertex (q1) at (1, -1);
		\vertex (q2) at (4, -1);
		\vertex (ph2) at (5, 0);
		\vertex (q-in) at (0, -3);
		\vertex (q-in1) at (0.5, -2);
		\vertex (q-out) at (5, -3);
		\vertex (q-out1) at (4.5, -2);

		\diagram *{
			(q-in) -- [fermion] (q-in1) -- [fermion] (q1) -- [fermion] (q2) -- [fermion] (q-out1) -- [fermion] (q-out);
			(ph1) -- [photon] (q1);
			(ph2) -- [photon] (q2);
			(q-in1) -- [gluon] (q-out1);
		};
	\end{feynman}
\end{tikzpicture}
\caption{The exemplary diagram $\mathcal{T}(z_1, z_2)$.}
\label{fig-T-no-gl}
\end{figure}
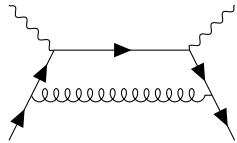

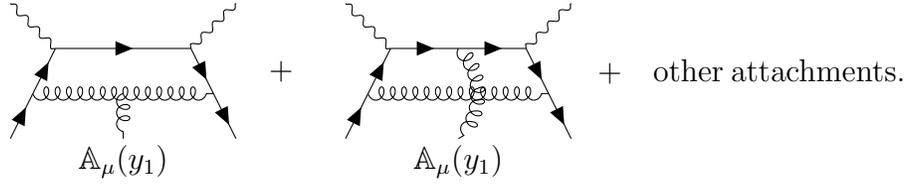
\begin{figure}[H]
\centering
\begin{tikzpicture}[scale = 0.6]
	\begin{feynman}
		\vertex (ph1) at (0,0);
		\vertex (q1) at (1, -1);
		\vertex (q2) at (4, -1);
		\vertex (ph2) at (5, 0);
		\vertex (q-in) at (0, -3);
		\vertex (q-in1) at (0.5, -2);
		\vertex (q-out) at (5, -3);
		\vertex (q-out1) at (4.5, -2);
		\vertex (gl-out) at (2.5, -3);
		\vertex (gl-aux) at (2.5, -2);

		\diagram *{
			(q-in) -- [fermion] (q-in1) -- [fermion] (q1) -- [fermion] (q2) -- [fermion] (q-out1) -- [fermion] (q-out);
			(ph1) -- [photon] (q1);
			(ph2) -- [photon] (q2);
			(q-in1) -- [gluon] (q-out1);
			(gl-aux) -- [gluon] (gl-out);
		};
	\end{feynman}
\node at (6, -1.5) {$+$};
\node at (2.5, -3.5) {$\mathbb{A}_\mu (y_1)$};
\end{tikzpicture}
\quad
\begin{tikzpicture}[scale = 0.6]
	\begin{feynman}
		\vertex (ph1) at (0,0);
		\vertex (q1) at (1, -1);
		\vertex (q3) at (2.5, -1);
		\vertex (q2) at (4, -1);
		\vertex (ph2) at (5, 0);
		\vertex (q-in) at (0, -3);
		\vertex (q-in1) at (0.5, -2);
		\vertex (q-out) at (5, -3);
		\vertex (q-out1) at (4.5, -2);
		\vertex (gl-out) at (2.5, -3);

		\diagram *{
			(q-in) -- [fermion] (q-in1) -- [fermion] (q1) -- [fermion] (q3) -- [fermion] (q2) -- [fermion] (q-out1) -- [fermion] (q-out);
			(ph1) -- [photon] (q1);
			(ph2) -- [photon] (q2);
			(q-in1) -- [gluon] (q-out1);
			(q3) -- [gluon, quarter left] (gl-out);
		};
	\end{feynman}
\node at (2.5, -3.5) {$\mathbb{A}_\mu (y_1)$};
\node at (9, -1.5) {$+\quad \mathrm{other} \:\mathrm{attachments.}$};
\end{tikzpicture}
\caption{The sum $\mathcal{T}_1^{\mu} (z_1, z_2; y)$ defined by  $\mathcal{T}(z_1, z_2)$ in Fig. \ref{fig-T-no-gl}.}
\label{fig-T-y}
\end{figure}

It turns out, that at higher orders in perturbative expansion the leading (in the hard scale) contributions stem not only from terms, in which there are just two field operators between the hadronic states. Equally important are graphs with an arbitrary number of gluons leaving the hard diagram and entering the nonperturbative part, presented graphically in Fig. \eqref{fig-dwa-jajeczka-z-gluonami}. In this part we argue, that these diagrams result in presence of the gauge link in operators defining GPDs.
To simplify these contributions, let us work on the matrix element from Eq. \eqref{eq-Gell-Mann-Low} in the coordinate space. Let $\mathcal{T}^{(i)}(z_1, z_2)$ denote a hard diagram with one parton $i$ incoming at the position $z_1$ and one (of the same kind) outgoing at $z_2$ (we do not explicitly write free spinor or vector indices that get contracted with their counterparts in the nonperturbative part). By $\mathcal{T}_n^{(i)\mu_1\dots\mu_n} (z_1, z_2; y_1, \dots, y_n)$ we denote the sum of all diagrams with $n$ outgoing gluons obtained by attaching gluon fields $\mathbb{A}_{\mu_i}(y_i)$ at positions $y_1, \dots, y_n$ on propagators lines inside of the diagram $\mathcal{T}^{(i)}(z_1, z_2)$. To make this statement more clear, let us present an example: let $\mathcal{T}(z_1, z_2)$ correspond to a diagram presented in Fig. \ref{fig-T-no-gl}.
Then, $\mathcal{T}_1^{\mu} (z_1, z_2; y_1)$ is a sum of diagrams shown in Fig. \ref{fig-T-y}.
For a given kind of parton $i$, summation over all possible hard part diagrams involving this parton and having an arbitrary number of outgoing gluon lines is equivalent to summation over all possible diagrams $\mathcal{T}^{(i)}$ and the corresponding $\mathcal{T}^{(i)}_n$. Inside of the time ordered product of operators in the matrix element \eqref{eq-Gell-Mann-Low} one finds the expression
\begin{equation}\label{eq-sum-T-y}
\sum_{\mathcal{T}^{(i)}} \bigg( \mathcal{T}^{(i)}(z_1, z_2) + \sum_{n=1}^{\infty} \Big( \prod_{j=1}^n \int d^4 y_j \mathbb{A}_{\mu_j}(y_j) \Big) \mathcal{T}_n^{(i)\mu_1\dots\mu_n} (z_1, z_2; y_1, \dots, y_n) \bigg).
\end{equation}
Note, that in the analysis presented in Section \ref{section-fact-lo} only the first term from \eqref{eq-sum-T-y} was present. The reasoning shown in the work of Efremov and Radyushkin \cite{Efremov} allows to prove, that for any diagram $\mathcal{T}^{(i)}$, the sum in \eqref{eq-sum-T-y} can be written as
\begin{equation}
 \eqref{eq-sum-T-y} = \mathcal{T}^{(i)}(z_1, z_2) P\: exp\Big( ig\int_{z_2}^{z_1} dz_\mu \mathbb{A}^\mu(z) \Big) + \quad \mathrm{remainder},
\end{equation}
where the remainder consists of operators which matrix elements yield sub-leading terms.

To show it, one has to analyze how attachments of gluon vertices modify propagators of the particles.
Consider the sum of quark propagators with additional attachments of gluons, as it is shown in Fig. \ref{fig-propagators}.
\begin{figure}[H] 
\centering
\begin{tikzpicture}[scale = 0.7]
\begin{feynman}
	\vertex (q1) at (0,0);
	\vertex (q2) at (3, 0);

\diagram *{
	(q1) -- [fermion] (q2);
};
\end{feynman}
	\node at (4, 0) {$+$};
\node at (0, -0.82) {$\:$};
\end{tikzpicture}
\quad
\begin{tikzpicture}[scale = 0.7]
\begin{feynman}
	\vertex (q1) at (0,0);
	\vertex (q2) at (1.5, 0);
	\vertex (q3) at (3, 0);
	\vertex (g) at (1.5, -1);

\diagram *{
	(q1) -- [fermion] (q2) -- [fermion] (q3);
	(g) -- [gluon] (q2);
};
\end{feynman}
	\node at (4, 0) {$+$};
\end{tikzpicture}
\quad
\begin{tikzpicture}[scale = 0.7]
\begin{feynman}
	\vertex (q1) at (0,0);
	\vertex (q2) at (1, 0);
	\vertex (q3) at (2, 0);
	\vertex (q4) at (3,0);
	\vertex (g) at (1, -1);
	\vertex (g2) at (2, -1);

\diagram *{
	(q1) -- [fermion] (q2) -- [fermion] (q3) -- [fermion] (q4);
	(g) -- [gluon] (q2);
	(g2) -- [gluon] (q3);
};
\end{feynman}
	\node at (4, 0) {$+\: \dots$};
\end{tikzpicture}
\label{fig-propagators}
\end{figure}
Sum of such diagrams can be written as
\begin{equation}
\mathcal{S}(x_1 - x_2) = S(x_1 - x_2) + g\int d^4 y_1 S(x_1 - y_1) \slashed{\mathbb{A}}(y_1) S(y_1 - x_2) + \dots
\end{equation}
Observe that for a massless quark the free propagator $S$ is the solution of the equation
\begin{equation}
i\slashed{\partial} S(x_1 - x_2) = -\delta^4(x_1 - x_2),
\end{equation}
while $\mathcal{S}$ is the propagator of the quark in the background gauge field, and 
\begin{equation}
i\slashed{D} \mathcal{S}(x_1 - x_2) = -\delta^4(x_1 - x_2),
\end{equation}
where $D_\mu = \partial_\mu + ig\mathbb{A}_\mu$ is the covariant derivative. It can be shown \cite{Efremov} that 
\begin{equation}
 \mathcal{S}(x_1 - x_2) = P exp\Big( ig\int_{x_2}^{x_1} dx_\mu \mathbb{A}^\mu(x) \Big) \big( S(x_1 - x_2)+ R(x_1, x_2) \big),
\end{equation}
where $P\: exp\Big( ig\int_{x_2}^{x_1} dx_\mu \mathbb{A}^\mu(x) \Big)$ is the path-ordered Wilson line (also know as the gauge link).
The ``remainder'' $R(x_1, x_2)$ depends on the gauge field by the field strength tensor only, which gives a sub-leading terms when inserted between the hadron states. 
The analogous reasoning applied to gluon and ghost propagators yields the same result. Hence, summation over all possible attachments of external gluon lines is (to the leading order in the hard scale)
equivalent to substituting of propagators inside of the diagram $\mathcal{T}^{(i)}(z_1, z_2)$ with ones multiplied by the gauge link.
That, on the other hand can be shown to give the overall factor 
$$
	P\: exp\Big( ig\int_{z_2}^{z_1} dz_\mu \mathbb{A}^\mu(z) \Big)
$$
standing by the amplitude $T^{(i)}(z_1, z_2)$. It is crucial that it does not depend on the parton type $i$ or the exact form of the diagram $\mathcal{T}^{(i)}$, and hence can included inside the nonperturbative part without any further complication. Therefore, after using the collinear approximation and applying the LSZ reduction formula to the vacuum Green's function of the product of parton fields, hadron operators $O$ and the gauge link one finally obtains the full definition of Generalised Parton Distributions shown in Formulas  \eqref{eq-gpd-q1}-\eqref{eq-gpd-g2}.

\subsection{UV and soft collinear divergences}\label{subsec-divergences}
Here we briefly describe, how divergences arise in all the considered amplitudes and how to classify them. It allows to clarify, why collinear divergences are present in our calculations,
and say in which diagram they occur.

\begin{figure}[H]
\centering
	\begin{tikzpicture}[scale = 0.7]
		\draw [->, thick] (0,0) -- (7, 0);
		\draw [->, thick] (1, -1) -- (1, 3);
		\draw (2.9, 0.4) -- (3.1, 0.6);
		\draw (2.9, 0.6) -- (3.1, 0.4);
		\draw [thick, color=red] (2, 0) arc (180:360:1);
		\draw [thick, color=red] (1, 0) -- (2, 0);
		\draw [thick, color=red] (4, 0) -- (6.5, 0);
		\node at (7, 0.5) {$\mathrm{Re} \: \alpha_j$};
	\end{tikzpicture}
\caption{An example of how one can change the contour avoiding integration in a neighborhood of the pole, which is marked by the cross slightly above the real axis.}
\label{fig-contour1}
\end{figure}
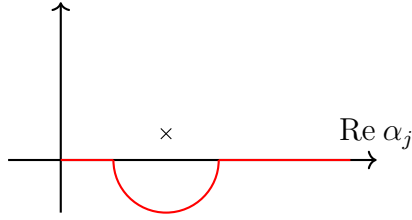

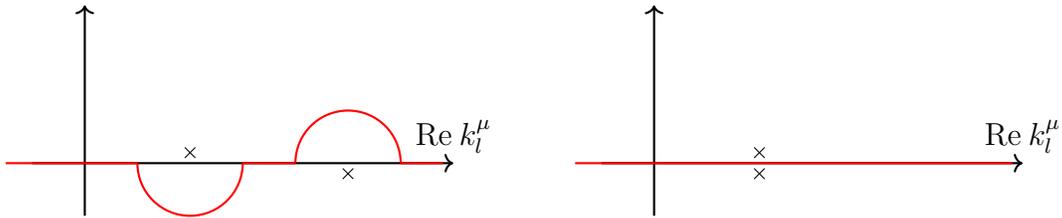
\begin{figure}[H]
\centering
	\begin{tikzpicture}[scale = 0.7]
		\draw [->, thick] (0,0) -- (8, 0);
		\draw [->, thick] (1, -1) -- (1, 3);
		\draw (2.9, 0.1) -- (3.1, 0.3);
		\draw (2.9, 0.3) -- (3.1, 0.1);
		\draw (5.9, -0.1) -- (6.1, -0.3);
		\draw (5.9, -0.3) -- (6.1, -0.1);
		\draw [thick, color=red] (2, 0) arc (180:360:1);
		\draw [thick, color=red] (7, 0) arc (0:180:1);
		\draw [thick, color=red] (-0.5, 0) -- (2, 0);
		\draw [thick, color=red] (4, 0) -- (5, 0);
		\draw [thick, color=red] (7, 0) -- (7.8, 0);
		\node at (8, 0.5) {$\mathrm{Re} \:k_l^\mu$};
	\end{tikzpicture}
\qquad
	\begin{tikzpicture}[scale = 0.7]
		\draw [->, thick] (0,0) -- (8, 0);
		\draw [->, thick] (1, -1) -- (1, 3);
		\draw (2.9, 0.1) -- (3.1, 0.3);
		\draw (2.9, 0.3) -- (3.1, 0.1);
		\draw (2.9, -0.1) -- (3.1, -0.3);
		\draw (2.9,-0.3) -- (3.1, -0.1);
		\draw [thick, color=red] (-0.5, 0) -- (7.8, 0);
		\node at (8, 0.5) {$\mathrm{Re} \:k_l^\mu$};
	\end{tikzpicture}
\caption{The illustration of a pinch singularity: in the case presented on the left hand side, one can modify the contour avoiding poles, while in the case on the right hand side it is impossible, because the initial contour is 'pinched' between two poles.}
\label{fig-contour2}
\end{figure}

A general form of a momentum integral encountered in a diagram with $L$ loops, which contain $N$ propagators (all of massless particles), can be written in the following form:
\begin{equation}\label{eq-general-L-loops}
	G\big( \{p_i \} \big) =  \prod_{l=1}^L \int \frac{d^dk_l}{(2\pi)^d} \frac{P\big( \{k_i \} \big)}{\prod_{n=1}^N \Big( Q_n^2 \big( \{ k_i \} \big) + i0 \Big) }~\!,
\end{equation}
where $\{p_i \}$ denotes the set of external momenta entering the loops, $Q_n^2 \big( \{ k_i \} \big)$ is the momentum squared in $n$-th propagator, which depends on momenta over which we integrate, and $P\big( \{k_i \} \big)$ denotes momenta-dependent polynomial. There are two classes of regions, which may lead to divergences. The first one, corresponding to the limit $k^2 \rightarrow \infty$ is the well-known ultraviolet divergence, which is fixed by the hamiltonian renormalization. The other source of divergence are poles of the integrand corresponding to $Q^2_n =0$ (particle on the mass shell). They can be studied using the method developed by Landau \cite{Landau}, which we describe now. Using the Feynman parametrization, one can write formula \eqref{eq-general-L-loops} as
\begin{equation}
	G\big( \{p_i\} \big) = \prod_{j=1}^N \int_0^1 d\alpha_j \delta \Big( 1 - \sum_{i=1}^N \alpha_i \Big) \prod_{l=1}^L \frac{d^d k_l}{(2\pi)^d} \frac{P\big( \{k_i\}, \{p_i\} \big) }{\Big[ \sum_{n=1}^N \alpha_n Q_n^2 \big( \{ k_i \} \big) + i 0 \Big]^N}.
\end{equation}
Let us denote the denominator as $D\big( \{ k_i \} \big) + i0$. The integrand is an analytical function of $\{ k_i \}$ and $\{\alpha_i\}$, so that one can modify the contour of the integration in a way that it avoids singular points. For example, assume that, for some fixed momenta, $D$ becomes $0$ and there is some $\alpha_j$ such that at this problematic point $Q_j^2 \neq 0$. Then, if $\alpha_j \neq 0$ (it is not at the endpoint), one can shift the contour to avoid the singularity, see Fig. \ref{fig-contour1}
Modification of the integration contour avoiding the pole is impossible, if that occurs at one of the endpoints ($0$ or $1$) or the term $Q_j^2$, which multiplies $\alpha_j$ is zero. In the case $\alpha_j = 1$, all other Feynman parameters are equal to $0$. It means that, if $D=0$, then $Q_j^2 = 0$. Hence, modification of $\alpha_j$ does not allow to avoid the singular point if
\begin{equation}\label{eq-landau-1}
	\alpha_j = 0 \quad \mathrm{or} \quad Q_j^2 = 0 \quad \mathrm{for} \: \mathrm{all} \: j.
\end{equation}
If that is the case, one can try to avoid the singularity by modifying the contour of integration over some component of the loop momentum $k_l^\mu$. Since the domain of the integration is the whole real axis, then there are no problems with the endpoints. However, there still can be a situation, when it is impossible to avoid the singularity -- because $D$ is a quadratic function of momenta $\{k_i\}$, then for each considered component of momentum there are 2 corresponding roots of $D$. If their real values are equal (and their corresponding imaginary parts are separated only by an infinitesimally small value $i0$), then the line of integration lays between these two poles and one cannot change the contour, see Fig. \ref{fig-contour2}. One can say, that these two poles 'pinch' the line of the integration -- hence the name 'pinch singularities'.
The pinch singularity corresponds to the situation, when all components of considered loop momenta $k^\mu_l$ are double roots of $D$, which translates into the condition
\begin{equation}\label{eq-pre-landau}
	\frac{\partial}{\partial k_l^\mu} D =0.
\end{equation}
If we assume that all loop momenta come with the positive sign into propagator momenta $Q$, we obtain that Eq. \eqref{eq-pre-landau} is equivalent to
\begin{equation}\label{eq-landau-2}
	\sum_{n=0}^N \alpha_n Q_n^\mu = 0.
\end{equation}
Equations \eqref{eq-landau-1} and \eqref{eq-landau-2} are known as Landau equations. We use them to analyze, what kind of divergences are present in the 1-loop hard diagrams in DVCS.
\begin{figure}[H]
\centering
\begin{tikzpicture}
\begin{feynman}
\vertex (p1) at (0,0);
\vertex (p2) at (3, 0);
\vertex (p3) at (5,0);
\vertex (p4) at (7,0);

\diagram *{
	(p1) -- [fermion, momentum =$p$] (p2) -- [fermion] (p3) -- [fermion] (p4);
	(p3) -- [gluon, half left, momentum = $k$] (p2);
};
\end{feynman}
\end{tikzpicture}
\end{figure}
Let us start from the simplest case, namely the self-energy correction, where the momentum of the quark $p$ is off-shell. In dimension $d\geq 4$ this diagram naturally has the UV divergence, but we want to focus only on the on-shell singularities. The corresponding denominator can be written using the Feynman parameters as 
\begin{equation}
	D = \alpha_1 k^2 + \alpha_2 (p+k)^2.
\end{equation}
Using the Landau equations we find, that there are two singular points, which cannot be avoided:
\begin{equation}
	\begin{aligned}
	& \alpha_1 = 0, \quad p^\mu + k^\mu =0,\\
	& \alpha_2 = 0, \quad k^\mu = 0.
	\end{aligned}
\end{equation}
However, note that this is a singularity at a single point, therefore, for a number of dimensions $d$ greater than $2$ it is integrable and there are no resulting infrared divergences. Hence, the considered self-energy diagram has only UV divergence for $d \geq 4$, is finite for $d=3$ and develops the infrared divergence for $d=2$.

Now let us consider a vertex diagram with the incoming quark line on-shell. Assume that $p = p^+ n_+^\mu$. 
\begin{figure}[H]
\centering
\begin{tikzpicture}
\begin{feynman}
\vertex (p1) at (0,0);
\vertex (p2) at (3, 0);
\vertex (p3) at (5,0);
\vertex (p4) at (7,0);
\vertex (qin) at (4, 1);
\vertex (q) at (4, 0);

\diagram *{
	(p1) -- [fermion, momentum =$p$] (p2) -- [fermion] (q) -- [fermion] (p3) -- [fermion] (p4);
	(p3) -- [gluon, half left, momentum = $k$] (p2);
	(qin) -- [photon, momentum=$q$] (q);
};
\end{feynman}
\end{tikzpicture}
\end{figure}
Landau equation again tell us that there are pinch singularities when $k$, $p+k$ or $p+q+k$ are zero, but in dimension $4$ it does not lead to IR divergences. However, there is another solution which corresponds to the situation, when the gluon's momentum is collinear with that of the incoming quark: $k^\mu = k^+ n_+^\mu$. In such situation the pinch singularity occurs at 
\begin{equation}
	\alpha_1 k^+ + \alpha_2 (p^+ + k^+) = 0, \quad \alpha_3=0, \quad \iff \quad \alpha_2 = -\frac{k^+}{p^+}.
\end{equation}
To obtain the second equation we used the fact that $\alpha_1 + \alpha_2 + \alpha_3 = 1$. That leads to so-called collinear divergences, which correspond to the situation when two or more momenta of massless particles become collinear and on-shell. Note, that if the momentum of the incoming quark was off-shell, then the vertex diagram would have only the UV divergence. Collinear divergence is present also in the case of the box diagram (i.e. the graph $(c)$ in Fig. \ref{fig-dvcs-loops}). As it was shown in the previous part, the key point of factorization theorem is proving that collinear divergences from the hard part can be absorbed in the divergences present in the non-renormalized parton distributions.

\section{Renormalization of non-local product of operators}\label{section-string-renormalization}
As it was discussed at the beginning of Section \ref{section-renormalization}, to gain control over divergences in parton distributions, one need to find renormalized
string operators used in the definition of GPDs. In this Section we describe the procedure of construction of these
 operators.
For further convenience, let use denote string operators according to the following formula:
\begin{align}
	 &\mathcal{O}^{(f)} (z_1 , z_2) := \bar{\psi}_{(f)}(z_1) \gamma^+ [z_1 , z_2 ] \psi_{(f)}(z_2), \label{eq-Oqq}\\
	&\Tilde{\mathcal{O}}^{(f)} (z_1 , z_2) := \bar{\psi}_{(f)}(z_1)\gamma^+ \gamma_5 [z_1 , z_2] \psi_{(f)}(z_2), \label{eq-OqqT}\\
	&\mathcal{O}^{gg}  (z_1 , z_2) := G^{+\mu} (z_1)  [z_1 , z_2] G^{\phantom{a}+}_\mu(z_2), \label{eq-Ogg}\\
	&\Tilde{\mathcal{O}}^{gg}  (z_1 , z_2) := G^{+\mu} (z_1)  [z_1 , z_2 ] \Tilde{G}^{\phantom{a}+}_\mu(z_2). \label{eq-OggT}
\end{align}
Our goal is to express them in terms of renormalized operators minus counter-terms. To do so, we follow the reasoning shown in Appendix G in \cite{Radyushkin}. It allows to write bare GPDs in terms of renormalized parton distributions plus divergent terms. 

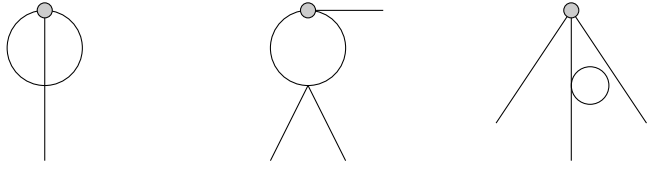
\begin{figure}[H]
\centering
\begin{tikzpicture}[scale = 0.5]
	\draw (0, -1) circle (1);
	\draw (0, 0) -- (0, -4);
	
	\draw (7, -1) circle (1);
	\draw (7, -2) -- (6, -4);
	\draw (7, -2) -- (8, -4);
	\draw (7, 0) -- (9, 0);

	\draw (14, 0) -- (12, -3);
	\draw (14, 0) -- (14, -4);
	\draw (14, 0) -- (16, -3);
	\draw (14.5, -2) circle (0.5);

	\filldraw [color = black, fill=gray!40] (0, 0) circle (0.2);
	\filldraw [color = black, fill=gray!40] (7, 0) circle (0.2);
	\filldraw [color = black, fill=gray!40] (14, 0) circle (0.2);

\end{tikzpicture}
\caption{Diagrams contributing to Green's function. The gray dots denote insertion of the operator $\phi^3$. The first diagram from the left represents $G^1(p)$ and two other: $G^3 (p_1, p_2, p_3)$ -- note that for each of them there are 3 possible permutations of insertions of momenta. The divergence from the self-energy correction in the third diagram is cancelled by the field-strength renormalization constant. Divergences present in the rest of diagrams are not cured by the lagrangian renormalization, so that additional counter-terms are required.}
\label{fig-phi3-ren}
\end{figure}

Let us start with description of methods used in renormalization of local operators. For example will serve the operator $\phi^3(0)$ in the scalar theory $\phi^4$ in dimension 4. We demand that all vacuum Green's function of this operator be finite to a given order in perturbation theory. At the first order in perturbation theory, the following two Green's functions become divergent:
\begin{align}
	&G^1(p) := \int d^4 x e^{-ipx} \bra{\Omega_{\mathrm{out} }} T  \phi^3_H(0) \phi_H(x)  \ket{\Omega_{\mathrm{in}} }, \\
	&G^3 (p_1, p_2, p_3) := \int d^4 x_1 \: d^4 x_2 \: d^4 x_3 \: e^{-ip_1x_1 -ip_2 x_2 - ip_3 x_3} \bra{\Omega_{\mathrm{out} }} T \phi^3_H(0) \phi_H(x_1) \phi_H (x_2) \phi_H (x_3) \ket{\Omega_{\mathrm{in}} },
\end{align}
Using the Gell-Mann and Low theorem \cite{Chank-LSZ} one can perform computation using standard Feynman rules.
The corresponding diagrams leading to divergences are shown in Fig. \ref{fig-phi3-ren}.
The renormalized operator $\phi^3(0)$ at the arbitrary order needs to be corrected by counter-terms proportional to renormalized operators $\phi_R$ and $\phi^3_R$:
\begin{equation}
	\big[ \phi^3(0)\big]_R = \phi^3_R (0) + c_1 \phi_R(0) + c_3 \phi^3_R (0),
\end{equation}
where $c_{1/3}$ are the renormalization constants. Equivalently, one can write it in terms of the bare operators $\phi_0$, using $\phi_R = \mathcal{Z}_\phi^{-1/2} \phi_0$, where $\mathcal{Z}_\phi$ is the renormalization constant.

In the more general case of two operators separated by a lightlike vector $z_1 - z_2$, one has to use a more complicated form of counter-terms, which become an product of operators integrated over positions:
\begin{equation}
	\mathrm{counterterm} \sim \int_0^1 d\alpha_1 d\alpha_2 K(\alpha_1, \alpha_2) \mathcal{O}\big( \alpha_1 z_1 + (1-\alpha_1) z_2, \alpha_2 z_2 + (1-\alpha_2) z_1 \big).
\end{equation}
Distributions $K(\alpha_1, \alpha_2)$ are called the evolution kernels.\\
It turns out, that the Green's functions of fields of different kinds do not vanish, hence, for example, the renormalized quark operator consists of gluon fields. Because of that, it is convenient to consider combinations of operators defined in Eqs \eqref{eq-Oqq}-\eqref{eq-OggT} of definite charge-parity. The non-singlet operators, which are $\mathcal{C}$-odd and do not mix under renormalization, are the following:
\begin{align}
	&\mathcal{O}^{(f)}_{NS} (z_1 , z_2) = \mathcal{O}^{(f)} (z_1 , z_2) + \mathcal{O}^{(f)} (z_2 , z_1),\\
	&\Tilde{\mathcal{O}}^{(f)}_{NS} (z_1 , z_2) = \Tilde{\mathcal{O}}^{(f)} (z_1 , z_2) - \Tilde{\mathcal{O}}^{(f)} (z_2 , z_1).
\end{align}
$\mathcal{C}$-even (singlet) combinations are:
\begin{align}
	& \mathcal{O}_S^{(f)}  (z_1 , z_2) = \mathcal{O}^{(f)} (z_1 , z_2) - \mathcal{O}^{(f)} (z_2 , z_1)\\
	& \mathcal{O}_S^{gg}  (z_1 , z_2) = \mathcal{O}^{gg}  (z_1 , z_2), \\
	& \Tilde{\mathcal{O}}_S^{(f)}  (z_1 , z_2) = \Tilde{\mathcal{O}}^{(f)} (z_1 , z_2) + \Tilde{\mathcal{O}}^{(f)} (z_2 , z_1)\\
	& \Tilde{\mathcal{O}}_S^{gg}  (z_1 , z_2) = \Tilde{\mathcal{O}}^{gg}  (z_1 , z_2),
\end{align}
(it is worth to stress, that gluon operators are always charge-even).

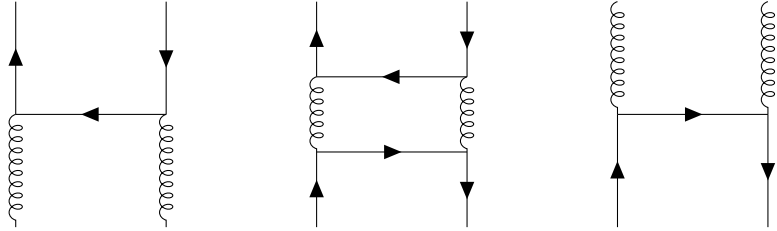
\begin{figure}[H]
\centering
\begin{tikzpicture}
\begin{feynman}
	\vertex (q1) at (0,0);
	\vertex (q2) at (0, -1.5);
	\vertex (q3) at (2, -1.5);
	\vertex (q4) at (2, 0);
	\vertex (g1) at (0, -3);
	\vertex (g2) at (2, -3);

	\vertex (q1p) at (4,0);
	\vertex (q2p) at (4, -1);
	\vertex (q3p) at (6, -1);
	\vertex (q4p) at (6, 0);
	\vertex (g1p) at (4, -2);
	\vertex (g2p) at (6, -2);
	\vertex (q1pp) at (4,-3);
	\vertex (q2pp) at (6, -3);

	\vertex (q1c) at (8,0);
	\vertex (q2c) at (8, -1.5);
	\vertex (q3c) at (10, -1.5);
	\vertex (q4c) at (10, 0);
	\vertex (g1c) at (8, -3);
	\vertex (g2c) at (10, -3);

	\diagram* {
	(q4) -- [fermion] (q3);
	(q3) -- [fermion] (q2);
	(q2) -- [fermion] (q1);
	(q2) -- [gluon] (g1);
	(q3) -- [gluon] (g2);

	(q4p) -- [fermion] (q3p);
	(q3p) -- [fermion] (q2p);
	(q2p) -- [fermion] (q1p);
	(q2p) -- [gluon] (g1p);
	(q3p) -- [gluon] (g2p);
	(q1pp) -- [fermion] (g1p);
	(g1p) -- [fermion] (g2p);
	(g2p) -- [fermion] (q2pp);

	(q4c) -- [gluon] (q3c);
	(q2c) -- [fermion] (q3c);
	(q1c) -- [gluon] (q2c);
	(g1c) -- [fermion] (q2c);
	(q3c) -- [fermion] (g2c);
    };
\end{feynman}
\end{tikzpicture}
\caption{Diagrams appearing in computation of Green's functions of string operators at higher orders, which contribute to the mixing of the operators. The diagram in the middle corresponds to mixing of different quark flavors, while the other two result in mixing of quarks and gluons.}
\label{fig-mixing}
\end{figure}

Similarly as in the simpler case of local operator, one needs to find counter-terms necessary to make the vacuum Green's functions of the time-ordered product of bare fields and operators $\bar{\psi}(z_1) [z_1, z_2] \psi(z_2)$, $G^{+\mu}(z_1) [z_1, z_2] G^{\phantom{a}+}_\mu(z_2)$ finite.
In the case of the quark string operator one considers two kinds of such Green's functions:
\begin{equation}
	\begin{aligned}
	&G^{ff'}(p_1, p_2, z_1, z_2) := \int d^4 x_1 \int d^4 x_2 e^{-ip_1 x_1 - ip_2 x_2} \bra{\Omega_{\mathrm{out} }} T \mathcal{O}^{(f)} (z_1 , z_2) \bar{\psi}_{(f')}(x_1) \psi_{(f')}(x_2) \big\} \ket{\Omega},\\
	&G^{fg}(p_1, p_2, z_1, z_2) := \int d^4 x_1 \int d^4 x_2 e^{-ip_1 x_1 - ip_2 x_2} \bra{\Omega_{\mathrm{out} }} T \mathcal{O}^{(f)} (z_1 , z_2) G^{+\mu}(z_1) G^{\phantom{a}+}_\mu (x_2)  \ket{\Omega},
	\end{aligned}
\end{equation}
in the limit when $p_{1,2}$ have only the "$+$" component of momentum (that is due to the fact, that we assume partons taking part in the hard process to be massless and nearly collinear with the hadron). The function $G^{fg}$ corresponds to the mixing of quark $f$ and gluon operators and is present only in the case of the singlet combination. At the $\alpha_S$ order, Green's functions with quark operators do not vanish only as $f=f'$, however, at higher orders there are also contributions responsible for mixing of quark flavors in the singlet sector, as it is presented in Fig. \ref{fig-mixing}.

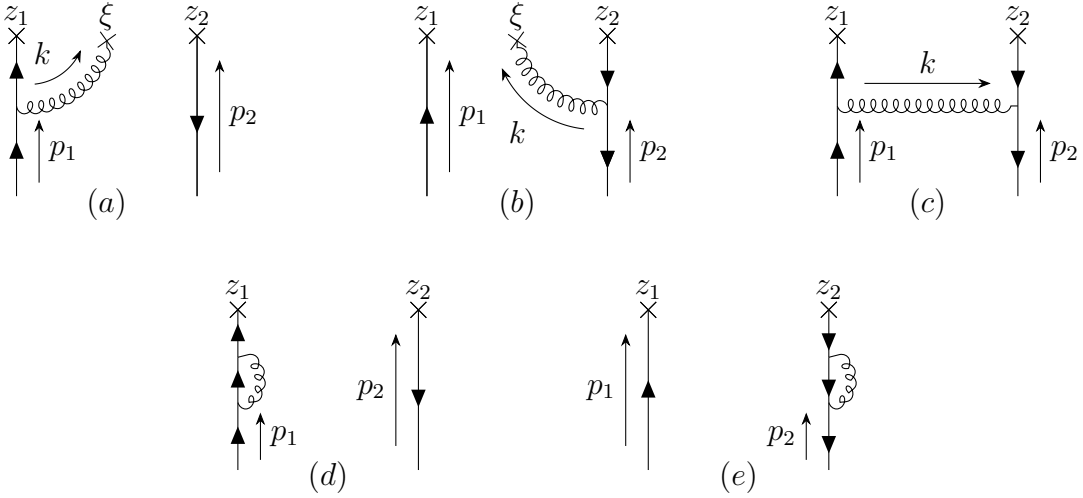
\begin{figure}
\centering
\begin{tikzpicture}
\begin{feynman}[scale = 0.6]
	\vertex (z1) at (0, 0) {$z_1$};
	\vertex (z2) at (4, 0) {$z_2$};
	\vertex (p1) at (0, -4);
	\vertex (p2) at (4, -4); 
	\vertex (g) at (2, 0) {$\xi$};
	\vertex (g-in) at (0, -2);

	\diagram *{
	(p1) -- [fermion, momentum'=\( p_1 \)] (g-in) -- [fermion, insertion=0.999] (z1);
	(g-in) -- [gluon, momentum ={$k$}, quarter right, insertion = 0.999] (g);
	(p2) -- [anti fermion, insertion = 1.] (z2);
	(p2) -- [anti fermion, momentum'=\(p_2 \)] (z2);
	};
\end{feynman}
\node at (1.2, -2.5) {$(a)$};
\end{tikzpicture}
\qquad \qquad
\begin{tikzpicture}
\begin{feynman}[scale = 0.6]
	\vertex (z1) at (0, 0) {$z_1$};
	\vertex (z2) at (4, 0) {$z_2$};
	\vertex (p1) at (0, -4);
	\vertex (p2) at (4, -4); 
	\vertex (g) at (2, 0) {$\xi$};
	\vertex (g-in) at (4, -2);

	\diagram *{
	(p1) -- [fermion, insertion=1.] (z1);
	(g-in) -- [gluon, momentum ={$k$}, quarter left, insertion = 0.999] (g);
	(p2) -- [anti fermion] (g-in) -- [anti fermion, insertion = 0.999] (z2);
	(p1) -- [fermion, momentum'=\( p_1 \)] (z1);
	(p2) -- [anti fermion, momentum'=\(p_2 \)] (g-in);
	};
\end{feynman}
\node at (1.2, -2.5) {$(b)$};
\end{tikzpicture}
\qquad \qquad
\begin{tikzpicture}
\begin{feynman}[scale = 0.6]
	\vertex (z1) at (0, 0) {$z_1$};
	\vertex (z2) at (4, 0) {$z_2$};
	\vertex (p1) at (0, -4);
	\vertex (p2) at (4, -4); 
	\vertex (g1) at (0, -2);
	\vertex (g2) at (4, -2);

	\diagram *{
	(g1) -- [gluon, momentum ={$k$}] (g2);
	(p1) -- [fermion, momentum'=\( p_1 \)] (g1) -- [fermion, insertion = 0.999] (z1);
	(p2) -- [anti fermion, momentum'=\(p_2 \)] (g2) -- [anti fermion, insertion = 0.999] (z2);
	};
\end{feynman}
\node at (1.2, -2.5) {$(c)$};
\end{tikzpicture}
\\[0.5 cm]
\begin{tikzpicture}
\begin{feynman}[scale = 0.6]
	\vertex (z1) at (0, 0) {$z_1$};
	\vertex (z2) at (4, 0) {$z_2$};
	\vertex (p1) at (0, -4);
	\vertex (p2) at (4, -4); 
	\vertex (g1) at (0, -2.5);
	\vertex (g2) at (0, -1.5);

	\diagram *{
	(g1) -- [gluon, half right] (g2);
	(p1) -- [fermion, momentum'=\( p_1 \)] (g1) -- [fermion] (g2) -- [fermion, insertion = 0.999] (z1);
	(p2) -- [anti fermion, insertion = 0.999, momentum = {$p_2$}] (z2);
	};
\end{feynman}
\node at (1.2, -2.5) {$(d)$};
\end{tikzpicture}
\qquad \qquad 
\begin{tikzpicture}
\begin{feynman}[scale = 0.6]
	\vertex (z1) at (0, 0) {$z_1$};
	\vertex (z2) at (4, 0) {$z_2$};
	\vertex (p1) at (0, -4);
	\vertex (p2) at (4, -4); 
	\vertex (g1) at (4, -2.5);
	\vertex (g2) at (4, -1.5);

	\diagram *{
	(g1) -- [gluon, half right] (g2);
	(p1) -- [fermion, momentum={$p_1$}, insertion = 0.999] (z1);
	(p2) -- [anti fermion, momentum = {$p_2$}] (g1) -- [anti fermion] (g2) -- [anti fermion, insertion = 0.999] (z2);
	};
\end{feynman}
\node at (1.2, -2.5) {$(e)$};
\end{tikzpicture}
\caption{Diagrams present at the 1-loop computation of $G^{ff}_{amp.}(p_1, p_2, z_1, z_2)$. Diagrams $(a)$ and $(b)$ contain insertion of a gluon field coming from expanding the gauge link up to the order $g$ (it is denoted by the cross labelled by $\xi$). Diagram $(c)$ corresponds to gluon exchange between two on-shell quarks. Diagrams $(d)$ and $(e)$ are the standard self-energy corrections. As it was said at the beginning of Section \ref{section-renormalization}, at the one-loop level one can use the unrenormalized perturbation theory, so that here we do not include diagrams with the self-energy counter-terms.}
 \label{fig-green-nlo}
\end{figure}

For simplicity, we consider amputated Green's functions, which are ordinary Green's functions, but with removed external propagators and spinors (or polarization vectors). For example, in the leading order in the strong coupling constant, the only non-vanishing Green's function can be represented by the following diagram:
\begin{equation}
    \begin{tikzpicture}
  \begin{feynman}
   	\vertex (z1) at (0, 0) {$z_1$};
	\vertex (z2) at (3, 0) {$z_2$};
	\vertex (p1) at (0, -2);
	\vertex (p2) at (3, -2);   
    \diagram* {
	(p1) -- [fermion, insertion=1.] (z1);
	(p2) -- [anti fermion, insertion = 1.] (z2);
	(p1) -- [fermion, momentum'=\( p_1 \)] (z1);
	(p2) -- [anti fermion, momentum'=\(p_2 \)] (z2);
    };
  \end{feynman}
\node at (-3, -1) {$G^{ff}_{amp.}(p_1, p_2, z_1, z_2)\Big|_{\alpha_S^0} = $};
\end{tikzpicture}
. \label{eq-green-lo}
\end{equation}
Crosses labeled $z_1$ and $z_2$ denote insertions of operators $\bar{\psi}(z_1)$ and $\psi(z_2)$. It is easy to find, that
\begin{equation}\label{eq-Green-LO}
\eqref{eq-green-lo} = \gamma^+ e^{-iz_1^-p_1^+ - iz_2^- p_2^+} .
\end{equation}
At the 1-loop order, when computing $G^{ff}_{amp.}(p_1, p_2, z_1, z_2)$, one has to consider diagrams presented in Fig. \ref{fig-green-nlo}.
Let us present a detailed computation of the diagram $(a)$. It allows to show methods used in computation of considered Green's functions and explain, how one obtains the necessary counter-terms. In the Sudakov basis, which is the most convenient for analysis of operators on the lightcone, usually one has to treat "$\pm$" and "$\perp$" components independently. It turns out, that the simplest way to achieve it, is to use the Schwinger parametrization of propagator:
\begin{equation}
	\frac{i}{k^2 + i0} = \int_0^\infty  e^{iy(k^2 + i0)}\: dy.
\end{equation}
When more propagators are present, it is useful to perform the following substitution in the integration:
\begin{equation}
	\prod_{j=1}^N \int_0^\infty dy_j \rightarrow \int_0^\infty d\rho \rho^{N-1} \prod_{j=1}^N \int_0^1 d\alpha_j \delta\Big( \sum_{j=1}^N \alpha_j -1 \Big).
\end{equation}
Because of divergences, we use the dimensional regularization. For gluon field we use the Feynman gauge. Averaging over colors we get that the contribution to the Green's function from the considered diagram, which we denote by $\mathcal{M}^{ff}_a$, is
\begin{equation}
	\mathcal{M}^{ff}_a = g^2 C_F \int_{z_2^-}^{z_1^-} d\xi^- \mu_R^{4-d} \int \frac{d^d k}{(2\pi)^d} \gamma^+ e^{-iz_1^-(p_1^+ - k^+) -iz_2^- p_2^+}e^{-i\xi^- k^+} \frac{\slashed{p}_1-\slashed{k}}{(p_1 -k)^2 +i0}\gamma^+ \frac{1}{k^2 + i0} ~\!, \label{eq-nlo-qq-b}
\end{equation}
where $C_F = \frac43$ is the color factor and $\mu_R$ is the renormalization scale added to keep the proper dimension of the amplitude. Phases are the result of insertion of fields with given momenta at points $z_1$, $z_2$ and $\xi$ -- the parameter $\xi$ is the position, at which one inserts the gluon field from the gauge link. Note that, since $(\gamma^+)^2 =0$, we can write
\begin{equation}
	\gamma^+ \big( \slashed{p}_1-\slashed{k} \big) \gamma^+ = 2 \big( p_1^+ - k^+ \big) \gamma^+ ~\!,
\end{equation}
so that the gamma matrix structure of \eqref{eq-nlo-qq-b} is the same as that of \eqref{eq-green-lo}. Performing the integral over $d\xi^-$ we get
\begin{equation}
	\eqref{eq-nlo-qq-b} = 2ig^2 C_F \gamma^+ \mu_R^{4-d} \int \frac{d^d k}{(2\pi)^d} \frac{ p_1^+ - k^+}{k^2 (p_1-k)^2} e^{-iz_1^-(p_1^+ - k^+) - iz_2^-p_2^+} \frac{1}{k^+} \Big( e^{-iz_1^-k^+} - e^{-iz_2^- k^+} \Big).
\end{equation}
Let us denote $\Delta z = z_2^- - z_1^-$ and use the following identity:
\begin{equation}
	\frac{1}{k^+} \Big( 1 - e^{-i\Delta z k^+} \Big) = i\Delta z \int_0^1 d\tau e^{-i\Delta z^- k^+ \tau}.
\end{equation}
The resulting integrals can be done using the previously discussed Schwinger parametrization:
\begin{equation}
\begin{aligned}
&\int \frac{d^d k}{(2\pi)^d} \int_0^1 d\tau e^{-i\Delta z k^+ \tau} \frac{ p_1^+ - k^+}{k^2 (p_1-k)^2} = \\
&-\int \frac{d^d k}{(2\pi)^d} \int_0^\infty d\rho \int_0^1 d\tau \int_0^1 d\alpha \big( (1-\alpha)p_1^+ - k^+ \big) e^{i\rho k^2 } e^{-i\tau\Delta z (k^+ +\alpha p_1^+ )\tau}. \label{eq-ren-momentum}
\end{aligned}
\end{equation}
Going to the second line we used the substitution $k \rightarrow k - \alpha p_1$.
Integration over $dk^-$ produces a delta distribution:
\begin{equation}
	\int \frac{dk^-}{2\pi} e^{2i\rho k^+ k^-} = \frac12 \rho^{-1} \delta(k^+) ~\!,
\end{equation}
which allows to easily perform the integral over $dk^+$. Keep in mind that $k^2 = 2k^+ k^- - k_\perp^2$. Integration over perpendicular momenta is just a standard Gaussian integral:
\begin{equation}
	\int \frac{d^{d-2}k_\perp }{(2\pi)^{d-2}} e^{-i\rho k^2_\perp} = \frac{1}{(4\pi)^{d/2-1}} \big(- i \rho \big)^{-d/2 + 1}.
\end{equation}
The resulting integral over $d\rho$ is ill-defined:
\begin{equation}\label{eq-drho}
	\frac{1}{(4\pi)^{d/2-1}}\int_0^\infty d\rho \: (-i \rho)^{-d/2 + 1}.
\end{equation}
To solve this problem, let us multiply the integrand by $e^{-i\rho \mu_F^2}$, where $\mu_F$ is called the factorization scale. It is equivalent with the assumption that all partons in the diagram have masses $\mu_F$ \cite{Radyushkin}. This way, instead of the integral in formula \eqref{eq-drho}, we have
\begin{equation}
	\frac{1}{(4\pi)^{d/2-1}}\int_0^\infty d\rho \: (-i \rho)^{-d/2 + 1} e^{-i\mu_F^2 \rho} = \frac{i}{4\pi} \Big( \frac{\mu_F^2}{4\pi} \Big)^{d/2-2} \Gamma\Big(2-\frac{d}{2} \Big),
\end{equation}
where $\Gamma(z)$ is the gamma function. It is clear, that in the limit of $d\rightarrow 4$, this expression becomes divergent. Now we perform the integration over the parameter $d\tau$:
\begin{equation}
	p_1^+ (1-\alpha) \int_0^1 d\tau e^{-i\alpha p_1^+ \Delta z \tau} = \frac{i}{\Delta z} \frac{1-\alpha}{\alpha} \big( e^{-i\alpha p_1^+ \Delta z} - 1 \big).
\end{equation}
All these integrals yield the following result:
\begin{equation}
\eqref{eq-ren-momentum} = - \frac{i}{(4\pi)^2} \Big( \frac{\mu_F^2}{4\pi} \Big)^{d/2-2} \Gamma\Big(2-\frac{d}{2}\Big) \frac{i}{\Delta z} \frac{1-x}{x} \big( e^{-ixp_1^+ \Delta z} - 1 \big).
\end{equation}
Hence, after multiplying by the remaining constants, we obtain
\begin{equation}\label{eq-nlo-qq-b-pre-final}
\eqref{eq-nlo-qq-b} = \frac{1}{8\pi^2}C_F g^2 \Big( \frac{\mu_F^2}{4\pi\mu_R^2} \Big)^{d/2-2} \Gamma\Big( 2 - \frac{d}{2} \Big)\gamma^+ \int_0^1 d\alpha \frac{1-\alpha}{\alpha} \Big( e^{-i(z_2^- - z_1^-) \alpha p_1^+} - 1 \Big) e^{-iz_1^- p_1^+ + iz_2^-p_2^+}.
\end{equation}
It can be conveniently written using so-called 'plus prescription':
\begin{equation}
	\int_0^1 d\alpha \: \Big[ \frac{1-\alpha }{\alpha } \Big]_+ f(\alpha) := \int_0^1 d\alpha  \frac{1-\alpha }{\alpha }  \big( f(\alpha) - f(0) \big).
\end{equation}
Moreover, using the Laurent expansion of the gamma function ($d=4-\varepsilon$),
\begin{equation}
	\Gamma\Big(2 - \frac{d}{2}  \Big) = \Gamma\Big( \frac{\varepsilon}{2} \Big) = \frac{2}{\varepsilon} - \gamma + \mathcal{O}(\epsilon),
\end{equation}
we can write
\begin{equation}
	\Big( \frac{\mu_F^2}{4\pi\mu_R^2} \Big)^{d/2-2} \Gamma\Big(2- \frac{d}{2} \Big) = \frac{2}{\varepsilon}\Big( \frac{e^\gamma \mu_F^2}{4\pi \mu_R^2} \Big)^{-\varepsilon/2}.
\end{equation}
Finally, using $\alpha_S = g^2/4\pi$, we can re-express the result in a compact form
\begin{equation}\label{eq-Ma-fin}
	\eqref{eq-nlo-qq-b-pre-final} = \frac{2}{\varepsilon}\frac{\alpha_S}{2\pi}C_F \Big( \frac{e^\gamma \mu_F^2}{4\pi \mu_R^2} \Big)^{-\varepsilon/2} \gamma^+ \int_0^1 d\alpha \: \Big[ \frac{1-\alpha}{\alpha} \Big]_+ e^{-ip_1^+ \big( (1-\alpha )z_1^- + \alpha z_2^- \big) - iz_2^- p_2^+}.
\end{equation} 
At the first sight, this expression looks complicated and finding the counter-term may appear a cumbersome task. However, let us notice, that Eq. \eqref{eq-Ma-fin} can be written in terms of the LO amputated Green's function, see Eq. \eqref{eq-Green-LO}:
\begin{equation}
	\eqref{eq-Ma-fin} = \frac{2}{\varepsilon}\frac{\alpha_S}{2\pi}C_F\Big( \frac{e^\gamma \mu_F^2}{4\pi \mu_R^2} \Big)^{-\varepsilon/2} \int_0^1 d\alpha\:  \Big[ \frac{1-\alpha}{\alpha} \Big]_+ G^{ff}_{amp.}(p_1, p_2, (1-\alpha)z_1^- + \alpha z_2^-, z_2)\Big|_{\alpha_S^0}.
\end{equation}
Using this result, one can easily notice, that the counter-term necessary to cancel divergence from the diagram $(a)$ at the order $\alpha_S^1$  is
\begin{equation}
	 -\frac{2}{\varepsilon}\Big( \frac{e^\gamma \mu_F^2}{4\pi \mu_R^2} \Big)^{-\varepsilon/2} \int_0^1 d\alpha \: \frac{\alpha_S}{2\pi} C_F \Big[ \frac{1-\alpha}{\alpha} \Big]_+ \mathcal{O}^{ff}\big( (1-\alpha)z_1^- + \alpha z_2^-, z_2 \big).
\end{equation}
After considering all diagrams presented in Fig. \ref{fig-green-nlo} one obtains the renormalized non-singlet operator:
\begin{equation}
\begin{aligned}
	\mathcal{O}^{(f)}_{NS, R} (z_1, z_2; \mu_F) =& \: \mathcal{O}^{(f)}_{NS} (z_1, z_2) \:+\\&- \frac{2}{\varepsilon}\Big( \frac{e^\gamma \mu_F^2}{4\pi \mu_R^2} \Big)^{-\varepsilon/2}  \int_0^1 d\alpha_1 \: d\alpha_2 \: K^{ff}_{NS} (\alpha_1, \alpha_2) \: \mathcal{O}^{(f)}_{NS} \big( \bar{\alpha}_1 z_1 + \alpha_1 z_2, \bar{\alpha}_2 z_2 + \alpha_2 z_1 \big)~\!,
\end{aligned}
\end{equation}
where we denote $\bar{\alpha}_i = 1- \alpha_i$ and the evolution kernel is given by
\begin{equation}\label{eq-kernel-K}
	K^{ff}_{NS} (\alpha_1, \alpha_2) = \frac{\alpha_S}{2\pi}C_F \bigg( 1 + \delta(\alpha_1) \Big[ \frac{\bar{\alpha}_2 }{\alpha_2} \Big]_+ + \delta(\alpha_2) \Big[ \frac{\bar{\alpha}_1 }{\alpha_1} \Big]_+ + \frac{3}{2} \delta(\alpha_1) \delta(\alpha_2) \bigg).
\end{equation}
Other evolution kernels in the position space can be found in \cite{Diehl, Radyushkin}. The relation between renormalized and bare operators can be inverted:
\begin{equation}\label{eq-Oqq-bare-vs-ren}
\begin{aligned}
	\mathcal{O}^{(f)}_{NS} (z_1, z_2) = & \: \mathcal{O}^{(f)}_{NS, R} (z_1, z_2; \mu_F)\:+ \\& + \frac{2}{\varepsilon}\Big( \frac{e^\gamma \mu_F^2}{4\pi \mu_R^2} \Big)^{-\varepsilon/2}  \int_0^1 d\alpha_1 \: d\alpha_2 \: K^{ff}_{NS} (\alpha_1,\alpha_2) \: \mathcal{O}^{(f)}_{NS, R} \big( \bar{\alpha}_1 z_1 + \alpha_1 z_2, \bar{\alpha}_2 z_2 + \alpha_2 z_1 ; \mu_F \big)~\!,
\end{aligned}
\end{equation}
where we used the fact that, at the order $\alpha_S$, in the counter-term we can use either renormalized operator, or the bare one (the difference is of order $\alpha_S^2$).
This result allows to write bare GPDs in terms of renormalized parton distributions and their convolutions with the corresponding evolution kernels.

\section{From evolution of operators in the position space to evolution of GPDs}
In this section we finally obtain the relation between bare and renormalized parton distributions and formulate their evolution equations.
We define renormalized GPDs in the same way as bare one, but using renormalized string operator. For example, the renormalized non-singlet vector quark distribution is
\begin{equation}
F^f_R(x,\xi,t; \mu_F):= \int \frac{dz^-}{2\pi} e^{ixp^+ z^-}\bra{N(\mathbf{p}_2, \sigma_2)_{\mathrm{out}} } \mathcal{O}^{(f), H}_{NS, R}(-z^-/2, z^-/2; \mu_F) \ket{N(\mathbf{p}_1, \sigma_1)_{\mathrm{in}} }.
\end{equation}
From the Formula \eqref{eq-Oqq-bare-vs-ren} we obtain the difference between bare and renormalized GPD:
\begin{equation}
\begin{aligned}
	&F^f_{NS,R}(x,\xi,t; \mu_F) - F^f_{NS}(x,\xi,t)= \\
	=&-\frac{2}{\varepsilon}\Big( \frac{e^\gamma \mu_F^2}{4\pi \mu_R^2} \Big)^{-\varepsilon/2}\int \frac{dz^-}{2\pi} e^{ixp^+ z^-} \int_0^1 d\alpha_1 \: d\alpha_2 \: K^{ff}_{NS} (\alpha_1, \alpha_2)\times \\& \bra{N(\mathbf{p}_2, \sigma_2)_{\mathrm{out}} } \mathcal{O}^{(f), H}_{NS, R} \big( -z^-(1/2 - \alpha_1), z^-(1/2 - \alpha_2); \mu_F \big) \ket{N(\mathbf{p}_1, \sigma_1)_{\mathrm{in}} }. \label{eq-gpd-kernel-1}
\end{aligned}
\end{equation}
To write it as a parton distribution convoluted with some kernel, let us first use the Poincar\'e covariance of the matrix element:
\begin{equation}
\begin{aligned}
	&\bra{N(\mathbf{p}_2, \sigma_2)_{\mathrm{out}} } \mathcal{O}^{(f)}_{NS, R} \big( -z^-(1/2 - \alpha_1), z^-(1/2 - \alpha_2) \big) \ket{N(\mathbf{p}_1, \sigma_1)_{\mathrm{in}} } =\\=& e^{-ip^+z^- \xi (\alpha_1 - \alpha_2)} \bra{N(\mathbf{p}_2, \sigma_2)_{\mathrm{out}} } \mathcal{O}^{(f)}_{NS, R} \big( -z^-(1 - \alpha_1 -\alpha_2)/2, z^-(1 - \alpha_1- \alpha_2)/2; \mu_F \big) \ket{N(\mathbf{p}_1, \sigma_1)_{\mathrm{in}} }.
\end{aligned}
\end{equation}
Using the substitution $z'= z(1-\alpha_1 - \alpha_2)$ we rewrite the integrals from the Eq. \eqref{eq-gpd-kernel-1} in the following form:
\begin{equation}
\int \frac{dz'^-}{2\pi} e^{ip^+ z'^-\frac{x-\xi(\alpha_1 - \alpha_2)}{1-\alpha_1 - \alpha_2}} \int_0^1 d\alpha_1 \: d\alpha_2 \: K^{ff}_{NS} (\alpha_1, \alpha_2) \bra{N(\mathbf{p}_2, \sigma_2)_{\mathrm{out}} }  \mathcal{O}^{(f)}_{NS, R} \big( -z'^-/2), z'^-/2; \mu_F \big) \ket{N(\mathbf{p}_1, \sigma_1)_{\mathrm{in}} }.
\end{equation}
It is easy to verify that it can be written as
\begin{equation}
\begin{aligned}
	&\int_{-1}^{1} dx' \: \int_0^1 d\alpha_1 d\alpha_2 \: \delta \Big( x - x'(1-\alpha_1 - \alpha_2) - \xi(\alpha_1 - \alpha_2) \Big)  K^{ff}_{NS} (\alpha_1, \alpha_2) F^f_{NS,R}(x', \xi, t) =\\=& \int_{-1}^{1} dx' \: \frac{1}{|\xi|}V_{NS}\Big( \frac{x}{\xi}, \frac{x'}{\xi} \Big) F_{NS, R}^f (x', \xi, t; \mu_F),
\end{aligned}
\end{equation}
where the evolution kernel for the non-singlet GPD is defined as
\begin{equation}
	\frac{1}{|\xi|}V_{NS}\Big( \frac{x}{\xi}, \frac{x'}{\xi} \Big) = \int_0^1 d\alpha_1 d\alpha_2  \: \delta \Big( x - x'(1-\alpha_1 - \alpha_2) - \xi(\alpha_1 - \alpha_2) \Big)  K^{ff}_{NS} (\alpha_1, \alpha_2) .
\end{equation}
The resulting formula takes the same form as the one in Eq. \eqref{eq-gpdB-gpdR}:
\begin{equation}
	F_{NS}^f (x, \xi, t) = F_{NS, R}^f (x, \xi, t; \mu_F) +\frac{2}{\varepsilon}\Big( \frac{e^\gamma \mu_F^2}{4\pi \mu_R^2} \Big)^{-\varepsilon/2}\int_{-1}^{1} dx' \: \frac{1}{|\xi|}V_{NS}\Big( \frac{x}{\xi}, \frac{x'}{\xi} \Big) F_{NS, R}^f (x', \xi, t; \mu_F).
\end{equation}
Demanding that the bare operators are independent of the factorization scale, one obtains the evolution equation:
\begin{equation}
	\mu_F^2 \frac{d}{d \mu_F^2} F_{NS, R}^f (x, \xi, t; \mu_F) = \int_{-1}^{1} dx' \: \frac{1}{|\xi|}V_{NS}\Big( \frac{x}{\xi}, \frac{x'}{\xi} \Big) F_{NS, R}^f (x', \xi, t; \mu_F).
\end{equation}
In the case of singlet GPDs all these relations take the same form, but with the evolution kernels in the form of a matrix with indices corresponding to quark and gluon distributions. All kernels can be found in \cite{Diehl, Radyushkin}.

\chapter{Photoproduction of photon pairs}\label{chapter-diphoton}
After introducing the collinear factorization, generalized parton distributions and methods used in the computation of higher-order QCD corrections, we present a detailed analysis of photoproduction of photon pair with large invariant mass on a nucleon, i.e. in the process 
$$\gamma(\mathbf{q}, \epsilon) \: N(\mathbf{p}_1, \sigma_1) \longrightarrow \gamma' (\mathbf{q}_1, \epsilon_1(\mathbf{q}_1) ) \: \gamma'' (\mathbf{q}_2, \epsilon(\mathbf{q}_2) ) \:  N(\mathbf{p}_2, \sigma_2) .$$
$\epsilon$ denotes polarization vectors of photons.
It is the simplest reaction, which allows to study the factorization in a $2\rightarrow 3$ process -- and hence a good starting point in studies of this general class of processes. Because of the charge conjugation symmetry, it is sensitive only to non-singlet combinations of GPDs (in contrary to e.g. DVCS, TCS or DVMP), so it might be useful in extraction of GPDs from experimental data. In Section \ref{section-kinematics} we describe kinematics of this process. In Section \ref{section-digamma-LO} we show the leading-order results \cite{cytuj-promotora1, cytuj-promotora11}. The remaining part of the work are my original results. Throughout Sections \ref{section-23-point} - \ref{section-5-point} we present a detailed computation of one loop QCD corrections to the hard-part. The full result can be found in Appendix \ref{Appdx-full-amplitude}. The proof of validity of the collinear factorization at NLO is shown in Section \ref{section-factorization}. 
\section{Kinematics, amplitude, and the differential cross section}\label{section-kinematics}
We parametrize particles' momenta and polarizations using the following light-cone four-vectors:
\begin{equation}
p^\mu = \frac{\sqrt{s}}{2}(1,0,0,1), \quad \quad n^\mu =  \frac{\sqrt{s}}{2}(1,0,0,-1),
\end{equation}
the parameter $s$ will be described in more detail later.
We use the Sudakov decomposition:
\begin{equation}
\begin{aligned}
&p_1^\mu = (1+\xi)p^\mu + \frac{m_N^2}{s(1+\xi)}n^\mu, \qquad p_2^\mu = (1-\xi)p^\mu + \frac{m_N^2 + \vec{\Delta}_t^2}{s(1-\xi)}n^\mu + \Delta_\perp^\mu, \qquad q^\mu = n^\mu,
\\
&q_1^\mu = \alpha n^\mu + \frac{(\vec{p}_t-\frac{1}{2}\vec{\Delta}_t)^2}{\alpha s} p^\mu + p_\perp^\mu - \frac{1}{2}\Delta_\perp^\mu,
\\
&q_2^\mu = \bar{\alpha} n^\mu + \frac{(\vec{p}_t+\frac{1}{2}\vec{\Delta}_t)^2}{\bar{\alpha} s} p^\mu - p_\perp^\mu - \frac{1}{2}\Delta_\perp^\mu.
\end{aligned}
\end{equation}
$\Delta^\mu = p_2^\mu - p_1^\mu$, $m_N$ is the mass of the nucleon, and the parameter $\xi$ is called skewedness. We denote the Euclidean product as $\vec{p}_t^{\phantom{a}2} = -p_\perp^2 \geq 0$.
We use the gauge $A^- = 0$ for the electromagnetic field. It implies that for each photon's polarization vector 
$$\epsilon \cdot p = 0.$$
Hence, the polarization vectors are the following:
\begin{equation}
\begin{aligned}
&\epsilon^\mu(\mathbf{q}_1) = \epsilon_{\perp}^\mu(\mathbf{q}_1)+ \frac{2\vec{\epsilon}_t(\mathbf{q}_1)\big{(}\vec{p}_t - \frac{1}{2}\vec{\Delta}_t\big{)}}{\alpha s}p^\mu,
\\
&\epsilon^\mu(\mathbf{q}_2) = \epsilon_{\perp}^\mu(\mathbf{q}_1) - \frac{2\vec{\epsilon}_t(\mathbf{q}_2)\big{(}\vec{p}_t + \frac{1}{2}\vec{\Delta}_t\big{)}}{\bar{\alpha} s}p^\mu,
\\
&\epsilon^\mu(\mathbf{q}) = \epsilon_\perp^\mu(\mathbf{q}).
\end{aligned}
\end{equation}
The process is described in terms of 4 Mandelstam variables:
\begin{equation}
S_{\gamma N} = (p_1 + q)^2, \quad t = (p_1 - p_2)^2, \quad M_{\gamma\gamma}^2 = (q_1 + q_2)^2, \quad u' = (q_2 - q)^2.
\end{equation}
For further convenience we relate them to other invariants:
\begin{equation}
\begin{aligned}
&S_{\gamma N} = (1+\xi) s + m_N^2 \: \implies \: s = \frac{S_{\gamma N} - m_N^2}{1+\xi},
\\
&t' = (q_1 - q)^2 = t - M_{\gamma\gamma}^2 - u'.
\end{aligned}
\end{equation}
Like in the work \cite{cytuj-promotora1}, we consider the situation when $M_{\gamma\gamma}^2$ is large and $t$ is small (with respect to the mass squared of the nucleon). It implies that transverse momenta of outgoing photons are large and approximately opposite. Let us discuss the case $\Delta_\perp^\mu = 0$. It is justified, since in the considered kinematics one has $|p_\perp^2| \gg |\Delta_\perp^2|$. The simplified kinematical relations used in the calculation of the hard part are:
\begin{equation}
\begin{aligned} \label{eq-kin-simplified}
& \xi = \frac{\tau}{2-\tau}, \qquad \tau = \frac{M_{\gamma\gamma}^2 - t}{S_{\gamma N} - m_N^2},\\
&-t' = \frac{\vec{p}_t^{\phantom{a}2}}{\alpha}, \qquad -u' = \frac{\vec{p}_t^{\phantom{a}2}}{\bar{\alpha}}, \qquad \alpha + \bar{\alpha} = 1.
\end{aligned}
\end{equation}
The second line of Eq. \eqref{eq-kin-simplified} yields
\begin{equation}
\vec{p}_t^{\phantom{a}2} = -\frac{u' t'}{u' + t'}, \qquad \alpha = \frac{u'}{u' + t'}, \qquad \bar{\alpha} = \frac{t'}{u' + t'}.
\end{equation}
Because of the charge conjugation symmetry, the amplitude of the photoproduction of two real photons on a proton is sensitive only to non-singlet (odd with respect to the charge conjugation) parton distributions. To avoid proliferation of indices, from now on by 
$$H^q (x, \xi, t; \mu_F) \quad \mathrm{and} \quad E^q (x, \xi, t; \mu_F)$$
we denote the \textit{renormalized}, \textit{non-singlet} vector GPDs corresponding to quark of flavour $q$, see Eq. \eqref{eq-gpd-q1GPD}. We neglect the axial GPDs, since the LO analysis \cite{cytuj-promotora1} has shown, that their contribution in this process is negligible. The scattering amplitude of the full process $\mathcal{M}$ can be written in the following way:
\begin{equation}
\mathcal{M} = \frac{1}{2s} \sum_q \Big[ \mathcal{H}^q \bar{u}(\mathbf{p}_2, \sigma_2) \gamma^+ u(\mathbf{p}_1, \sigma_1) + \mathcal{E}^q \bar{u}(\mathbf{p}_2, \sigma_2) \frac{i\sigma^{\mu+}\Delta_\mu}{2m_N} u(\mathbf{p}_1, \sigma_1) \Big],
\end{equation}
where $u(\mathbf{p}, \sigma)$ denotes the bispinor of the nucleon, and
\begin{equation}
\begin{aligned}
&\mathcal{H}^q = \int_{-1}^1 dx \: H^q (x, \xi, t; \mu_F)  \mathcal{T}^q_R (x, \dots; \mu_F), \\
&\mathcal{E}^q = \int_{-1}^1 dx \: E^q  (x, \xi, t; \mu_F)  \mathcal{T}^q_R (x, \dots; \mu_F), 
\end{aligned}
\end{equation}
where $\mathcal{T}^q_R (x, \dots; \mu_F)$ denotes the ``renormalized'' amplitude of the hard sub-process:
\begin{equation}
    \mathcal{T}_R^i (x, \xi, \dots; \mu_F) \equiv \mathcal{C}^i_0 (x, \xi, \dots) +  \mathcal{C}^i_1(x,\xi,\dots) + \log\Big(\frac{s}{\mu_F^2}\Big)  \mathcal{C}^i_{coll.}(x,\xi, \dots),
\end{equation}
as it is described by the Formula \eqref{eq-final-factorization-nlo}. Ellipsis denotes all parameters other than $x$, on which depends the hard amplitude. Here, the hard scale is provided by $s$ instead of $|Q^2|$ used in Chapter \ref{chapter-factorization}.

The amplitude squared, summed (averaged) over the final (initial) nucleon spins $\sigma_1$, $\sigma_2$ reads:
\begin{equation}
\frac{1}{2}\sum_{\sigma_1, \sigma_2} \big| \mathcal{T} \big|^2 = \frac{1}{4} \Big[ (1-\xi^2) \big| \sum_q \mathcal{H}^q \big|^2 - 2\xi^2 \mathrm{Re} \sum_{q} \mathcal{H}^q \sum_{q'} \mathcal{E}^{q'\: *} + \frac{\xi^4}{1-\xi^2}  \big| \sum_q \mathcal{E}^q \big|^2 \Big].
\end{equation}
The differential cross section, averaged over the initial photon polarization and summed over the final polarizations, has the following form \cite{cytuj-promotora1}:
\begin{equation}
\begin{aligned}
&\frac{d\sigma}{dM_{\gamma\gamma}^2 dt d(-u')} = \frac{1}{2}\frac{1}{32(2\pi)^3 S_{\gamma N}^2 M_{\gamma\gamma}^2} \sum_{\mathrm{pol.}}\sum_{\sigma_1, \sigma_2} \frac{1}{4} \big| \mathcal{T} \big|^2 =\\
& = \frac{1}{(16\pi)^3 S_{\gamma N}^2 M_{\gamma\gamma}^2 } \sum_{\mathrm{pol.}}\Big[ (1-\xi^2) \big| \sum_q \mathcal{H}^q \big|^2 - 2\xi^2 \mathrm{Re} \sum_{q} \mathcal{H}^q \sum_{q'} \mathcal{E}^{q\: *} + \frac{\xi^4}{1-\xi^2}  \big| \sum_q \mathcal{E}^q \big|^2 \Big].
\end{aligned}
\end{equation}
In Section \ref{section-digamma-LO} we recall results of \cite{cytuj-promotora1} and \cite{cytuj-promotora11} concerning the LO amplitude, in particular the term $\mathcal{C}^q_0 (x, \xi, \dots)$.
In the subsequent part we present the computation of the NLO amplitude of the hard part, i.e. $\mathcal{C}^q_1 (x, \xi, \dots)$ and $\mathcal{C}^q_{coll.}(x,\xi, \dots)$. 

To reduce the number of considered NLO diagrams (originally 48), we consider a process with 3 \textit{incoming} photons with momenta $(k_1,k_2,k_3)$ and polarisation vectors $(\epsilon_1, \epsilon_2, \epsilon_3)$, and then we sum the amplitudes over all combinations
$$\big{(}(k_1,\epsilon_1), (k_2, \epsilon_2) ,(k_3,\epsilon_3)\big{)}$$
corresponding to all permutations of 
$$\big{(}(-q_1,\epsilon^* (\mathbf{q}_1)), (-q_2, \epsilon^* (\mathbf{q}_2) ) ,(q,\epsilon(\mathbf{q}) )\big{)}.$$
We introduce the following parametrization of these invariants:
\begin{equation}
2pk_i := \beta_i s, \quad 2k_1 k_2 := \kappa_3 s, \quad 2k_2 k_3 := \kappa_1  s, \quad 2k_1 k_3 := \kappa_2  s.
\end{equation}
It is easy to check that
$$
\kappa_i = -2\xi \beta_i.
$$
Scalar products of the corresponding momenta are:
\begin{equation}
\begin{aligned}
&2pq_1 = \alpha s, \qquad 2pq_2 = \bar{\alpha} s, \qquad 2pq = s;
\\
&2q_1 q_2 = M_{\gamma\gamma}^2, \qquad -2qq_2 = u', \qquad -2qq_1 = t'.
\end{aligned}
\end{equation}
Hence, parameters $\beta_i$ and $\kappa_i$ are the following permutations:
\begin{equation}
\begin{aligned}
&\{\beta_i\} = \: \mathrm{permutations} \: \mathrm{of} \: \Big\{ -\alpha, - \bar{\alpha}, 1 \Big\}, 
\\
&\{\kappa_i\} = \: \mathrm{permutations} \: \mathrm{of} \: \Big\{ \frac{M_{\gamma\gamma}^2}{s}, \frac{u'}{s}, \frac{t'}{s} \Big\}.
\end{aligned}
\end{equation}
We often use the following abbreviations: $\bar{p} := (x+\xi) p$, $\underline{p} := (x-\xi) p$.

\section{The LO amplitude}\label{section-digamma-LO}
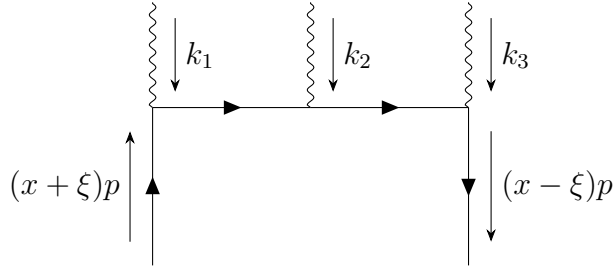
\begin{figure}[H]
    \centering
    \begin{tikzpicture}[scale = 0.7]
  \begin{feynman}
    \vertex (q1) at (0, 0);
	\vertex (q1a) at (0.5, 0);
	\vertex (q1b) at (2.5, 0);
    \vertex (q3) at (3, 0);
	\vertex (q4) at (3, 0);
    \vertex (q5) at (6, 0);
    \vertex (in) at (0, -3);
    \vertex (out) at (6, -3);
    \vertex (k1) at (0, 2);
    \vertex (k2) at (3, 2);
    \vertex (k3) at (6, 2);
    \diagram* {
      (q1) -- [fermion] (q3) -- [fermion]  (q5);
      (k1) -- [photon, momentum=\(k_1\)] (q1);
      (k2) -- [photon, momentum=\(k_2\)] (q3);
      (k3) -- [photon, momentum=\(k_3\)] (q5);
      (in) -- [fermion, momentum=\((x+\xi)p\)] (q1);
      (q5) -- [fermion, momentum=\((x-\xi)p\)] (out);
    };
  \end{feynman}
\end{tikzpicture}
    \caption{The general form of the LO diagram. As it was described in Sec. \ref{section-kinematics}, to simplify the computation we first consider a diagram with 3 incoming photons, and at the end of the computations we use appropriate substitution and sum over permutations.}
    \label{fig-lo-diagram}
\end{figure}

At the leading order there are 6 tree diagrams, each corresponding to a different permutation of photons attached to the fermion line, as it is shown in Fig. \ref{fig-lo-diagram}.
The hard-part amplitude corresponding to a given permutation reads
\begin{equation}\label{eq-M-LO-0}
	Tr\big[ i\mathcal{M}^0_{1,2,3}~ \slashed{p} \big] = ie_q^3 \frac{1}{2(x+\xi)pk_1 + i0}\frac{1}{-2(x-\xi)pk_3 + i0} Tr \Big( \slashed{\epsilon}_3 ( \underline{\slashed{p}} - \slashed{k}_3 ) \slashed{\epsilon}_2  ( \bar{\slashed{p}} + \slashed{k}_1 ) \slashed{\epsilon}_1 \slashed{p} \Big).
\end{equation}
Let us denote $i0_i = i0 \times sgn(\beta_i)$. Using definitions from Sec. \ref{section-kinematics} and $\slashed{p}\slashed{p}=0$, we rewrite \eqref{eq-M-LO-0} in a compact form:
\begin{equation}\label{eq-amp-lo-1}
	Tr\big[ i\mathcal{M}^0_{1,2,3}~ \slashed{p} \big] = ie_q^3 s^{-2} \frac{1}{\beta_1 \beta_3} \frac{1}{x+\xi + i0_1} \frac{1}{x-\xi -i0_3} Tr \Big( \slashed{\epsilon}_3\slashed{k}_3  \slashed{\epsilon}_2  \slashed{k}_1 \slashed{\epsilon}_1 \slashed{p} \Big).
\end{equation}
The "$-$" sign standing by $\slashed{k}_3$ in the trace got absorbed by the "$-$" in the propagator $\frac{1}{-(x-\xi) +i0_3}$.
It turns out that the trace structure present here often appears in loop calculations, and because of that we write
\begin{equation}
4\mathcal{A}_{1,2,3} := - Tr \Big( \slashed{\epsilon}_3 \slashed{k}_3 \slashed{\epsilon}_2 \slashed{k}_1 \slashed{\epsilon}_1 \slashed{p} \Big).
\end{equation}
Let us observe that the interchange of indices $1\leftrightarrow 3$ in Eq. \eqref{eq-amp-lo-1} does not change that expression if $sgn(\beta_1) = sgn(\beta_3)$, and results in complex conjugation of the term 
$$\frac{1}{x+\xi + i0_1} \frac{1}{x-\xi -i0_3}$$
if $sgn(\beta_1) = -sgn(\beta_3)$. The full amplitude of the hard sub-process is obtained by summing over all permutations of photons $k_i$, so that the only terms proportional to the imaginary part of propagators which survive, are those from diagrams in which the incoming photon is the middle one (so that $k_2 = q$). It turns out, that terms proportional to the real part of propagators cancel as well, since in the considered kinematics
\begin{equation}
	\sum_{perm.} \frac{1}{\beta_1 \beta_3}  Tr \Big( \slashed{\epsilon}_3\slashed{k}_3  \slashed{\epsilon}_2  \slashed{k}_1 \slashed{\epsilon}_1 \slashed{p} \Big) = \frac{1}{\beta_1 \beta_2 \beta_3} \sum_{perm.} \beta_2  Tr \Big( \slashed{\epsilon}_3\slashed{k}_3  \slashed{\epsilon}_2  \slashed{k}_1 \slashed{\epsilon}_1 \slashed{p} \Big)=0.
\end{equation}
That can be checked by a straightforward computation, for reference see \cite{cytuj-promotora11} (note that the original formula in \cite{cytuj-promotora1} contained a mistake). Hence, after summing over all the possible photons' permutations, one arrives at the following hard part:
\begin{equation}
\begin{aligned}
	&\mathcal{C}^q_0 (x, \xi, \dots) = \sum_{\mathrm{perm.}}Tr\big[ \mathcal{M}^0_{1,2,3}~ \slashed{p} \big]=  \frac{e_q^3}{ s^2 \alpha \bar{\alpha} } \bigg( \frac{1}{x+\xi - i0} \frac{1}{x-\xi +i0} - c.c. \bigg) Tr \Big( \slashed{\epsilon}^* (\mathbf{q}_2) \slashed{q}_2  \slashed{\epsilon}(\mathbf{q})  \slashed{q}_1 \slashed{\epsilon}^* (\mathbf{q}_1) \slashed{p} \Big)\\
&\phantom{\mathcal{C}^q_0 (x, \xi, \dots) } = -i\frac{2\pi e_q^3}{s\alpha \bar{\alpha}} \frac{1}{\xi} \Big( \delta(x-\xi) + \delta(x+\xi) \Big) \times \\ & \qquad \bigg[ (\alpha - \bar{\alpha} )\big( \vec{\epsilon^*}_t(\mathbf{q}_1) \vec{\epsilon^*}_t(\mathbf{q}_2) \big) \big( \vec{p}_t \vec{\epsilon}_t (\mathbf{q}) \big) - \big(  \vec{p}_t \vec{\epsilon^*}_t(\mathbf{q}_1) \big) \big(  \vec{\epsilon}_t (\mathbf{q}) \vec{\epsilon^*}_t(\mathbf{q}_2) \big) + \big(  \vec{p}_t \vec{\epsilon^*}_t(\mathbf{q}_2) \big) \big(  \vec{\epsilon}_t(\mathbf{q}) \vec{\epsilon^*}_t(\mathbf{q}_1) \big) \bigg].
\end{aligned}
\end{equation}
\section{Self-energy corrections}\label{section-23-point}
Before presenting the computation of NLO diagrams, let us recapitulate our conventions. We use the dimensional regularization with the number of dimensions $d = 4 - \varepsilon$. It results in the modification of the strong coupling $g \rightarrow g \mu^{2-d/2}$, which ensures the correct dimensionality of the amplitude. For gamma matrices in $d$ dimensions one has
\begin{equation}
\begin{aligned}
	&\gamma^\mu \gamma^\nu \gamma_\mu = (2-d)\gamma^\nu = (-2 + \varepsilon)\gamma^\nu,\\
	& \gamma^\mu \gamma^\alpha \gamma^\beta \gamma^\delta \gamma_\mu = -2 \gamma^\delta \gamma^\beta \gamma^\alpha + \varepsilon \gamma^\alpha \gamma^\beta \gamma^\delta.
\end{aligned}
\end{equation}
We use the Feynman formula for combining denominators
\begin{equation}
	\frac{1}{a_1 \dots a_n } = \Gamma(n) \int_0^1 du_1 \dots \int_0^1 du_n \delta \big( 1 - u_1 - \dots - u_n \big) \frac{1}{\big( u_1 a_1 + \dots u_n a_n \big)^n } .
\end{equation}
Finally, we perform momentum integration according to the following formulas:
\begin{equation}\label{eq-loop-int}
\begin{aligned}
	&\int \frac{d^d k}{(2\pi)^d} \frac{1}{\Big( k^2 - \Delta + i0 \Big)^n } = \frac{i (-1)^n}{(4\pi)^{d/2}} \Big( \Delta - i0 \Big)^{d/2 - n} \frac{\Gamma\big( n-d/2 \big)}{\Gamma(n) },\\
	&\int \frac{d^d k}{(2\pi)^d}\frac{k^\mu k^\nu}{(k^2 + \Delta + i0)^n} = -\frac{i}{2}g^{\mu\nu}\frac{(-1)^n}{(4\pi)^{d/2}}\frac{\Gamma(n-1-d/2)}{\Gamma(n)}\Big{(} -\Delta - i0 \Big{)}^{d/2+1-n}.
\end{aligned}
\end{equation}
Let us emphasise that, since the hard amplitude is integrated over the parameter $x$ and may contain singularities, it is important to keep the $i0$ terms through all the calculation.\\
We use the following naming convention: a diagram with a loop at left/right/middle involving $n$ propagators will be denoted by superscript $n.L$/ $n.R$/ $n.M$. Let us recall one more time, that we often use the short-hand notation $\bar{p} := (x+\xi) p$, $\underline{p} := (x-\xi) p$.

\begin{figure}[H]
    \centering
    \begin{tikzpicture}[scale = 0.7]
  \begin{feynman}
    \vertex (q1) at (0, 0);
	\vertex (q1a) at (0.5, 0);
	\vertex (q1b) at (2.5, 0);
    \vertex (q3) at (3, 0);
	\vertex (q4) at (3, 0);
    \vertex (q5) at (6, 0);
    \vertex (in) at (0, -3);
    \vertex (out) at (6, -3);
    \vertex (k1) at (0, 2);
    \vertex (k2) at (3, 2);
    \vertex (k3) at (6, 2);
    \diagram* {
      (q1) -- [fermion] (q3) -- [fermion]  (q5);
      (k1) -- [photon, momentum=\(k_1\)] (q1);
      (k2) -- [photon, momentum=\(k_2\)] (q3);
      (k3) -- [photon, momentum=\(k_3\)] (q5);
      (in) -- [fermion, momentum=\((x+\xi)p\)] (q1);
      (q5) -- [fermion, momentum=\((x-\xi)p\)] (out);
	(q1b) -- [gluon, half left, momentum = $k$] (q1a);
    };
  \end{feynman}
\end{tikzpicture}
\quad
    \begin{tikzpicture}[scale = 0.7]
  \begin{feynman}
    \vertex (q1) at (0, 0);
	\vertex (q1a) at (3.5, 0);
	\vertex (q1b) at (5.5, 0);
    \vertex (q3) at (3, 0);
	\vertex (q4) at (3, 0);
    \vertex (q5) at (6, 0);
    \vertex (in) at (0, -3);
    \vertex (out) at (6, -3);
    \vertex (k1) at (0, 2);
    \vertex (k2) at (3, 2);
    \vertex (k3) at (6, 2);
    \diagram* {
      (q1) -- [fermion] (q3) -- [fermion]  (q5);
      (k1) -- [photon, momentum=\(k_1\)] (q1);
      (k2) -- [photon, momentum=\(k_2\)] (q3);
      (k3) -- [photon, momentum=\(k_3\)] (q5);
      (in) -- [fermion, momentum=\((x+\xi)p\)] (q1);
      (q5) -- [fermion, momentum=\((x-\xi)p\)] (out);
	(q1b) -- [gluon, half left, momentum = $k$] (q1a);
    };
  \end{feynman}
\end{tikzpicture}
    \caption{Diagrams $2.L$ and $2.R$.}
    \label{2LR}
\end{figure}
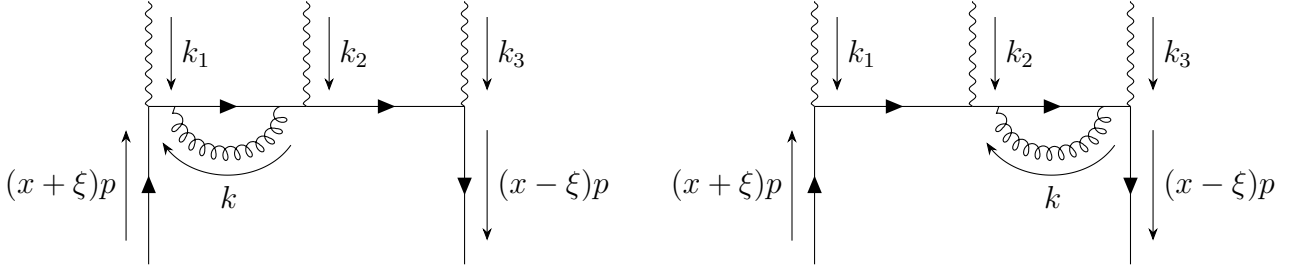

After this discussion, we can proceed with the computation of loop corrections. The simplest ones, self-energy diagrams, are shown in Fig. \ref{2LR}.
\noindent Amplitude corresponding to $2.L$ reads
\begin{equation}
\begin{aligned}
&Tr\Big[ i\mathcal{M}^{2.L}_{1,2,3} ~\slashed{p}\Big] = -C_F(ie_q)^3 (ig)^2\Big{(} \frac{i}{\big{(}(x+\xi)p + k_1 \big{)}^2 + i0} \Big{)}^2 \frac{i}{\big{(}(x-\xi)p - k_3 \big{)}^2 + i0} \times
\\
&\phantom{Tr\Big[ i\mathcal{M}^{2.L}_{1,2,3} ~\slashed{p}\Big] =} Tr \Big{\{} \slashed{\epsilon}_3 \big{[} (x-\xi)\slashed{p} - \slashed{k}_3 \big{]}  \slashed{\epsilon}_2 \big{[} (x+\xi)\slashed{p} + \slashed{k}_1 \big{]}\times \\& \phantom{Tr\Big[ i\mathcal{M}^{2.L}_{1,2,3} ~\slashed{p}\Big] =}  \mu^{4-d} \int \frac{i^2 d^dk}{(2\pi)^d}\frac{\gamma^\mu[(x+\xi)\slashed{p} + \slashed{k}_1 + \slashed{k}]\gamma_\mu}{(k^2 +i0)(((x+\xi)p+k_1 +k)^2 + i0)} \big{[} (x+\xi)\slashed{p} + \slashed{k}_1 \big{]}  \slashed{\epsilon}_1   \slashed{p} \Big{ \} }.
\end{aligned}
\end{equation}
\subsection{Integral}
Using the discussed integration formula \eqref{eq-loop-int}, we first employ the Feynman parametrization and shift momentum to complete the square in denominator:
\begin{equation}
I = \mu^{4-d} \int \frac{d^dk}{(2\pi)^d} \frac{\gamma^\mu(\slashed{P} + \slashed{k})\gamma_\mu}{(k^2 +i0)((P+k)^2 + i0)} = \mu^{4-d} \int \frac{d^dk}{(2\pi)^d} \int_0^1 dx \frac{(2-d)(1-x)\slashed{P}}{[k^2 + x(1-x)P^2 + i0]^2},
\end{equation}
and then integrate over the loop momentum $k$:
\begin{equation}
I= \mu^{4-d}(2-d)\int_0^1 dx \frac{i\slashed{P}(1-x)}{(4\pi)^{d/2}}\big{(} -x(1-x)P^2 -i0 \big{)}^{d/2 - 2} \Gamma(2 - d/2).
\end{equation}
In the next step we use the fact that $x(1-x) \geq 0$, so that it can be taken outside the parenthesis without changing the sign of $i0$
\begin{equation}
I=(-2 + \varepsilon)\frac{i\slashed{P}}{(4\pi)^2}\int_0^1 dx \big{(} -P^2/4\pi\mu^2 -i0 \big{)}^{-\varepsilon/2} x^{-\varepsilon/2}(1-x)^{1-\varepsilon/2}\Gamma(\varepsilon/2),
\end{equation}
and finally we perform the integral over $dx$ obtaining the result expressed in terms of Euler gamma functions:
\begin{equation}
I= (-2 + \varepsilon)\frac{i\slashed{P}}{(4\pi)^2} \big{(} -P^2/4\pi\mu^2 -i0 \big{)}^{-\varepsilon/2} \Gamma(\varepsilon/2)\frac{\Gamma(1-\varepsilon/2)\Gamma(2-\varepsilon/2)}{\Gamma(3-\varepsilon)}.
\end{equation}
Taking the limit $\varepsilon \rightarrow 0$ we get
\begin{equation}
(-2+\varepsilon)\Gamma(\varepsilon/2)\frac{\Gamma(1-\varepsilon/2)\Gamma(2-\varepsilon/2)}{\Gamma(3-\varepsilon)} = -\frac{2}{\varepsilon} - 1 + \gamma + \mathcal{O}(\varepsilon).
\end{equation}
Finally, we have
\begin{equation}
I = \frac{i\slashed{p}}{(4\pi)^2}\big{(} -\frac{2}{\varepsilon} + log(-P^2/4\pi\mu^2 -i0) - 1 + \gamma \big{)}.
\end{equation}
\subsection{Traces}
In the case of the self-energy contributions we 
have the following traces:
\begin{equation}
\begin{aligned}
&Tr \Big{\{}\slashed{\epsilon}_3\big{[} (x-\xi)\slashed{p} - \slashed{k}_3 \big{]} \slashed{\epsilon}_2   \big{[} (x+\xi)\slashed{p} + \slashed{k}_1 \big{]}^3 \slashed{\epsilon}_1 \slashed{p} \Big{ \} },
\\
&Tr \Big{\{}\slashed{\epsilon}_3\big{[} (x-\xi)\slashed{p} - \slashed{k}_3 \big{]}^3 \slashed{\epsilon}_2   \big{[} (x+\xi)\slashed{p} + \slashed{k}_1 \big{]} \slashed{\epsilon}_1 \slashed{p} \Big{ \} }.
\end{aligned}
\end{equation}
They are simplified using $\slashed{P}\slashed{P} = P^2$, and the problem reduces to computation of
\begin{equation}
Tr \Big{\{}\slashed{\epsilon}_3\big{[} (x-\xi)\slashed{p} - \slashed{k}_3 \big{]} \slashed{\epsilon}_2   \big{[} (x+\xi)\slashed{p} + \slashed{k}_1 \big{]} \slashed{\epsilon}_1 \slashed{p} \Big{ \} }.
\end{equation}
Using $\slashed{p}\slashed{p} =0$ we obtain
\begin{equation}
- Tr \Big{\{} \slashed{\epsilon}_3 \slashed{k}_3 \slashed{\epsilon}_2 \slashed{k}_1 \slashed{\epsilon}_1 \slashed{p} \Big{ \} } = 4\mathcal{A}_{1,2,3}.
\end{equation}
Putting together the results of the integration and the trace yields
$$
\begin{aligned}
Tr \big[ i\mathcal{M}^{2.L}_{1,2,3}~ \slashed{p} \big] = &-C_F e_q^3 g^2 \frac{i}{(4\pi)^2}\frac{1}{(x+\xi)pk_1 +i0}\frac{1}{(x-\xi)pk_3 -i0}\mathcal{A}_{1,2,3} \times \\ &\Big{(} -\frac{2}{\varepsilon} + log(-(x+\xi)(pk_1)/2\pi\mu^2 -i0) - 1 +\gamma \Big{)}.
\end{aligned}
$$
It can be further simplified using $pk_1 / 2\pi = s\beta_1 / 4\pi$ and
\begin{equation}
	-\frac{2}{\varepsilon} + log(-(x+\xi)s\beta_1/4\pi\mu^2 -i0) +\gamma = \Big( \frac{s e^\gamma}{4\pi\mu^2} \Big)^{-\frac{\varepsilon}{2}}\bigg{(} -\frac{2}{\varepsilon} + log\Big(-(x+\xi)\beta_1 -i0\Big) \bigg{)} + \mathcal{O}(\varepsilon),
\end{equation}
so that
\begin{equation}
\begin{aligned}
&Tr \big[ i\mathcal{M}^{2.L}_{1,2,3} ~\slashed{p} \big]= -C_F e_q^3 g^2 \frac{i}{(4\pi)^2}\frac{1}{x+\xi + i0_1}\frac{1}{x-\xi-i0_3} \frac{4}{s^2 \beta_1 \beta_3}\mathcal{A}_{1,2,3}\times \\  & \phantom{Tr \big[ i\mathcal{M}^{2.L}_{1,2,3} ~\slashed{p} \big] = }\Big( \frac{s e^\gamma}{4\pi\mu^2} \Big)^{-\frac{\varepsilon}{2}}\bigg{(} 
-\frac{2}{\varepsilon} + log\Big(-(x+\xi)\beta_1 -i0\Big)- 1 \bigg{)}\\
&\phantom{Tr \big[ i\mathcal{M}^{2.L}_{1,2,3} ~\slashed{p} \big]}= Tr[i\mathcal{M}^{0}_{1,2,3}~\slashed{p}]\cdot
\frac{g^2C_F}{(4\pi)^2}\Big( \frac{s e^\gamma}{4\pi\mu^2} \Big)^{-\frac{\varepsilon}{2}}
\bigg{(} 
-\frac{2}{\varepsilon} + log\Big(-(x+\xi)\beta_1 -i0\Big)- 1 
\bigg{)}
\end{aligned}
\end{equation}

Using this results one can write the second amplitude in a straightforward way:
\begin{equation}
\begin{aligned}
&Tr \big[ i\mathcal{M}^{2.R}_{1,2,3} ~\slashed{p} \big] = -C_F (ie_q)^3 (ig)^2 \frac{i}{\big{(}(x+\xi)p + k_1 \big{)}^2 + i0}  \bigg{(}\frac{i}{\big{(}(x-\xi)p - k_3 \big{)}^2 + i0}\bigg{)}^2 \times
\\
&\phantom{Tr \big[ i\mathcal{M}^{2.R}_{1,2,3} ~\slashed{p} \big] =} Tr \Big{\{} \slashed{\epsilon}_3 \big{[} (x-\xi)\slashed{p} - \slashed{k}_3 \big{]} \mu^{4-d} \int \frac{i^2 d^dk}{(2\pi)^d} \frac{\gamma^\mu[(x-\xi)\slashed{p} - \slashed{k}_3 + \slashed{k}]\gamma_\mu}{(k^2 +i0)(((x-\xi)p-k_3 +k)^2 + i0)}\times \\&\phantom{Tr \big[ i\mathcal{M}^{2.R}_{1,2,3} ~\slashed{p} \big] =} \big{[} (x-\xi)\slashed{p} - \slashed{k}_3 \big{]}  \slashed{\epsilon}_2 \big{[} (x+\xi)\slashed{p} + \slashed{k}_1 \big{]} \slashed{\epsilon}_1   \slashed{p} \Big{ \} } =
\\
&=-C_F e_q^3 g^2 \frac{i}{(4\pi)^2}\frac{1}{x+\xi + i0_1}\frac{1}{x-\xi-i0_3} \frac{4}{s^2 \beta_1 \beta_3}\mathcal{A}_{1,2,3}\Big{(} -\frac{2}{\varepsilon} + log\big( (x-\xi)(pk_3)/2\pi\mu^2 -i0\big) - 1 +\gamma \Big{)}
\\
&=\phantom{-} Tr[i\mathcal{M}^{0}_{1,2,3}~\slashed{p}]\cdot
\frac{g^2C_F}{(4\pi)^2}\Big( \frac{s e^\gamma}{4\pi\mu^2} \Big)^{-\frac{\varepsilon}{2}}
\bigg{(} 
-\frac{2}{\varepsilon} + log\Big((x-\xi)\beta_3 -i0\Big)- 4 
\bigg{)}.
\end{aligned}
\end{equation}
\newpage
\section{Vertex corrections - 3.L and 3.R}
\begin{figure}[H]
    \centering
    \begin{tikzpicture}[scale = 0.7]
  \begin{feynman}
    \vertex (q1) at (0, 0);
	\vertex (q1a) at (0, -1.5);
	\vertex (q1b) at (1.5, 0);
    \vertex (q3) at (3, 0);
	\vertex (q4) at (3, 0);
    \vertex (q5) at (6, 0);
    \vertex (in) at (0, -3);
    \vertex (out) at (6, -3);
    \vertex (k1) at (0, 2);
    \vertex (k2) at (3, 2);
    \vertex (k3) at (6, 2);
    \diagram* {
      (q1) -- [fermion] (q1b) -- [fermion] (q3) -- [fermion]  (q5);
      (k1) -- [photon, momentum=\(k_1\)] (q1);
      (k2) -- [photon, momentum=\(k_2\)] (q3);
      (k3) -- [photon, momentum=\(k_3\)] (q5);
      (in) -- [fermion, momentum=\((x+\xi)p\)] (q1a) -- [fermion] (q1);
      (q5) -- [fermion, momentum=\((x-\xi)p\)] (out);
	(q1b) -- [gluon, quarter left, momentum = $k$] (q1a);
    };
  \end{feynman}
\end{tikzpicture}
\quad
    \begin{tikzpicture}[scale = 0.7]
  \begin{feynman}
    \vertex (q1) at (0, 0);
	\vertex (q1a) at (6, -1.5);
	\vertex (q1b) at (4.5, 0);
    \vertex (q3) at (3, 0);
	\vertex (q4) at (3, 0);
    \vertex (q5) at (6, 0);
    \vertex (in) at (0, -3);
    \vertex (out) at (6, -3);
    \vertex (k1) at (0, 2);
    \vertex (k2) at (3, 2);
    \vertex (k3) at (6, 2);
    \diagram* {
      (q1) -- [fermion] (q3) -- [fermion] (q1b) -- [fermion]  (q5);
      (k1) -- [photon, momentum=\(k_1\)] (q1);
      (k2) -- [photon, momentum=\(k_2\)] (q3);
      (k3) -- [photon, momentum=\(k_3\)] (q5);
      (in) -- [fermion, momentum=\((x+\xi)p\)] (q1);
      (q5) --[fermion] (q1a) -- [fermion, momentum=\((x-\xi)p\)] (out);
	(q1a) -- [gluon, quarter left, momentum = $k$] (q1b);
    };
  \end{feynman}
\end{tikzpicture}
    \caption{Diagrams 3.L and 3.R.}
    \label{3LR}
\end{figure}
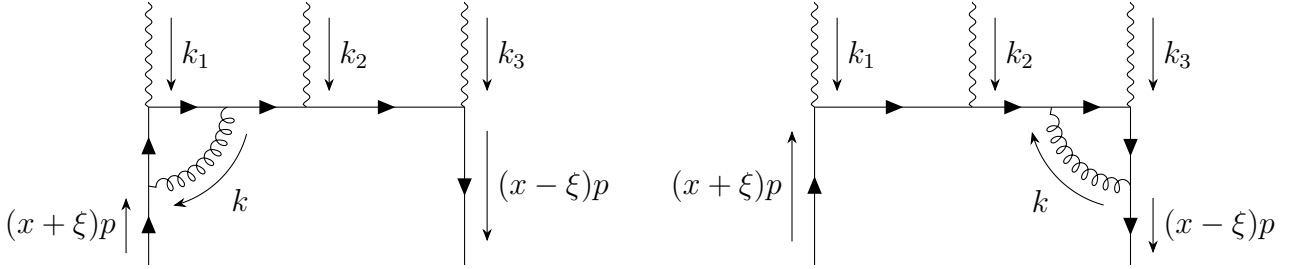

The vertex corrections with the gluon line attached to the on-shell quark develop both the UV and collinear divergence, as it was explained in Subsec. \ref{subsec-divergences}.
The considered diagrams are shown in Fig. \ref{3LR}
\noindent The full form of the amplitude corresponding to $3.L$ is
\begin{equation}\label{eq-m3l-0}
\begin{aligned}
&Tr\big[ i\mathcal{M}^{3.L}_{1,2,3}~ \slashed{p} \big] = -C_F (ie_q)^3(ig)^2 \frac{i}{\big{(}(x+\xi)p + k_1 \big{)}^2 + i0} \frac{i}{\big{(}(x-\xi)p - k_3 \big{)}^2 + i0} Tr \Big{\{} \slashed{\epsilon}_3 \big{[} (x-\xi)\slashed{p} - \slashed{k}_3 \big{]} \slashed{\epsilon}_2 \times \\
& \big{[} (x+\xi)\slashed{p} + \slashed{k}_1 \big{]}
 \mu^{4-d} \int \frac{d^dk}{(2\pi)^d} \frac{i^3 \gamma_\mu \big{(}(x+\xi)\slashed{p} + \slashed{k}_1 + \slashed{k} \big{)} \slashed{\epsilon}_1 \big{(}(x+\xi)\slashed{p} + \slashed{k} \big{)}\gamma^\mu }{\big{(}((x+\xi)p + k)^2 + i0 \big{)}\big{(}((x+\xi)p + k_1 + k)^2 + i0 \big{)}\big{(}k^2 + i0 \big{)}}\slashed{p} \Big{\}}.
\end{aligned}
\end{equation}
First, let us first simplify the traces present in Eq. \eqref{eq-m3l-0}. Using the formula 
$$\gamma_\mu \slashed{A}\slashed{B} \slashed{C}\gamma^\mu = -2 \slashed{C}\slashed{B}\slashed{A} + \varepsilon \slashed{A}\slashed{B} \slashed{C},$$ 
for arbitrary numbers $a,b,c,d$ we obtain
\begin{equation}\label{Tr3L}
\begin{aligned}
&Tr \Big{\{} \slashed{\epsilon}_3 \big{[} (x-\xi)\slashed{p} - \slashed{k}_3 \big{]} \slashed{\epsilon}_2 \big{[} (x+\xi)\slashed{p} + \slashed{k}_1 \big{]} \big{[} a\slashed{p} + b\slashed{k}_1 \big{]} \slashed{\epsilon}_1 \big{[} c\slashed{p} + d\slashed{k}_1 \big{]} \slashed{p}  \Big{\}}= -ad Tr \Big{\{} \slashed{\epsilon}_3 \slashed{k}_3 \slashed{\epsilon}_2 \slashed{k}_1 \slashed{p}  \slashed{\epsilon}_1 \slashed{k}_1 \slashed{p}  \Big{\}}=
\\
&= -2ad (pk_1) Tr\big{(} \slashed{\epsilon}_3 \slashed{k}_3 \slashed{\epsilon}_2 \slashed{k}_1 \slashed{p}  \slashed{\epsilon}_1 \big{)} = %-2ad (pk_1) \Big{(} (pk_1) Tr\big{(} \slashed{\epsilon}_3 \slashed{k}_3 \slashed{\epsilon}_2  \slashed{\epsilon}_1 \big{)} + (pk_3) Tr\big{(} \slashed{\epsilon}_3 \slashed{\epsilon}_2 \slashed{k}_1 \slashed{\epsilon}_1 \big{)}  \Big{)}=
%\\
%&=2ad(pk_1)(4-\varepsilon)\Big{(} (pk_1)\big{(} (\epsilon_3\epsilon_2)(\epsilon_1 k_3) - (\epsilon_3\epsilon_1)(k_3 \epsilon_2)\big{)} + (pk_3)\big{(} (\epsilon_2\epsilon_1)(\epsilon_3 k_1) - (\epsilon_3\epsilon_1)(k_1 \epsilon_2)\big{)}  \Big{)}=
%\\
 -8ad(pk_1)\mathcal{A}_{1,2,3}.
\end{aligned}
\end{equation}
We proceed with the computation by applying the Feynman parametrization to the integral:
\begin{equation}
\begin{aligned}
&\mu^{4-d} \int \frac{d^dk}{(2\pi)^d} \frac{ \gamma_\mu \big{(}(x+\xi)\slashed{p} + \slashed{k}_1 + \slashed{k} \big{)} \slashed{\epsilon}_1 \big{(}(x+\xi)\slashed{p} + \slashed{k} \big{)}\gamma^\mu }{\big{(}((x+\xi)p + k)^2 + i0 \big{)}\big{(}((x+\xi)p + k_1 + k)^2 + i0 \big{)}\big{(}k^2 + i0 \big{)}} 
\\
&=2\mu^{4-d}\int_0^1 dy \int_0^{1-y}dz \int \frac{d^dk}{(2\pi)^d} \frac{\gamma_\mu \big{(}(x+\xi)\slashed{p} + \slashed{k}_1 + \slashed{k} \big{)} \slashed{\epsilon}_1 \big{(}(x+\xi)\slashed{p} + \slashed{k} \big{)}\gamma^\mu }{\Big{(} \big{(} k + (y+z)(x+\xi)p + yk_1 \big{)}^2 + 2(x+\xi)(pk_1)(1-y-z)y  + i0 \Big{)}^3}.
\end{aligned}
\end{equation}
Substituting $k' = k + (y+z)(x+\xi)p + yk_1$ and using the trace formula \eqref{Tr3L} we reduce the problem of computation of the integral inside the trace to the following one:
\begin{equation}
I=2\mu^{4-d}\int_0^1 dy \int_0^{1-y}dz \int \frac{d^dk}{(2\pi)^d} \frac{-2(1-y-z)(x+\xi)(1+(\varepsilon/2 -1) y) \slashed{p}\slashed{\epsilon}_1\slashed{k}_1 + (-2 + \varepsilon) k^\mu k^\nu \gamma_\mu \slashed{\epsilon}_1 \gamma_\nu}{\Big{(} k^2 + 2(x+\xi)(pk_1)(1-y-z)y + i0 \Big{)}^3}.
\end{equation}
It breaks into 2 term which are proportional to different gamma structures:
\begin{equation}
    I \equiv (x+\xi)\slashed{p}\slashed{\epsilon}_1 \slashed{k}_1 I_1 + I_2 \slashed{\epsilon}_1.
\end{equation}
After performing the momentum integration, $I_1$ is found to be
\begin{equation}
I_1 = \frac{2i\mu^{4-d}}{(4\pi)^{d/2}}\Gamma\big{(}3-\frac{d}{2} \big{)}\int_0^1 dy \int_0^{1-y}dz\frac{ (1-y-z)(1+(\varepsilon/2 -1) y)}{ \big{(} -2(x+\xi)(pk_1)(1-y-z)y - i0 \big{)}^{3-d/2}}.
\end{equation}
The integration over Feynman parameters $y$ and $z$ can be simplified by the substitution:
\begin{equation}\label{eq-substitution-3l}
y = \bar{z}(1-\bar{y}), \quad \quad z = \bar{y}\bar{z}, \quad \quad \int_0^1 dy \int_0^{1-y}dz \rightarrow \int_0^1 d\bar{y} \int_0^1 \bar{z}d\bar{z}.
\end{equation}
$I_1$ has now the following form (for simplicity we omit the lines over the reparametrized variables $\bar{y}, \bar{z}$):
$$
I_1 = \Gamma\big{(}3-\frac{d}{2} \big{)} \frac{2i\mu^{4-d}}{(4\pi)^{d/2}}\int_0^1 dy \int_0^1 zdz \frac{ (1-z) \big{(}1+(\varepsilon/2 -1) z(1-y)\big{)}}{ \big{(} -2(x+\xi)(pk_1)(1-z)(1-y)z - i0 \big{)}^{3-d/2} }.
$$
The integration over $dx$ and $dz$ can be now factorized into a product of 2 integrals:
$$
\begin{aligned}
I_1 =&\frac{2i\mu^{4-d}}{(4\pi)^{d/2}} \big{(} -2(x+\xi)(pk_1) - i0 \big{)}^{d/2-3}\Gamma\big{(}3-\frac{d}{2} \big{)}\Big{(} \int_0^1 (1-y)^{d/2-3}dy \int_0^1 z^{d/2-2}(1-z)^{d/2-2} 
\\
&+ (\varepsilon/2 - 1) \int_0^1 (1-y)^{d/2-2} \int_0^1 z^{d/2-1}(1-z)^{d/2-2} dz \Big{)}.
\end{aligned}
$$
These integral are easily performed and one obtains
\begin{equation}
\begin{aligned}
I_1 = &\frac{2i\mu^{4-d}}{(4\pi)^{d/2}} \big{(} -2(x+\xi)(pk_1) - i0 \big{)}^{d/2-3}\times \\ &\Big{(} \frac{\Gamma\big{(}3-\frac{d}{2} \big{)}}{d/2-2}\frac{\Gamma(d/2-1)\Gamma(d/2-1)}{\Gamma(d-2)} + (\varepsilon/2 - 1) \frac{\Gamma\big{(}3-\frac{d}{2} \big{)}}{d/2-1}\frac{\Gamma(d/2)\Gamma(d/2-1)}{\Gamma(d-1)} \Big{)}.
\end{aligned}
\end{equation}
For $d= 4-\varepsilon$, in the limit $\varepsilon \rightarrow 0$ one has
\begin{equation}
\begin{aligned}
&\frac{\Gamma\big{(}3-\frac{d}{2} \big{)}}{d/2-2}\frac{\Gamma(d/2-1)\Gamma(d/2-1)}{\Gamma(d-2)} \approx -\frac{2}{\varepsilon} - 2 + \gamma,
\\
&(\varepsilon/2 - 1) \frac{\Gamma\big{(}3-\frac{d}{2} \big{)}}{d/2-1}\frac{\Gamma(d/2)\Gamma(d/2-1)}{\Gamma(d-1)} \approx -\frac{1}{2}.
\end{aligned}
\end{equation}
In this way, we obtain the final form of $I_1$:
\begin{equation}
\begin{aligned}
I_1 &= \frac{2i}{(4\pi)^{2}} \big{(}-2(x+\xi)(pk_1) - i0 \big{)}^{-1}\Big{(} 1 - \frac{\varepsilon}{2}log\big{(}-(x+\xi)(pk_1)/2\pi\mu^2 - i0 \big{)} \Big{)}\Big{(} -\frac{2}{\varepsilon} +\gamma -\frac{5}{2} \Big{)} 
\\
&= \frac{2i}{(4\pi)^{2}} \big{(}-2(x+\xi)(pk_1)  - i0 \big{)}^{-1}\Big{(} -\frac{2}{\varepsilon} + log\big{(}-(x+\xi)(pk_1)/2\pi\mu^2 - i0 \big{)} + \gamma - \frac{5}{2} \Big{)}.
\end{aligned}
\end{equation}
Let us note, that the loop momentum integral inside of $I_1$ was UV safe, from the argument of power counting, and hence the divergence present therein is the result of the region of integration, where $k$ and $(x+\xi)p +k$ become collinear and lightlike (i.e. the collinear divergence). In such case, the divergence manifested in the integration over the Feynman parameters -- one can easily check, that if one assumed a non-zero mass of partons, the relevant integral would be finite.

The second integral, after using $\gamma^\mu \slashed{\epsilon}_1 \gamma_\mu = (-2+\varepsilon)\slashed{\epsilon}_1$ and integrating over momenta becomes
$$
I_2 \slashed{\epsilon}_1 = \frac{i\mu^{4-d}}{2(4\pi)^{d/2}}\Gamma\big{(}2- d/2 \big{)}(2-\varepsilon)^2 \slashed{\epsilon}_1\int_0^1 dy \int_0^{1-y}dz\frac{ 1 }{ \big{(} -2(x+\xi)(pk_1)(1-y-z)y - i0 \big{)}^{2-d/2}}.
$$
We see that this integral is not UV-safe and there is a divergent term $\Gamma\big{(}2- d/2 \big{)}$ resulting from the integration over the loop momentum. Now we use the same substitution as in Eq. \eqref{eq-substitution-3l}:
$$
I_2 \slashed{\epsilon}_1=\frac{i\mu^{4-d}}{2(4\pi)^{d/2}}\Gamma\big{(}2- d/2 \big{)}(2-\varepsilon)^2 \slashed{\epsilon}_1\int_0^1 dy \int_0^{1}zdz \big{(} -2(x+\xi)(pk_1)(1-z)(1-y)z - i0 \big{)}^{d/2-2},
$$
because $(1-z)(1-y)z\geq0$, we can bring that term outside the parenthesis without changing the sign of $i0$, and as a result we get
$$
I_2 \slashed{\epsilon}_1= \frac{i\mu^{4-d}}{2(4\pi)^{d/2}}\Gamma\big{(}2- d/2 \big{)}(2-\varepsilon)^2 \slashed{\epsilon}_1 \big{(} -2(x+\xi)(pk_1) - i0 \big{)}^{d/2-2} \frac{\Gamma(d/2)\Gamma(d/2-1)}{\Gamma(d-1)}\frac{1}{d/2-1}.
$$
We expand with respect to $\varepsilon$:
$$
\Gamma\big{(}2- d/2 \big{)}(2-\varepsilon)^2 \frac{\Gamma(d/2)\Gamma(d/2-1)}{\Gamma(d-1)}\frac{1}{d/2-1} \approx 2\big{(} \frac{2}{\varepsilon} + 1 - \gamma \big{)}.
$$
In this case, the integrals over Feynman parameters are finite, since the $k^2$ from the denominator cancels with the $k^2$ from the trace, and the integrand does not have a collinear singularity. The final expression for the second integral is then
\begin{equation}
\begin{aligned}
I_2 \slashed{\epsilon}_1 &= \frac{i\mu^{4-d}}{2(4\pi)^{d/2}} \slashed{\epsilon}_1 \big{(} -2(x+\xi)(pk_1) - i0 \big{)}^{d/2-2} 2\big{(} \frac{2}{\varepsilon} + 1 - \gamma \big{)} =
\\
&= \frac{i}{(4\pi)^{2}} \slashed{\epsilon}_1 \Big{(} \frac{2}{\varepsilon} + 1 - \gamma - log\big{(}-(x+\xi)pk_1/2\pi\mu^2 - i0 \big{)} \Big{)}.
\end{aligned}
\end{equation}
Let us recall that
$$
I = (x+\xi)\slashed{p}\slashed{\epsilon}_1\slashed{k}_1 I_1 +  I_2 \slashed{\epsilon}_2.
$$
Using the previous results concerning traces we obtain 
\begin{equation}
\begin{aligned}
Tr\big[ i\mathcal{M}^{3.L}_{1,2,3}~ \slashed{p} \big]
&=-C_F e_q^3 g^2 \frac{i}{(4\pi)^2} \frac{1}{\bar{p}k_1 + i0}\frac{1}{\underline{p}k_3 - i0}\mathcal{A}_{1,2,3} \Big{(} -\frac{2}{\varepsilon} + log\big{(} -(x+\xi)(pk_1)/2\pi\mu^2 - i0 \big{)} + \gamma - 4 \Big{)}
\\
&= Tr[i\mathcal{M}^{0}_{1,2,3}~\slashed{p}]\cdot
\frac{g^2C_F}{(4\pi)^2}\Big( \frac{s e^\gamma}{4\pi\mu^2} \Big)^{-\frac{\varepsilon}{2}}
\bigg{(} 
-\frac{2}{\varepsilon} + log\Big(-(x+\xi)\beta_1 -i0\Big)- 4 
\bigg{)}
\end{aligned}
\end{equation}
\\[0.2cm]
The computation of the diagram $3.R$ is done in the same way, and it yields
\begin{equation}
\begin{aligned}
Tr\big[ i\mathcal{M}^{3.R}_{1,2,3}~\slashed{p} \big] 
&=-C_F e_q^3 g^2 \frac{i}{(4\pi)^2} \frac{1}{\bar{p}k_1 + i0}\frac{1}{\underline{p}k_3 - i0}\mathcal{A}_{1,2,3} \Big{(} -\frac{2}{\varepsilon} + log\big{(} (x-\xi)(pk_3)/2\pi\mu^2 - i0 \big{)} + \gamma - 4 \Big{)}
\\
	&= Tr[i\mathcal{M}^{0}_{1,2,3}~\slashed{p}]\cdot
\frac{g^2C_F}{(4\pi)^2}\Big( \frac{s e^\gamma}{4\pi\mu^2} \Big)^{-\frac{\varepsilon}{2}}
\bigg{(} 
-\frac{2}{\varepsilon} + log\Big((x-\xi)\beta_3 -i0\Big)- 4 
\bigg{)}.
\end{aligned}
\end{equation}
\section{Vertex corrections - 3.M}
\begin{figure}[H]
    \centering
    \begin{tikzpicture}[scale = 0.7]
  \begin{feynman}
    \vertex (q1);
    \vertex [right=of q1] (q2);
    \vertex [right=of q2] (q3);
    \vertex [right=of q3] (q4);
    \vertex [right=of q4] (q5);
    \vertex [below=of q1] (qb1);
    \vertex [below=of q5] (qb5);
    \vertex [below=of qb1] (in);
    \vertex [below=of qb5] (out);
    \vertex [above=of q1] (k1);
    \vertex [above=of q3] (k2);
    \vertex [above=of q5] (k3);
    \diagram* {
      (q1) -- [fermion] (q2) -- [fermion] (q3) -- [fermion] (q4) -- [fermion] (q5);
      (k1) -- [photon, momentum=\(k_1\)] (q1);
      (k2) -- [photon, momentum=\(k_2\)] (q3);
      (k3) -- [photon, momentum=\(k_3\)] (q5);
      (in) -- [fermion, momentum=\((x+\xi)p\)] (q1);
      (q5) -- [fermion, momentum=\((x-\xi)p\)] (out);
      (q4) -- [gluon, half left, momentum=\(k\)] (q2);
    };
  \end{feynman}
\end{tikzpicture}
    \caption{Diagram 3.M.}
    \label{3M}
\end{figure}
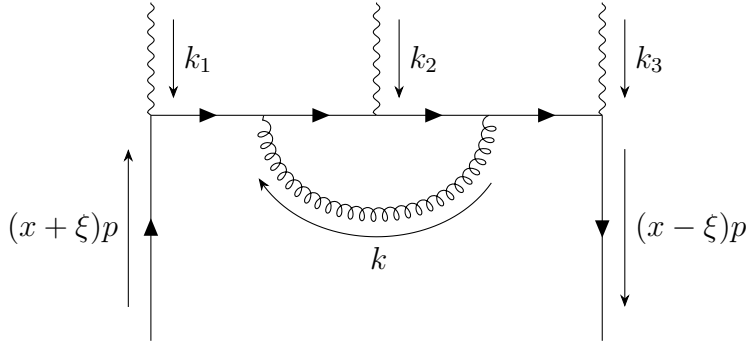

The considered diagram is shown in fig. \ref{3M}
The fact that the gluon line in the diagram $3.M$ is connected to quarks that are off-shell implies that in this term there is no collinear divergence here. The integration over the Feynman parameters becomes more complicated, but it turns out that an appropriate substitution allows to simplify it significantly.\\
\noindent The considered amplitude reads
\begin{equation}\label{eq-3m-0}
\begin{aligned}
&Tr\big[ i\mathcal{M}_{3.M}~\slashed{p} \big] = -C_F(ie_q)^3(ig)^2 \frac{i}{\big{(}(x+\xi)p + k_1 \big{)}^2 + i0} \frac{i}{\big{(}(x-\xi)p - k_3 \big{)}^2 + i0}Tr\Big{\{} \slashed{\epsilon}_3 \big{[} (x-\xi)\slashed{p} - \slashed{k}_3 \big{]} \times
\\
 & \mu^{4-d} \int \frac{d^d k}{(2\pi)^d} \frac{i^3\gamma_\mu\big{[} (x+\xi)\slashed{p} + \slashed{k}_1 + \slashed{k}_2 + \slashed{k} \big{]}\slashed{\epsilon}_2 \big{[} (x+\xi)\slashed{p} + \slashed{k}_1 + \slashed{k} \big{]}\gamma^\mu }{\big{(}((x+\xi)p + k_1 + k_2+ k)^2 + i0 \big{)}\big{(}((x+\xi)p + k_1 + k)^2 + i0 \big{)}\big{(}k^2 + i0 \big{)}}\big{[} (x+\xi)\slashed{p} + \slashed{k}_1 \big{]} \slashed{\epsilon}_1 \slashed{p} \Big{\}}.
\end{aligned}
\end{equation}
First, we simplify the trace:
\begin{equation}\label{eq-tr-3m}
\begin{aligned}
&Tr\Big{\{} \slashed{\epsilon}_3 (-\slashed{k}_3) \big{[} a(x+\xi)\slashed{p} + a\slashed{k}_1 + b\slashed{k}_2 \big{]}\slashed{\epsilon}_2 \big{[} a(x+\xi)\slashed{p} + a\slashed{k}_1 + c\slashed{k}_2 \big{]}\slashed{k}_1 \slashed{\epsilon}_1\slashed{p}  \Big{\}}=
\\
&= 2pk_1 Tr\Big{\{} \slashed{\epsilon}_3 \slashed{k}_3 \big{[} a(x+\xi)\slashed{p} + a\slashed{k}_1 + b\slashed{k}_2 \big{]}\slashed{\epsilon}_2 \big{[} a(x+\xi)\slashed{p} + c\slashed{k}_2 \big{]} \slashed{\epsilon}_1 \Big{\}}+
\\
&- Tr\Big{\{} \slashed{\epsilon}_3 \slashed{k}_3 \big{[} a(x+\xi)\slashed{p} + a\slashed{k}_1 + b\slashed{k}_2 \big{]}\slashed{\epsilon}_2  c\slashed{k}_2 \slashed{p} \slashed{k}_1 \slashed{\epsilon}_1 \Big{\}}.
\end{aligned}
\end{equation}
Using identities $(x+\xi)p + k_1  = (x-\xi)p - k_2 - k_3$ and $\slashed{k}_i \slashed{k}_i = 0$ we obtain
\begin{equation}
\begin{aligned}
&\eqref{eq-tr-3m} = 2pk_1 Tr\Big{\{} \slashed{\epsilon}_3 \slashed{k}_3 \big{[} a(x-\xi)\slashed{p} + (b-a)\slashed{k}_2 \big{]}\slashed{\epsilon}_2 \big{[} a(x+\xi)\slashed{p} + c\slashed{k}_2 \big{]} \slashed{\epsilon}_1 \Big{\}}
- Tr\Big{\{} \slashed{\epsilon}_3 \slashed{k}_3 a(x-\xi)\slashed{p} \slashed{\epsilon}_2  c\slashed{k}_2 \slashed{p} \slashed{k}_1 \slashed{\epsilon}_1 \Big{\}}
\\
&=2a(x-\xi)c(pk_1)Tr\Big{\{} \slashed{\epsilon}_3 \slashed{k}_3 \slashed{p} \slashed{\epsilon}_2\slashed{k}_2 \slashed{\epsilon}_1 \Big{\}} + 2a(x+\xi)(b-a)(pk_1) Tr\Big{\{} \slashed{\epsilon}_3 \slashed{k}_3 \slashed{k}_2 \slashed{\epsilon}_2\slashed{p} \slashed{\epsilon}_1 \Big{\}}+
\\
&-2ac(x-\xi)(pk_2)Tr\Big{\{} \slashed{\epsilon}_3 \slashed{k}_3\slashed{p}\slashed{\epsilon}_2  \slashed{k}_1 \slashed{\epsilon}_1 \Big{\}}
\equiv a\big{(} \mathcal{B}(b-a) + \mathcal{C}c \big{)}.
\end{aligned}
\end{equation}
$\mathcal{B}$ and $\mathcal{C}$ are auxiliary parameters, which do not appear in the final expression for the amplitude. Let us further simplify $\mathcal{C}$. First, apply $pk_2 = -p(k_1 + k_3)$ to the second trace inside the parenthesis:
$$
\begin{aligned}
\mathcal{C} &= 2(x-\xi) \bigg{(} (pk_1)Tr\Big{\{} \slashed{\epsilon}_3 \slashed{k}_3 \slashed{p} \slashed{\epsilon}_2\slashed{k}_2 \slashed{\epsilon}_1 \Big{\}} - (pk_2)Tr\Big{\{} \slashed{\epsilon}_3 \slashed{k}_3\slashed{p}\slashed{\epsilon}_2  \slashed{k}_1 \slashed{\epsilon}_1 \Big{\}} \bigg{)}=
\\
&= 2(x-\xi) \bigg{(} (pk_1)Tr\Big{\{} \slashed{\epsilon}_3 \slashed{k}_3 \slashed{p} \slashed{\epsilon}_2(\slashed{k}_1 + \slashed{k}_2) \slashed{\epsilon}_1 \Big{\}} + (pk_3)Tr\Big{\{} \slashed{\epsilon}_3 \slashed{k}_3\slashed{p}\slashed{\epsilon}_2  \slashed{k}_1 \slashed{\epsilon}_1 \Big{\}} \bigg{)}.
\end{aligned}
$$
Then, from the relation $\slashed{p} \slashed{\epsilon}_2 (\slashed{k}_1 + \slashed{k}_2) = -\slashed{p} \slashed{\epsilon}_2\slashed{k}_3$ follows that
$$
\mathcal{C}= 2(x-\xi) \bigg{(} -(pk_1)Tr\Big{\{} \slashed{\epsilon}_3 \slashed{k}_3 \slashed{p} \slashed{\epsilon}_2 \slashed{k}_3 \slashed{\epsilon}_1 \Big{\}} + (pk_3)Tr\Big{\{} \slashed{\epsilon}_3 \slashed{k}_3\slashed{p}\slashed{\epsilon}_2  \slashed{k}_1 \slashed{\epsilon}_1 \Big{\}} \bigg{)}.
$$
Finally, using a straightforward trace algebra one verifies that
\begin{equation}
\begin{aligned}
\mathcal{C}&= 2(x-\xi) \bigg{(} -8(pk_1)(pk_3)(k_3\epsilon_1)(\epsilon_2\epsilon_3) + (pk_3)Tr\Big{\{} \slashed{\epsilon}_3 \slashed{k}_3\slashed{p}\slashed{\epsilon}_2  \slashed{k}_1 \slashed{\epsilon}_1 \Big{\}} \bigg{)}=
\\
    &= 2(x-\xi)pk_3 Tr\Big{\{} \slashed{\epsilon}_3 \slashed{p}\slashed{\epsilon}_2  \slashed{k}_1 \slashed{k}_3\slashed{\epsilon}_1 \Big{\}}.
\end{aligned}
\end{equation}

To perform the integration, let us denote $P = (x+\xi)p + k_1$, $Q= P+k_2$. Obviously $(P-Q)^2=0$, so that $P^2 + Q^2 = 2PQ$. We take the integral \eqref{eq-3m-0}, employ the Feynman parametrization and shift the loop momentum to complete the square. The part of the resulting integral without the $k^2$ term in numerator, after computing the trace according to the previous results, reads
\begin{equation}
I_1 = -2i\mu^{4-d}\int\frac{d^d k}{(2\pi)^d} \int_0^1 dy \int_0^{1-y}dz \frac{-2(1-y-z)\big{(} (\mathcal{C} - \mathcal{B}) + (1-\varepsilon/2)\mathcal{B} y - (1-\varepsilon/2)\mathcal{C}z \big{)} }{\Big{(} k^2 + (1-y-z)(yP^2 + zQ^2) + i0 \Big{)}^3}.
\end{equation}
Performing the momenta integration and making the substitution \eqref{eq-substitution-3l} one obtains
\begin{equation}
\begin{aligned}
I_1 &= \frac{2\mu^{4-d}}{(4\pi)^{d/2}}\int_0^1 dy \int_0^1 z dz \Big{(} - (1-z)z\big{(}(1-y)P^2 + yQ^2\big{)} - i0 \Big{)}^{d/2-3}\Gamma(3-d/2)\times
\\
& (1-z) \Big{(} (\mathcal{C} - \mathcal{B}) + z\big{(} 1-\frac{\varepsilon}{2} \big{)}\big{(} \mathcal{B}(1-y) - y\mathcal{C} \big{)} \Big{)}.
\end{aligned}
\end{equation}
The integrals over $dz$ are finite:
\begin{equation}\label{eq-3m-dy-int}
\begin{aligned}
I_1 &=\frac{2\mu^{4-d}}{(4\pi)^{d/2}}\Gamma(3-d/2) \bigg{(} \frac{\Gamma(d/2-1)\Gamma(d/2-1)}{\Gamma(d-2)}\int_0^1 dy \Big{(} y(P^2 - Q^2) -(P^2+i0) \Big{)}^{d/2-3}(\mathcal{C} - \mathcal{B}) +
\\
&+\frac{\Gamma(d/2)\Gamma(d/2-1)}{\Gamma(d-1)}(1-\frac{\varepsilon}{2})\Big{[} \mathcal{B}\int_0^1 dy \Big{(} y(P^2 - Q^2) -(P^2+i0) \Big{)}^{d/2-3} +
\\
&-(\mathcal{C}+\mathcal{B})\int_0^1 dy  \Big{(} y(P^2 - Q^2) -(P^2+i0) \Big{)}^{d/2-3}y\Big{]}\bigg{)}.
\end{aligned}
\end{equation}
Now let us calculate integrals over $dy$. The first one (second line of Eq. \eqref{eq-3m-dy-int}) is
$$
\begin{aligned}
&\frac{\mu^{4-d}}{{(4\pi)^{d/2}}}\int_0^1 dy \Big{(} y(P^2 - Q^2) -(P^2+i0) \Big{)}^{d/2-3} =
\\
&=\frac{\mu^{4-d}}{{(4\pi)^{d/2}}}\frac{1}{d/2-2}\frac{1}{P^2 - Q^2}\Big{(} \big{(}-Q^2 - i0 \big{)}^{d/2-2} - \big{(}-P^2 - i0 \big{)}^{d/2-2} \Big{)}.
\end{aligned}
$$
In the limit $d\rightarrow 4$ this expression is finite and equal to
\begin{equation}
 \frac{1}{(4\pi)^2} \frac{1}{P^2 - Q^2}\Big{(} log\big{(}-Q^2/4\pi\mu^2 - i0 \big{)} - log\big{(}-P^2/4\pi\mu^2 - i0 \big{)}^{d/2-2} \Big{)}.
\end{equation}
The second integral (in the last line of Eq. \eqref{eq-3m-dy-int}) yields
\begin{equation}
\begin{aligned}
&\frac{\mu^{4-d}}{{(4\pi)^{d/2}}}\int_0^1 ydy \Big{(} y(P^2 - Q^2) -(P^2+i0) \Big{)}^{d/2-3} =
\\
&= \frac{\mu^{4-d}}{{(4\pi)^{d/2}}}\frac{1}{d/2-2}\frac{1}{P^2 - Q^2} \Big{(} (-Q^2 - i0)^{d/2-2} - \frac{1}{d/2-1}\frac{1}{P^2 - Q^2} \big{(} ( -Q^2 - i0 )^{d/2-1} - ( -P^2 - i0 )^{d/2-1} \big{)} \Big{)}
\\
&= \frac{1}{(4\pi)^2}\frac{1}{(P^2-Q^2)^2}\Big{(} P^2 \log\frac{Q^2 + i0}{P^2 + i0} + P^2 - Q^2  \Big{)}.
\end{aligned}
\end{equation}
Notice that both values are finite. Finally, we get:
\begin{equation}
\begin{aligned}
    I_1 =& \frac{4}{(4\pi)^2} \bigg{(} 2(\mathcal{C} - \frac{1}{2}\mathcal{B}) \frac{log(-Q^2/4\pi\mu^2 - i0) - log(-P^2/4\pi\mu^2 - i0)}{P^2 - Q^2}+
\\
& - (\mathcal{B} + \mathcal{C}) \frac{1}{(P^2-Q^2)^2}\Big{(} P^2 \log\frac{Q^2 + i0}{P^2 + i0} + P^2 - Q^2  \Big{)} \bigg{)}.
\end{aligned}
\end{equation}
Now let us compute the second part of the loop momentum integral with the term $k^2$ in the numerator.
\begin{equation}
\begin{aligned}
&I_2 = 2\mu^{4-d} \int\frac{d^d k}{(2\pi)^d} \int_0^1 dy \int_0^{1-y} dz \frac{\gamma_\mu \slashed{k}\slashed{\epsilon}_2 \slashed{k} \gamma^\mu}{\big{(} k^2 + (1-y-z)(yP^2 + zQ^2) + i0 \big{)}^3} 
\\
&= \slashed{\epsilon}_2\frac{i\mu^{4-d}}{(4\pi)^{d/2}} \frac{\Gamma(2-d/2)}{2}(-2+\varepsilon)^2 \int_0^1 dy \int_0^1 z dz \slashed{\epsilon}_2 \big{(} -(1-z)z((1-y)P^2 + yQ^2) - i0 \big{)}^{d/2-2}
\\
&= \slashed{\epsilon}_2\frac{i\mu^{4-d}}{(4\pi)^{d/2}} \frac{\Gamma(2-d/2)}{2}(-2+\varepsilon)^2 \frac{\Gamma(d/2)\Gamma(d/2-1)}{\Gamma(d-1)} \frac{1}{d/2-1} \frac{(-Q^2 - i0)^{d/2-1} - (-P^2 - i0)^{d/2-1}}{P^2 - Q^2}
\\
&= \slashed{\epsilon}_2\frac{i}{(4\pi)^2} \Big{(} \frac{2}{\varepsilon} + 1 - \gamma + \frac{Q^2 log(-Q^2/4\pi\mu^2 - i0) - P^2 log(-P^2/4\pi\mu^2 - i0)}{P^2 - Q^2} \Big{)}.
\end{aligned}
\end{equation}
After taking the traces one obtains
\begin{equation}
\begin{aligned}
&Tr\big[ i\mathcal{M}^{3.M}_{1,2,3} ~\slashed{p}\big] =  -C_F e_q^3 g^2 \frac{i}{(4\pi)^2}\frac{1}{\bar{p}k_1 + i0} \frac{1}{\underline{p}k_3 - i0} \times
\\
& \Bigg{(} -\frac{P^2}{(P^2-Q^2)^2} Tr\Big{\{} \slashed{\epsilon}_3 \slashed{k}_3 \slashed{k}_1 \slashed{\epsilon}_2\slashed{p} \slashed{\epsilon}_1 \Big{\}} \Big{(} (Q^2 - 2P^2)\log\frac{Q^2 + i0}{P^2 + i0} + Q^2 - P^2  \Big{)} +
\\
&- \frac{Q^2}{(P^2-Q^2)^2} Tr\Big{\{} \slashed{\epsilon}_3 \slashed{p}\slashed{\epsilon}_2  \slashed{k}_1 \slashed{k}_3\slashed{\epsilon}_1 \Big{\}} \Big{(} (P^2 - 2Q^2)\log\frac{Q^2 + i0}{P^2 + i0} + Q^2 - P^2  \Big{)}+
\\
    &+ \mathcal{A}_{1,2,3} \Big{(} \frac{2}{\varepsilon} + 1 - \gamma + \frac{Q^2 log(-Q^2/4\pi\mu^2 - i0) - P^2 log(-P^2/4\pi\mu^2 - i0)}{P^2 - Q^2} \Big{)} \Bigg{)}.
\end{aligned}
\end{equation}

To obtain a more compact and readable expression, let us define
\begin{equation}
\begin{aligned}
	&D_1 = P^2 / s = (x+\xi)\beta_1, \qquad D_3 = Q^2 / s = -(x-\xi) \beta_3, \\
	& \mathcal{B}_{1,2,3} = Tr\Big{\{} \slashed{\epsilon}_3 \slashed{k}_3 \slashed{k}_1 \slashed{\epsilon}_2\slashed{p} \slashed{\epsilon}_1 \Big{\}}, \\
	& \mathcal{C}_{1,23} = Tr\Big{\{} \slashed{\epsilon}_3 \slashed{p}\slashed{\epsilon}_2  \slashed{k}_1 \slashed{k}_3\slashed{\epsilon}_1 \Big{\}}.
\end{aligned}
\end{equation}
Finally, we get
\begin{equation}
\begin{aligned}
Tr[i\mathcal{M}^{3.M}_{1,2,3}~\slashed{p}]
=&
Tr[i\mathcal{M}^{0}_{1,2,3}~\slashed{p}]\cdot
\frac{g^2C_F}{(4\pi)^2}\Bigg\{\Big( \frac{s e^\gamma}{4\pi\mu^2} \Big)^{-\frac{\varepsilon}{2}}
\bigg{(} 
+\frac{2}{\varepsilon} +1- \frac{D_1log\Big(-D_1 -i0\Big) -  D_3log\Big(-D_3 -i0\Big)}{D_1-D_3}
\bigg{)} 
\\
&
\phantom{Tr[i\mathcal{M}^{0}_{1,2,3}~\slashed{p}]\cdot
\frac{g^2C_F}{(4\pi)^2}\Bigg\{}
+\bigg(\frac{\mathcal{B}_{1,2,3}}{\mathcal{A}_{1,2,3}}\bigg)
\frac{D_1}{D_1-D_3}
\left(
1+\frac{-2D_1+D_3}{D_1-D_3}\log\frac{D_1+i0}{D_3+i0}
\right)
\\
&
\phantom{Tr[i\mathcal{M}^{0}_{1,2,3}~\slashed{p}]\cdot
\frac{g^2C_F}{(4\pi)^2}\Bigg\{}
+\bigg(\frac{\mathcal{C}_{1,2,3}}{\mathcal{A}_{1,2,3}}\bigg)
\frac{D_3}{D_1-D_3}
\left(
1+\frac{D_1-2D_3}{D_1-D_3}\log\frac{D_1+i0}{D_3+i0}
\right)\Bigg\} .
\end{aligned}
\end{equation}
%\begin{equation}
%\begin{aligned}
%&Tr\big[ i\mathcal{M}^{3.M}_{1,2,3} ~\slashed{p}\big] = -C_F e_q^3 g^2 \frac{i}{(4\pi)^2}\frac{1}{\bar{p}k_1 + i0} \frac{1}{\underline{p}k_3 - i0}\Big( \frac{s e^\gamma}{4\pi\mu^2} \Big)^{-\frac{\varepsilon}{2}} \times
%\\
% &\Bigg{(} \frac{(x+\xi)\beta_1}{\big( (x+\xi)\beta_1 + (x-\xi)\beta_3 \big)^2} Tr\Big{\{} \slashed{\epsilon}_3 \slashed{k}_3 \slashed{k}_1 \slashed{\epsilon}_2\slashed{p} \slashed{\epsilon}_1 \Big{\}}\times \\ &\times \Big{(} \big( 2(x+\xi)\beta_1 + (x-\xi)\beta_3 \big)\log\frac{-(x-\xi)\beta_3 + i0}{(x+\xi)\beta_1 + i0} +(x+\xi)\beta_1 + (x-\xi)\beta_3  \Big{)} +\\
%&-\frac{(x-\xi)\beta_3}{\big( (x+\xi)\beta_1 + (x-\xi)\beta_3 \big)^2} Tr\Big{\{} \slashed{\epsilon}_3 \slashed{p}\slashed{\epsilon}_2  \slashed{k}_1 \slashed{k}_3\slashed{\epsilon}_1 \Big{\}}\times \\ &\times \Big{(} \big( (x+\xi)\beta_1 + 2(x-\xi)\beta_3 \big)\log\frac{(x+\xi)\beta_1 + i0}{-(x-\xi)\beta_3 + i0} +(x+\xi)\beta_1 + (x-\xi)\beta_3  \Big{)}+
%\\
%    &+ \mathcal{A}_{1,2,3} \Big{(} \frac{2}{\varepsilon} + 1 - \frac{(x-\xi)\beta_3 log\big( (x-\xi)\beta_3 - i0\big) + (x+\xi)\beta_1 log\big((x+\xi)\beta_1 - i0\big)}{(x+\xi)\beta_1 + (x-\xi)\beta_3} \Big{)} \Bigg{)}.
%\end{aligned}
%\end{equation}
\section{Box diagrams - 4.L and 4.R}\label{Section-M4}
\begin{figure}[H]
    \centering
    \begin{tikzpicture}[scale = 0.7]
  \begin{feynman}
    \vertex (q1) at (0, 0);
	\vertex (q1a) at (0, -1.5);
	\vertex (q1b) at (4.5, 0);
    \vertex (q3) at (3, 0);
	\vertex (q4) at (3, 0);
    \vertex (q5) at (6, 0);
    \vertex (in) at (0, -3);
    \vertex (out) at (6, -3);
    \vertex (k1) at (0, 2);
    \vertex (k2) at (3, 2);
    \vertex (k3) at (6, 2);
    \diagram* {
      (q1) -- [fermion] (q3) -- [fermion] (q1b) -- [fermion] (q5);
      (k1) -- [photon, momentum=\(k_1\)] (q1);
      (k2) -- [photon, momentum=\(k_2\)] (q3);
      (k3) -- [photon, momentum=\(k_3\)] (q5);
      (in) -- [fermion, momentum=\((x+\xi)p\)] (q1a) -- [fermion] (q1);
      (q5) -- [fermion, momentum=\((x-\xi)p\)] (out);
	(q1b) -- [gluon, quarter left, momentum = $k$] (q1a);
    };
  \end{feynman}
\end{tikzpicture}
\quad
    \begin{tikzpicture}[scale = 0.7]
  \begin{feynman}
    \vertex (q1) at (0, 0);
	\vertex (q1a) at (6, -1.5);
	\vertex (q1b) at (1.5, 0);
    \vertex (q3) at (3, 0);
	\vertex (q4) at (3, 0);
    \vertex (q5) at (6, 0);
    \vertex (in) at (0, -3);
    \vertex (out) at (6, -3);
    \vertex (k1) at (0, 2);
    \vertex (k2) at (3, 2);
    \vertex (k3) at (6, 2);
    \diagram* {
      (q1) -- [fermion] (q1b) -- [fermion] (q3) -- [fermion]  (q5);
      (k1) -- [photon, momentum=\(k_1\)] (q1);
      (k2) -- [photon, momentum=\(k_2\)] (q3);
      (k3) -- [photon, momentum=\(k_3\)] (q5);
      (in) -- [fermion, momentum=\((x+\xi)p\)] (q1);
      (q5) --[fermion] (q1a) -- [fermion, momentum=\((x-\xi)p\)] (out);
	(q1a) -- [gluon, quarter left, momentum = $k$] (q1b);
    };
  \end{feynman}
\end{tikzpicture}
    \caption{Diagrams 4.L and 4.R.}
    \label{4L}
\end{figure}
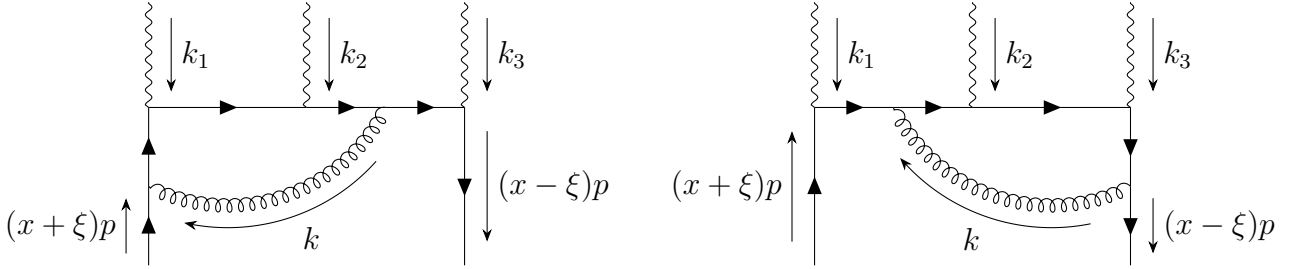

4-point diagrams, also called box diagrams, contain collinear divergence only. Because of more complicated structure of the corresponding traces and integrals, as compared to 2- and 3- point diagrams, we first focus only on the divergent part, and show how it can be handled analytically. Full derivation of the finite part of the amplitude is presented in Appendix \ref{Appdx-derivation-4}.
The box diagrams are presented in Fig. \ref{4L}.
Amplitude corresponding to 4.L takes the following form:
\begin{equation}\label{eq-4L-0}
\begin{aligned}
&Tr\big[i\mathcal{M}^{4.L}_{1,2,3} ~\slashed{p} \big] = -C_F (ie_q)^3(ig)^2 \frac{i}{-2\underline{p}k_3 + i0} \mu^{4-d} \int\frac{d^d k}{(2\pi)^d} Tr \Big{\{} \gamma_\mu \slashed{p} \slashed{\epsilon}_3 \big{(} \underline{\slashed{p}} - \slashed{k}_3 \big{)}\gamma^\mu \times
\\
&\phantom{Tr\big[i\mathcal{M}^{4.L}_{1,2,3}~\slashed{p} \big] =} \frac{i^4 \big{(} \bar{\slashed{p}} + \slashed{k}_1 + \slashed{k}_2 + \slashed{k} \big{)} \slashed{\epsilon}_2 \big{(} \bar{\slashed{p}} + \slashed{k}_1 + \slashed{k} \big{)} \slashed{\epsilon}_1 \big{(} \bar{\slashed{p}} + \slashed{k} \big{)}}{(k^2 + i0) \big{(} (\bar{p} + k)^2 + i0 \big{)}  \big{(} (\bar{p} + k + k_1)^2 + i0 \big{)}  \big{(}  (\bar{p} + k + k_1 + k_2)^2 + i0 \big{)} }  \Big{\}}.
\end{aligned}
\end{equation}
\subsection{Localizing divergences inside the integral}\label{subsec-int-4L}
Let use denote by $Tr(k)$ the trace present in Formula \eqref{eq-4L-0}. Using the Feynman parametrization we write the considered integral in the following form:
\begin{equation}\label{eq-integral-4L}
\begin{aligned}
&\int\frac{d^d k}{(2\pi)^d} \frac{Tr(k)}{(k^2 + i0) \big{(} (\bar{p} + k)^2 + i0 \big{)}  \big{(} (\bar{p} + k + k_1)^2 + i0 \big{)}  \big{(}  (\bar{p} + k + k_1 + k_2)^2 + i0 \big{)} }
\\
&= 6 \int\frac{d^d k}{(2\pi)^d} \int_{\substack{r,y,z \geq 0 \\ r+y+z \leq 1}} dr dy dz \:Tr(k) \times 
\\
& \bigg{(} k^2 + 2k\Big{(} (r+y+z)\bar{p} + (y+z) k_1 + z k_2 \Big{)} + 2(z+y)\bar{p}k_1 + 2z(\bar{p} + k_1)k_2 + i0  \bigg{)}^{-4}.
\end{aligned}
\end{equation}
The expression in the last line can be written as
$$
\Big{(} k + (r+y+z)\bar{p} + (y+z) k_1 + z k_2 \Big{)}^2 + 2(1-r-y-z)(y+z)\bar{p}k_1 + 2(1-r-y-z)z \bar{p}k_2 - 2(1-y-z)zk_1 k_2.
$$
We use the following substitution for Feynman parameters:
\begin{equation}\label{eq-4-subst}
  r = \bar{y}(1-\bar{z}), \quad  y = (1-\bar{r})(1-\bar{y}), \quad z = \bar{r}(1-\bar{y}).
\end{equation}
Note that
$$
y+z = 1 - \bar{y}, \quad r+y+z = 1 - \bar{z}\bar{y}.
$$
After reparametrization, the considered integral takes the following form (I omit the overline in $r,y,z$):
\begin{equation}
6 \int\frac{d^d k'}{(2\pi)^d} \int_0^1 y(1-y) dr dy dz \: Tr(k) \bigg{(} k'^2 + 2y(1-y)\Big{(} z\bar{p}k_1 + zr \bar{p}k_2 + r k_1 k_2 \Big{)} + i0  \bigg{)}^{-4},
\end{equation}
where
$$
k' = k + (1-yz)\bar{p} + (1-y) k_1 + r(1-y) k_2.
$$
Analogously to the case of diagrams $3.L$ and $3.R$, the collinear divergence manifests after performing the integration over Feynman parameters -- let us localize where it does occur.
The general form of the momentum integral reads
\begin{equation}
\begin{aligned}
&6 \int\frac{d^d k'}{(2\pi)^d} k'^{2n} \bigg{(} k'^2 + 2y(1-y)\Big{(} z\bar{p}k_1 + zr \bar{p}k_2 + r k_1 k_2 \Big{)} + i0  \bigg{)}^{-4} =
\\
& \frac{(-1)^ni }{(4\pi)^{d/2}} \Big{(} -2y(1-y)\big{(} z\bar{p}k_1 + zr \bar{p}k_2 + r k_1 k_2 \big{)} - i0  \Big{)}^{d/2 - 4 + n} \frac{\Gamma(4-n-d/2) \Gamma(n+d/2)}{\Gamma(d/2)},
\end{aligned}
\end{equation}
where $n$ can be equal to 0 or 1. To see it, note that from the trace one can get terms which are at most of power $3$ in $k'$. However, since we have shifted the momentum variable to complete the square in denominator, terms with $k^{'\alpha}$ and $k^{'\alpha}k^{'\beta}k^{'\gamma}$ vanish. The term with the even power of $k'$ in the numerator, i.e. $k^{'\alpha}k^{'\beta}$, is proportional to the metric tensor $g^{\alpha\beta}$ times the momentum integral with $k'^2$ in numerator, see Eq. \eqref{eq-loop-int}. Reparametrization of the Feynman parameters \eqref{eq-4-subst} allows to factor-out the integral over $dy$, which becomes
\begin{equation}\label{eq-int-over-dy}
\int_0^1 dy\: y^{d/2 - 3 + n + n_1} (1-y)^{d/2 - 3 + n + n_2} = \frac{\Gamma(d/2 - 2 + n + n_1) \Gamma(d/2 - 2 + n + n_2)}{\Gamma(d - 4 + 2n + n_1 + n_2)}.
\end{equation}
Indices $n_{1,2}$ are different for various terms in the trace, and will be discussed later. We see that the divergence in \eqref{eq-int-over-dy} occurs if both conditions: $n=0$ and ($n_1$ or $n_2$ = 0) hold.\\[0.2cm]
Integrals over $dr$ and $dz$ are more complicated and take the following general form:
\begin{equation}\label{eq-int-drdz-0}
\int_0^1 dr \int_0^1 dz\: r^{a}z^{b}\Big{(} -2\big{(} z\bar{p}k_1 + zr \bar{p}k_2 + r k_1 k_2 \big{)} - i0  \Big{)}^{d/2 - 4 + n}.
\end{equation}
$a, b, n$ are non-negative integers. It can be checked that \eqref{eq-int-drdz-0} is divergent only if $a=b=n=0$. Further discussion of divergences requires analysis of the trace.
\subsection{Trace - terms leading to divergences}
From the previous part, we see that divergences come from terms in the trace without the loop momentum components $k'^\mu$ in numerator, so that to find the divergent part of the amplitude we need to consider only the trace with all terms $\slashed{k}'$ removed:
\begin{equation}
Tr\Big{\{} \gamma_\mu \slashed{p} \slashed{\epsilon}_3 (-\slashed{k}_3) \gamma^\mu \big{(} yz\bar{\slashed{p}} + y\slashed{k}_1 + (1-r(1-y))\slashed{k}_2 \big{)} \slashed{\epsilon}_2 \big{(} yz \bar{\slashed{p}} + y \slashed{k}_1 - r(1-y)\slashed{k}_2 \big{)} \slashed{\epsilon}_1 \big{(} yz\bar{\slashed{p}} -(1-y)\slashed{k}_1 - r(1-y)\slashed{k}_2 \big{)} \Big{\}}.
\end{equation}
After contraction of matrices $\gamma^\mu$ in dimension $d=4-\varepsilon$ we get two terms:
\begin{equation}\label{eq-4L-tr}
\begin{aligned}
&2 Tr\Big{\{} \slashed{k}_3 \slashed{\epsilon}_3 \slashed{p} \big{(} y\slashed{k}_1 + (1-r(1-y))\slashed{k}_2 \big{)} \slashed{\epsilon}_2 \big{(} yz \bar{\slashed{p}} + y \slashed{k}_1 - r(1-y)\slashed{k}_2 \big{)} \slashed{\epsilon}_1 \big{(} yz\bar{\slashed{p}} -(1-y)\slashed{k}_1 - r(1-y)\slashed{k}_2 \big{)} \Big{\}}+
\\
&-\varepsilon Tr\Big{\{} \slashed{p} \slashed{\epsilon}_3 \slashed{k}_3 \big{(} yz\bar{\slashed{p}} + y\slashed{k}_1 + (1-r(1-y))\slashed{k}_2 \big{)} \slashed{\epsilon}_2 \big{(} yz \bar{\slashed{p}} + y \slashed{k}_1 - r(1-y)\slashed{k}_2 \big{)} \slashed{\epsilon}_1 \big{(} -(1-y)\slashed{k}_1 - r(1-y)\slashed{k}_2 \big{)} \Big{\}}.
\end{aligned}
\end{equation}
After expanding the traces we conclude that there are only 2 combinations which do not vanish in the considered kinematics (due to $\slashed{k}_i\slashed{k}_i = \slashed{p}\slashed{p}=0$) and lead to divergent integral over $dy$: 
\begin{equation}\label{eq-4l-div-0}
2y^2 z Tr\Big{\{} \slashed{k}_3 \slashed{\epsilon}_3 \slashed{p} \slashed{k}_2 \slashed{\epsilon}_2 \slashed{k}_1 \slashed{\epsilon}_1 \bar{\slashed{p}} \Big{\}} + 2 y^3 z Tr\Big{\{} \slashed{k}_3 \slashed{\epsilon}_3 \slashed{p} \slashed{k}_1 \slashed{\epsilon}_2 \slashed{k}_1 \slashed{\epsilon}_1 \bar{\slashed{p}} \Big{\}}.
\end{equation}
They are present only in the first trace, while all terms inside the trace proportional to $\varepsilon$ yield finite integrals, so that the second line of \eqref{eq-4L-tr} can be neglected. There are no terms corresponding to $n=a=b=0$ in Eq. \eqref{eq-int-drdz-0}, so that integrals over $dz$ and $dr$ are always convergent.  

The divergent part of the amplitude $4.L$ is hence proportional to
\begin{equation}\label{eq-4L-div}
\begin{aligned}
&\frac{2i}{(4\pi)^{d/2}}\Gamma(4-d/2) \int_0^1 dy\phantom{-} y^{d/2-1}(1-y)^{d/2-3}\times \\& \int_0^1 dr \int_0^1 dz z\Big{(} -2\big{(} z\bar{p}k_1 + zr \bar{p}k_2 + r k_1 k_2 \big{)} - i0  \Big{)}^{d/2 - 4} Tr\Big{\{} \slashed{k}_3 \slashed{\epsilon}_3 \slashed{p} \slashed{k}_2 \slashed{\epsilon}_2 \slashed{k}_1 \slashed{\epsilon}_1 \bar{\slashed{p}} \Big{\}}+
\\
&+\frac{2i}{(4\pi)^{d/2}}\Gamma(4-d/2) \int_0^1 dy y^{d/2}(1-y)^{d/2-3} \times \\ & \int_0^1 dr \int_0^1 dz z\Big{(} -2\big{(} z\bar{p}k_1 + zr \bar{p}k_2 + r k_1 k_2 \big{)} - i0  \Big{)}^{d/2 - 4} Tr\Big{\{} \slashed{k}_3 \slashed{\epsilon}_3 \slashed{p} \slashed{k}_1 \slashed{\epsilon}_2 \slashed{k}_1 \slashed{\epsilon}_1 \bar{\slashed{p}} \Big{\}}.
\end{aligned}
\end{equation}
For now, we focus on the divergent part of this expression only, so that we neglect terms in the expansion around $\varepsilon=0$ which are not divergent. They contribute to the finite part, which is computed later. We obtain
\begin{equation}\label{eq-4L-div-int}
  \eqref{eq-4L-div}_{div.} =  -\frac{2}{\varepsilon}\frac{i}{8\pi^2} Tr\Big{\{} \slashed{k}_3 \slashed{\epsilon}_3 \slashed{p} (\slashed{k}_1 + \slashed{k}_2) \slashed{\epsilon}_2 \slashed{k}_1 \slashed{\epsilon}_1 \bar{\slashed{p}} \Big{\}} \int_0^1 dr dz \: z\Big{(} -2\big{(} z\bar{p}k_1 + zr \bar{p}k_2 + r k_1 k_2 \big{)} - i0  \Big{)}^{-2}.
\end{equation}
Let us observe that we can equivalently write the expression in the integrand as:
$$
    -2\big{(} z(x+\xi)pk_1 + zr (x+\xi)pk_2 + r k_1 k_2 \big{)} - i0 \rightarrow
$$
\begin{equation}    
    \rightarrow  -2\big{(} z(x+\xi + i0 / pk_1)pk_1 + zr (x+\xi + i0/pk_2)pk_2 + r k_1 k_2 \big{)}
\end{equation}
Denote 
$$(x+\xi + i0 / pk_i)pk_i = \bar{p}k_i'.$$
The integral yields
\begin{equation}\label{eq-4L-div-0}
\begin{aligned}
&\int_0^1 dr dz \: z\Big{(} -2\big{(} z\bar{p}k_1' + zr \bar{p}k_2' + r k_1 k_2 \big{)} \Big{)}^{-2}=
\\
    &= \frac{1}{4(\bar{p}k_1' + \bar{p}k_2')\bar{p}k_1'}\log \bigg{(} \frac{\bar{p}k_1' + \bar{p}k_2' + k_1 k_2}{k_1 k_2 } \bigg{)}.
\end{aligned}
\end{equation}
Let us simplify Formula \eqref{eq-4L-div-0}.  Using $\bar{p}k_1' + \bar{p}k_2' = -\bar{p}k_3 +  i0$ one obtains
\begin{equation}
\begin{aligned}
&\bar{p}k_1' + \bar{p}k_2' + k_1 k_2 = (x+\xi) p (k_1+k_2) + k_1 k_2 + i0 =
\\
&=\frac{1}{2}\big{(} (x+\xi) p + k_1 + k_2 \big{)}^2 + i0 = \frac{1}{2} \big{(} (x-\xi) p - k_3 \big{)}^2 + i0 = -(x-\xi) pk_3 + i0.
\end{aligned}
\end{equation}
Moreover,
$$
(x+\xi) p + k_1 + k_2 = (x-\xi) p - k_3 \implies 2\xi p + k_3 = -(k_1 + k_2) \implies 2\xi pk_3 = k_1 k_2,
$$
so that
\begin{equation}
   \frac{\bar{p}k_1' + \bar{p}k_2' + k_1 k_2}{k_1 k_2 } = \frac{-(x-\xi)p k_3 + i0}{2\xi p k_3} = \frac{1}{2\xi}\big{(}\xi - x + i0\cdot sgn(pk_3)\big{)}. 
\end{equation}
Hence, the divergent part of the integral \eqref{eq-integral-4L} reads
\begin{equation}
    -\frac{2}{\varepsilon}\frac{i}{32\pi^2} Tr\Big{\{} \slashed{k}_3 \slashed{\epsilon}_3 \slashed{p} (\slashed{k}_1 + \slashed{k}_2) \slashed{\epsilon}_2 \slashed{k}_1 \slashed{\epsilon}_1 \bar{\slashed{p}} \Big{\}}\frac{-1}{(\bar{p}k_1 + i0)(\bar{p}k_3 + i0)} \log \Big{(} \frac{1}{2\xi}\big{(}\xi - x + i0\cdot sgn(pk_3)\big{)} \Big{)}.
\end{equation}
Using $(x+\xi)p + k_1 + k_2 = (x-\xi)p - k_3$ and anti-commuting $\slashed{p}\slashed{k}_3$ we can write it in a simpler way
\begin{equation}
    -\frac{2}{\varepsilon}\frac{i}{16\pi^2} Tr\Big{\{} \slashed{k}_3 \slashed{\epsilon}_3 \slashed{\epsilon}_2 \slashed{k}_1 \slashed{\epsilon}_1 \slashed{p} \Big{\}} \frac{1}{\bar{p}k_1 + i0} \log \Big{(} \frac{1}{2\xi}\big{(}\xi - x + i0\cdot sgn(pk_3)\big{)} \Big{)}.
\end{equation}
We recognize that the resulting trace is $\mathcal{A}_{1,2,3}$. Finally, the divergent part of the amplitude $4.L$ 
\begin{equation}
\begin{aligned}
    Tr\big[i\mathcal{M}^{4.L}_{1,2,3}~ \slashed{p} \big]_{\mathrm{div.}}&= iC_F e_q^3 g^2 \frac{1}{(4\pi)^2} \mathcal{A}_{1,2,3} \frac{1}{\bar{p}k_1 + i0}\frac{1}{\underline{p}k_3 - i0} \frac{4}{\epsilon}\log \Big{(} \frac{1}{2\xi}\big{(}\xi - x + i0\cdot sgn(pk_3)\big{)} \Big{)}\\
&= -\frac{4}{\varepsilon} \frac{g^2 C_F}{(4\pi)^2} Tr\big[i\mathcal{M}^{0}_{1,2,3}~ \slashed{p} \big] \log \Big{(} \frac{1}{2\xi}\big{(}\xi - x + i0\cdot sgn(pk_3)\big{)} \Big{)}.
\end{aligned}
\end{equation}
Amplitude corresponding to the diagram 4.R reads
\begin{equation}
\begin{aligned}
&Tr\big[i\mathcal{M}^{4.R}_{1,2,3}~ \slashed{p}\big] = -C_F (ie_q)^3(ig)^2 \frac{i}{2\bar{p}k_1 + i0} \mu^{4-d} \int\frac{d^d k}{(2\pi)^d} Tr \Big{\{} \gamma_\mu (\bar{\slashed{p}}+\slashed{k}_1) \slashed{\epsilon}_1 \slashed{p} \gamma^\mu \times
\\
&\phantom{Tr\big[i\mathcal{M}^{4.R}_{1,2,3}~ \slashed{p}\big] =} \frac{i^4 (\underline{\slashed{p}} + \slashed{k})\slashed{\epsilon}_3 (\underline{\slashed{p}} - \slashed{k}_3 + \slashed{k})\slashed{\epsilon}_2 (\underline{\slashed{p}}-\slashed{k}_3-\slashed{k}_2+\slashed{k}) }{(k^2 + i0) \big{(} (\underline{p} + k)^2 + i0 \big{)}  \big{(} (\underline{p} + k - k_3)^2 + i0 \big{)}  \big{(}  (\underline{p} + k - k_3 - k_2)^2 + i0 \big{)} }  \Big{\}}.
\end{aligned}
\end{equation}
Let us notice, that this expression can be obtained by taking the following substitutions: 
$$(\xi\rightarrow -\xi, \quad k_1\rightarrow -k_3, \quad k_2 \rightarrow -k_2)$$
in the Formula \eqref{eq-4L-0}. It allows to easily obtain the divergent part of the second box diagram:
\begin{equation}
\begin{aligned}
    Tr\big[i\mathcal{M}^{4.R}_{1,2,3}~\slashed{p}\big]_{\mathrm{div}}&= iC_F e_q^3 g^2 \frac{1}{(4\pi)^2} \mathcal{A}_{1,2,3} \frac{1}{\bar{p}k_1 + i0}\frac{1}{\underline{p}k_3 - i0} \frac{4}{\epsilon}\log \Big{(} \frac{1}{2\xi}\big{(}\xi + x + i0\cdot sgn(pk_1)\big{)} \Big{)}\\
&= -\frac{4}{\varepsilon} \frac{g^2 C_F}{(4\pi)^2} Tr\big[i\mathcal{M}^{0}_{1,2,3}~ \slashed{p} \big] \log \Big{(} \frac{1}{2\xi}\big{(}\xi + x + i0\cdot sgn(pk_1)\big{)} \Big{)}.
\end{aligned}
\end{equation}

\subsection{The finite part of the amplitude}
Contributions to the finite part can be divided into 2 classes: the first one will be terms in traces that yield finite integrals over Feynman parameters. From the discussion concerning the loop integrals in Subsec. \ref{subsec-int-4L} one concludes, that they are proportional to the following functions:
\begin{equation}\label{eq-F-def}
\begin{aligned}
&\int_0^1 dr dz \: r^{a}z^{b}\Big{(} -2\big{(} z\bar{p}k_1 + zr \bar{p}k_2 + r k_1 k_2 \big{)} - i0  \Big{)}^{-n} =\\
&= (-1)^n s^{-n} \int_0^1 dr dz r^{a}z^{b}\Big{(} \big{(} z(x+\xi)\beta_1 + zr (x+\xi)\beta_2 + r \kappa_3 \big{)} + i0  \Big{)}^{-n} = \\
&=: (-1)^n s^{-n} \mathcal{F}\big(n,a,b; x, \xi, \{ \beta_i \} \big).
\end{aligned}
\end{equation}
In the definition we used the fact that $\kappa_3 = -2\xi \beta_3$.
All the functions $\mathcal{F}$ present in the considered amplitude are collected in Appendix \ref{Appdx-full-amplitude}.

The second class of finite terms are those from the finite part of divergent integrals. In particular, since the integral 
$$\int_0^1 dr \int_0^1 dz \:z\Big{(} -2\big{(} z\bar{p}k_1 + zr \bar{p}k_2 + r k_1 k_2 \big{)} - i0  \Big{)}^{d/2 - 4}$$
is multiplied by a divergent factor $1/\varepsilon$, to finite part will contain a contribution resulting from expansion of the integrand in $\varepsilon$ according to 
\begin{equation}
X^{-2-\varepsilon/2} = X^{-2} -\frac{\varepsilon}{2}X^{-2} \log X + \mathcal{O}(\varepsilon^2).
\end{equation}
To account for this term, let us define the function
\begin{equation}\label{eq-G-def}
\begin{aligned}
	&\mathcal{G}(x,\xi, \{ \beta_i\}) := \\&\int_0^1 dr \int_0^1 dz z\Big{(} \big{(} z(x+\xi)\beta_1 + zr(x+\xi)\beta_2 + r\kappa_3 + i0  \Big{)}^{-2}\ln \big(- z(x+\xi)\beta_1 - zr(x+\xi)\beta_2 - r\kappa_3 - i0 \big).
	\end{aligned}
\end{equation}
For the exact form of \eqref{eq-G-def}, see Appendix \ref{Appdx-full-amplitude}. Using Formulas \eqref{eq-F-def} and \eqref{eq-G-def} we can write the finite part of the amplitude 4.L:
\begin{equation}\label{eq-4L-finite}
\begin{aligned}
&Tr\big[\mathcal{M}^{4.L}_{1,2,3}~\slashed{p} \big]_{\mathrm{fin.}} = \: -C_F e_q^3 g^2 \frac{i}{(4\pi)^2} \frac{1}{2\underline{p}k_3 - i0} s^{-2} \bigg\{ \\
&\mathcal{F} \big(2,1,0;x, \xi, \{ \beta_i \}\big) \Big( Tr\big( \slashed{k}_3 \slashed{\epsilon}_3 \slashed{p} \slashed{k}_1 \slashed{\epsilon}_2 \slashed{k}_2 \slashed{\epsilon}_1 \slashed{k}_1 \big) - Tr\big( \slashed{k}_3 \slashed{\epsilon}_3 \slashed{p} \slashed{k}_1 \slashed{\epsilon}_2 \slashed{k}_1 \slashed{\epsilon}_1 \slashed{k}_2 \big) -2Tr\big( \slashed{k}_3 \slashed{\epsilon}_3 \slashed{p} \slashed{k}_2 \slashed{\epsilon}_2 \slashed{k}_1 \slashed{\epsilon}_1 \slashed{k}_2 \big) \Big)+\\
&\mathcal{F} \big(2,0,1;x, \xi, \{ \beta_i \}\big) \Big( - Tr\big( \slashed{k}_3 \slashed{\epsilon}_3 \slashed{p} \slashed{k}_1 \slashed{\epsilon}_2 \bar{\slashed{p}} \slashed{\epsilon}_1 \slashed{k}_1 \big) -2Tr\big( \slashed{k}_3 \slashed{\epsilon}_3 \slashed{p} \slashed{k}_2 \slashed{\epsilon}_2 \bar{\slashed{p}} \slashed{\epsilon}_1 \slashed{k}_1 \big) +\\
&\quad - 4 Tr\big( \slashed{k}_3 \slashed{\epsilon}_3 \slashed{p} \slashed{k}_2 \slashed{\epsilon}_2 \slashed{k}_1 \slashed{\epsilon}_1 \bar{\slashed{p}} \big) - 5 Tr\big( \slashed{k}_3 \slashed{\epsilon}_3 \slashed{p} \slashed{k}_1 \slashed{\epsilon}_2 \slashed{k}_1 \slashed{\epsilon}_1 \bar{\slashed{p}} \big) \Big)\\
&\mathcal{F} \big(2,1,1;x, \xi, \{ \beta_i \}\big) \Big( -Tr\big( \slashed{k}_3 \slashed{\epsilon}_3 \slashed{p} \slashed{k}_1 \slashed{\epsilon}_2 \bar{\slashed{p}} \slashed{\epsilon}_1 \slashed{k}_2 \big) - Tr\big( \slashed{k}_3 \slashed{\epsilon}_3 \slashed{p} \slashed{k}_1\slashed{\epsilon}_2 \slashed{k}_2 \slashed{\epsilon}_1 \bar{\slashed{p}} \big) +\\
&\quad + Tr\big( \slashed{k}_3 \slashed{\epsilon}_3 \slashed{p} \slashed{k}_2 \slashed{\epsilon}_2 \bar{\slashed{p}} \slashed{\epsilon}_1 \slashed{k}_1 \big) -2 Tr\big( \slashed{k}_3 \slashed{\epsilon}_3 \slashed{p} \slashed{k}_2 \slashed{\epsilon}_2 \bar{\slashed{p}} \slashed{\epsilon}_1 \slashed{k}_2 \big) -  Tr\big( \slashed{k}_3 \slashed{\epsilon}_3 \slashed{p} \slashed{k}_2 \slashed{\epsilon}_2 \slashed{k}_1 \slashed{\epsilon}_1 \bar{\slashed{p}} \big) \Big) +\\
&\mathcal{F} \big(2,2,0;x, \xi, \{ \beta_i \}\big) \Big( Tr\big( \slashed{k}_3 \slashed{\epsilon}_3 \slashed{p} \slashed{k}_1 \slashed{\epsilon}_2 \slashed{k}_2 \slashed{\epsilon}_1 \slashed{k}_2 \big) + Tr\big( \slashed{k}_3 \slashed{\epsilon}_3 \slashed{p} \slashed{k}_2 \slashed{\epsilon}_2 \slashed{k}_1 \slashed{\epsilon}_1 \slashed{k}_2 \big) \Big) + \\
&  \mathcal{F} \big(2,2,1;x, \xi, \{ \beta_i \}\big) Tr\big( \slashed{k}_3 \slashed{\epsilon}_3 \slashed{p} \slashed{k}_2 \slashed{\epsilon}_2 \bar{\slashed{p}} \slashed{\epsilon}_1 \slashed{k}_2 \big)+\\
&s \mathcal{F}(1,1,0;x, \xi \{ \beta_i \}) \Big( Tr\big(\slashed{k}_3 \slashed{\epsilon}_3 \slashed{p} \slashed{\epsilon}_2 \slashed{\epsilon}_1 \slashed{k}_2 \big) + Tr\big(\slashed{p}  \slashed{\epsilon}_3 \slashed{k}_3 \slashed{\epsilon}_2 \slashed{k}_2 \slashed{\epsilon}_1  \big)+ Tr\big(\slashed{k}_3 \slashed{\epsilon}_3 \slashed{p} \slashed{k}_2 \slashed{\epsilon}_2 \slashed{\epsilon}_1 \big) \Big)+\\
& s \mathcal{F}(1,0,0; x, \xi, \{ \beta_i \}) \Big( Tr\big(\slashed{k}_3 \slashed{\epsilon}_3 \slashed{p} \slashed{\epsilon}_2 \slashed{\epsilon}_1 \slashed{k}_1 \big) -  Tr\big(\slashed{p}  \slashed{\epsilon}_3 \slashed{k}_3 \slashed{\epsilon}_2 \slashed{k}_1 \slashed{\epsilon}_1  \big) - Tr\big(\slashed{k}_3 \slashed{\epsilon}_3 \slashed{p} \slashed{k}_1 \slashed{\epsilon}_2 \slashed{\epsilon}_1 \big) -2Tr\big(\slashed{k}_3 \slashed{\epsilon}_3 \slashed{p} \slashed{k}_2 \slashed{\epsilon}_2 \slashed{\epsilon}_1 \big) \Big) + \\
& - 2 \mathcal{G}(x, \xi, \{ \beta_i \}) Tr\big( \slashed{k}_3 \slashed{\epsilon}_3 \slashed{p} \slashed{k}_3 \slashed{\epsilon}_2 \slashed{k}_1 \slashed{\epsilon}_1 \bar{\slashed{p}} \big) \Big) \bigg\}.
\end{aligned}
\end{equation}
The expression for the finite part of $Tr\big[ \mathcal{M}^{4.R}_{1,2,3}~ \slashed{p} \big] $ is obtained by changing 
$$\xi\rightarrow -\xi, \quad k_1 \rightarrow -k_3, \quad k_2 \rightarrow -k_2, \quad k_3 \rightarrow -k_1,$$
according to the previous observation. The full derivation of \eqref{eq-4L-finite} is presented in the Appendix \ref{Appdx-derivation-4}. Traces present in the Formula \eqref{eq-4L-finite} can be further simplified using Mathematica -- the resulting form of the amplitude is shown in Appendix \ref{Appdx-full-amplitude}.

\section{The 5-point diagram}\label{section-5-point}
\begin{figure}[H]
    \centering
    \begin{tikzpicture}[scale = 0.6]
  \begin{feynman}
    \vertex (q1);
    \vertex [right=of q1] (q2);
    \vertex [right=of q2] (q3);
    \vertex [right=of q3] (q4);
    \vertex [right=of q4] (q5);
    \vertex [below=of q1] (qb1);
    \vertex [below=of q5] (qb5);
    \vertex [below=of qb1] (in);
    \vertex [below=of qb5] (out);
    \vertex [above=of q1] (k1);
    \vertex [above=of q3] (k2);
    \vertex [above=of q5] (k3);
    \diagram* {
      (q1) -- [fermion] (q2) -- [fermion] (q3) -- [fermion] (q4) -- [fermion] (q5);
      (k1) -- [photon, momentum=\(k_1\)] (q1);
      (k2) -- [photon, momentum=\(k_2\)] (q3);
      (k3) -- [photon, momentum=\(k_3\)] (q5);
      (in) -- [fermion, momentum=\((x+\xi)p\)] (qb1);
      (qb1) -- [fermion] (q1);
      (q5) -- [fermion] (qb5);
      (qb5) -- [fermion, momentum=\((x-\xi)p\)] (out);
      (qb5) -- [gluon, momentum=\(k\)] (qb1);
    };
  \end{feynman}
\end{tikzpicture}
    \caption{The 5-point diagram.}
    \label{5M}
\end{figure}
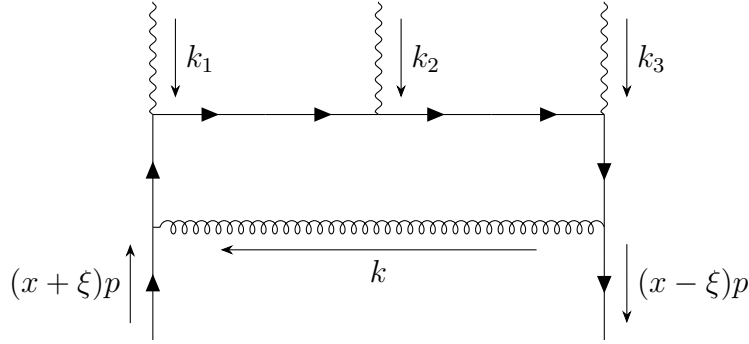

The last integral, which corresponds to the diagram shown in Fig. \ref{5M}, in which the loop contains 5 propagators, might (a priori) develop divergences $1/\varepsilon^2$, since 3 momenta can be simultaneously collinear and lightlike ($\bar{p} +k$, $k$, and $\underline{p} - k$). Fortunately, it turns out that the trace contains terms which cancel those in the denominator. 
The amplitude reads
\begin{equation}
Tr\big[ i\mathcal{M}^5_{1,2,3}~ \slashed{p} \big] = C_F e_q^3 g^2 \mu^{4-d} \int \frac{d^d k}{(2\pi)^d} \frac{Tr\Big{\{} \gamma_\mu \slashed{p} \gamma^\mu (\slashed{\underline{p}} + \slashed{k})\slashed{\epsilon}_3(\slashed{\bar{p}} + \slashed{k}_1 + \slashed{k}_2 + \slashed{k})\slashed{\epsilon}_2 (\slashed{\bar{p}} + \slashed{k}_1 + \slashed{k}) \slashed{\epsilon}_1 (\slashed{\bar{p}} + \slashed{k}) \Big{\}}}{(k^2+i0)(D_+ +i0)(D_{-} +i0)(D_1+i0) (D_{12}+i0)},
\end{equation}
where
$$
D_{\pm} = \big{(} (x\pm \xi)p + k  \big{)}^2, \quad D_{1} = \big{(} (x + \xi)p + k_1 + k  \big{)}^2, \quad D_{12} = \big{(} (x + \xi)p + k_1 + k_2 + k  \big{)}^2.
$$
We use the following gamma-identities inside the trace:
$$
\gamma^\mu \slashed{p} \gamma_\mu = (-2 + \varepsilon) \slashed{p}, \quad \slashed{k} \slashed{p} \slashed{k} = 2(pk) \slashed{k} - k^2 \slashed{p}.
$$
Let us observe, that
$$
2(pk) = \frac{1}{2\xi} (D_+ - D_{-} ), \quad k^2 = \frac{1}{2\xi} \big{(} (x+\xi)D_{-} - (x-\xi)D_+ \big{)}.
$$
It follows that
\begin{equation}
2(pk) \slashed{k} - k^2 \slashed{p} = \frac{D_+}{2\xi} (\slashed{k}+\slashed{\underline{p}}) - \frac{D_{-}}{2\xi} (\slashed{k} + \slashed{\bar{p}}).
\end{equation}
The amplitude breaks into two 4-point integrals:
\begin{equation}
Tr\big[ i\mathcal{M}^5_{1,2,3}~\slashed{p}\big] = Tr\big[ i\mathcal{M}^{5.L}_{1,2,3}~\slashed{p} \big] + Tr\big[ i\mathcal{M}^{5.R}_{1,2,3}~\slashed{p} \big] ,
\end{equation}
where
\begin{align}
&Tr\big[ i\mathcal{M}^{5.L}_{1,2,3}~ \slashed{p}\big] := - \frac{-2+\varepsilon}{2\xi}C_F e_q^3 g^2 \mu^{4-d} \int \frac{d^d k}{(2\pi)^d} \frac{Tr\Big{\{} (\slashed{\bar{p}} + \slashed{k})\slashed{\epsilon}_3(\slashed{\bar{p}} + \slashed{k}_1 + \slashed{k}_2 + \slashed{k})\slashed{\epsilon}_2 (\slashed{\bar{p}} + \slashed{k}_1 + \slashed{k}) \slashed{\epsilon}_1 \Big{\}}}{(k^2+i0)(D_{+} +i0)(D_1+i0) (D_{12}+i0)}, \label{eq-M-5-split}
\\
&Tr\big[ i\mathcal{M}^{5.R}_{1,2,3}~\slashed{p} \big] := \frac{-2+\varepsilon}{2\xi}C_F e_q^3 g^2 \mu^{4-d} \int \frac{d^d k}{(2\pi)^d} \frac{Tr\Big{\{} (\slashed{\underline{p}} + \slashed{k})\slashed{\epsilon}_3(\slashed{\bar{p}} + \slashed{k}_1 + \slashed{k}_2 + \slashed{k})\slashed{\epsilon}_2 (\slashed{\bar{p}} + \slashed{k}_1 + \slashed{k}) \slashed{\epsilon}_1 \Big{\}}}{(k^2+i0)(D_{-} +i0)(D_1+i0) (D_{12}+i0)}.
\end{align}
The names $5.L$ and $5.R$ come from the structure of denominators in integrands, which are the same as in amplitudes $4.L$ and $4.R$.

\subsection{The divergent part}
Let us start from the term $5.L$: performing steps analogous to those in the case of $4.L$ we obtain
\begin{equation}
\begin{aligned}
&\int \frac{d^d k}{(2\pi)^d} \frac{Tr\Big{\{} (\slashed{\bar{p}} + \slashed{k})\slashed{\epsilon}_3(\slashed{\bar{p}} + \slashed{k}_1 + \slashed{k}_2 + \slashed{k})\slashed{\epsilon}_2 (\slashed{\bar{p}} + \slashed{k}_1 + \slashed{k}) \slashed{\epsilon}_1 \Big{\}}}{(k^2+i0)(D_{+} +i0)(D_1+i0) (D_{12}+i0)} = 
\\
 & 6 \int\frac{d^d k'}{(2\pi)^d} \int_0^1 y(1-y) dr dy dz \frac{Tr\Big{\{} (\slashed{\bar{p}} + \slashed{k})\slashed{\epsilon}_3(\slashed{\bar{p}} + \slashed{k}_1 + \slashed{k}_2 + \slashed{k})\slashed{\epsilon}_2 (\slashed{\bar{p}} + \slashed{k}_1 + \slashed{k}) \slashed{\epsilon}_1 \Big{\}}}{\Big{(} k'^2 + 2y(1-y)\Big{(} z\bar{p}k_1 + zr \bar{p}k_2 + r k_1 k_2 \Big{)} + i0  \Big{)}^4},
    \label{eq-5-4r-1}
\end{aligned}
\end{equation}
where
\begin{equation}
k' = k + (1-yz)\bar{p} + (1-y) k_1 + r(1-y) k_2.
\end{equation}
Following the same reasoning as in the case of $4.L$, we analyze the following trace
\begin{equation}
 Tr\Big{\{} (yz\slashed{\bar{p}} -(1-y)\slashed{k}_1 -r(1-y)\slashed{k}_2)\slashed{\epsilon}_3(yz\slashed{\bar{p}} + y\slashed{k}_1 +\big{(}1 -r(1-y)\big{)}\slashed{k}_2\big{)}\slashed{\epsilon}_2 (yz\slashed{\bar{p}} + y\slashed{k}_1 -r(1-y)\slashed{k}_2) \slashed{\epsilon}_1 \Big{\}} .  
\end{equation}
Divergent integrals over the parameter $y$ come from sum of the following terms:
\begin{equation}
    y^3 z Tr\Big{\{} \slashed{\bar{p}} \slashed{\epsilon}_3 \slashed{k}_1 \slashed{\epsilon}_2 \slashed{k}_1 \slashed{\epsilon}_1 \Big{\}} + y^2 z Tr\Big{\{} \slashed{\bar{p}} \slashed{\epsilon}_3 \slashed{k}_2 \slashed{\epsilon}_2 \slashed{k}_1 \slashed{\epsilon}_1 \Big{\}}.
\end{equation}
As in the Section \ref{Section-M4}, all resulting integrals over $dz$ and $dr$ are convergent. The divergent part of \eqref{eq-5-4r-1} can be written as
\begin{equation}
-\frac{i}{(4\pi)^2}\frac{2}{\varepsilon} \Big( -(x+\xi)Tr\big\{\slashed{p}\slashed{\epsilon}_3 \slashed{k}_3 \slashed{\epsilon}_2 \slashed{k}_1 \slashed{\epsilon}_1 \big\} \Big) \int_0^1 dr dz\: z\Big{(} -2\big{(} z\bar{p}k_1 + zr \bar{p}k_2 + r k_1 k_2 \big{)} - i0  \Big{)}^{-2}.
\end{equation}
It is the same integral as the one in Eq. \eqref{eq-4L-div-int}. Using results of the previous section, we obtain the divergent part of Eq. \eqref{eq-5-4r-1} is
\begin{equation}
    \frac{i}{(4\pi)^2}\frac{2}{\varepsilon}\mathcal{A}_{1,2,3}\frac{x-\xi}{(\underline{p}k_3-i0)(\bar{p}k_1+i0)}\log \Big{(} \frac{1}{2\xi}\big{(}\xi - x + i0\cdot sgn(pk_3)\big{)} \Big{)}
\end{equation}
After multiplying it by $-\frac{-2+\epsilon}{2\xi}C_F e_q^3 g^2$ we get the divergent part of $Tr\big[ i\mathcal{M}^{5.L}_{1,2,3} ~\slashed{p}\big]$
\begin{equation}
  Tr\big[ i\mathcal{M}^{5.L}_{1,2,3} ~\slashed{p} \big]_{\mathrm{div.}} = \frac{i}{(4\pi)^2}C_F e_q^3 g^2\mathcal{A}_{1,2,3} \frac{1}{(\underline{p}k_3-i0)(\bar{p}k_1+i0)}\frac{2}{\varepsilon} \frac{x-\xi}{\xi}\log \Big{(} \frac{1}{2\xi}\big{(}\xi - x + i0\cdot sgn(pk_3)\big{)} \Big{)} \label{eq-M5-L-div}
\end{equation}
The divergent part of $Tr\big[ i\mathcal{M}^{5.R}_{1,2,3}~ \slashed{p} \big]$ by analogous steps is found to be equal to
\begin{equation}
      Tr\big[ i\mathcal{M}^{5.R}_{1,2,3}~ \slashed{p} \big]_{\mathrm{div.}} = -\frac{i}{(4\pi)^2}C_F e_q^3 g^2\mathcal{A}_{1,2,3} \frac{1}{(\underline{p}k_3-i0)(\bar{p}k_1+i0)}\frac{2}{\varepsilon} \frac{x+\xi}{\xi}\log \Big{(} \frac{1}{2\xi}\big{(}\xi + x + i0\cdot sgn(pk_1)\big{)} \Big{)}
\end{equation}
Hence the divergent part of amplitude $Tr\big[ i\mathcal{M}^5_{1,2,3}~\slashed{p}\big]$:
\begin{equation}
\begin{aligned}
	&Tr\big[ i\mathcal{M}^5_{1,2,3}~\slashed{p}\big]_{\mathrm{div.}} =\frac{i}{(4\pi)^2}C_F e_q^3 g^2\mathcal{A}_{1,2,3} \frac{1}{(\underline{p}k_3-i0)(\bar{p}k_1+i0)} \times \\
& \frac{2}{\varepsilon} \bigg(  \frac{x-\xi}{\xi}\log \Big{(} \frac{1}{2\xi}\big{(}\xi - x + i0\cdot sgn(pk_3)\big{)} \Big{)} - \frac{x+\xi}{\xi}\log \Big{(} \frac{1}{2\xi}\big{(}\xi + x + i0\cdot sgn(pk_1)\big{)} \Big{)} \bigg)\\
& =- \frac{g^2 C_F}{(4\pi)^2} Tr\big[ i\mathcal{M}^0_{1,2,3}~\slashed{p}\big] \times \\
& \frac{2}{\varepsilon} \bigg(  \frac{x-\xi}{\xi}\log \Big{(} \frac{1}{2\xi}\big{(}\xi - x + i0\cdot sgn(pk_3)\big{)} \Big{)} - \frac{x+\xi}{\xi}\log \Big{(} \frac{1}{2\xi}\big{(}\xi + x + i0\cdot sgn(pk_1)\big{)} \Big{)} \bigg)
\end{aligned}
\end{equation}
It concludes the computation of the divergent part of the amplitude at the one loop order. The divergence is proportional to the same trace structure, that is present in the LO amplitude -- as it will be discussed in Section \ref{section-factorization}, this feature will be crucial in the proof of the QCD collinear factorization for the considered process at NLO.

\subsection{The finite part}
Before proceeding with the proof of factorization, let us write the expression for the remaining finite part of $Tr\big[ i\mathcal{M}^{5.L}_{1,2,3}~\slashed{p}\big]$, which is
\begin{equation}
\begin{aligned}
&Tr\big[ i\mathcal{M}^{5.L}_{1,2,3}~\slashed{p}\big]_{\mathrm{fin.}} = \: -\frac{i}{(4\pi)^2}C_F e_q^3 g^2 \frac{1}{\xi}s^{-2} \bigg\{ \\
&\frac{1}{2}\mathcal{F}\big(2,0,1;x, \xi, \{ \beta_i \}\big)\Big( Tr\big( \slashed{k}_1 \slashed{\epsilon}_3 \bar{\slashed{p}} \slashed{\epsilon}_2 \slashed{k}_1 \slashed{\epsilon}_1  \big) + Tr\big( \slashed{k}_1 \slashed{\epsilon}_3 \slashed{k}_1 \slashed{\epsilon}_2  \bar{\slashed{p}} \slashed{\epsilon}_1  \big) 
	+ 2 Tr\big( \slashed{k}_1 \slashed{\epsilon}_3 \slashed{k}_2 \slashed{\epsilon}_2  \bar{\slashed{p}} \slashed{\epsilon}_1  \big)  + \\
& + 5 Tr\big( \bar{\slashed{p}} \slashed{\epsilon}_3 \slashed{k}_1 \slashed{\epsilon}_2 \slashed{k}_1 \slashed{\epsilon}_1 \big) + 4 Tr\big( \bar{\slashed{p}} \slashed{\epsilon}_3 \slashed{k}_2 \slashed{\epsilon}_2 \slashed{k}_1 \slashed{\epsilon}_1 \big) \Big) + \\
& \frac{1}{2}\mathcal{F}\big(2,1,0;x, \xi, \{ \beta_i \}\big)\Big( -Tr\big( \slashed{k}_1 \slashed{\epsilon}_3 \slashed{k}_1 \slashed{\epsilon}_2  \slashed{k}_2 \slashed{\epsilon}_1  \big) +Tr\big( \slashed{k}_2 \slashed{\epsilon}_3 \slashed{k}_1 \slashed{\epsilon}_2 \slashed{k}_1 \slashed{\epsilon}_1  \big) 
	+2Tr\big( \slashed{k}_2 \slashed{\epsilon}_3 \slashed{k}_2 \slashed{\epsilon}_2 \slashed{k}_1 \slashed{\epsilon}_1  \big) \Big) + \\
&\frac{1}{2} \mathcal{F}\big(2,1,1;x, \xi, \{ \beta_i \}\big) \Big( Tr\big( \bar{\slashed{p}} \slashed{\epsilon}_3 \slashed{k}_1 \slashed{\epsilon}_2 \slashed{k}_1 \slashed{\epsilon}_1  \big) + Tr\big( \bar{\slashed{p}} \slashed{\epsilon}_3 \slashed{k}_2 \slashed{\epsilon}_2 \slashed{k}_1 \slashed{\epsilon}_1  \big) - Tr\big( \slashed{k}_1 \slashed{\epsilon}_3 \bar{\slashed{p}} \slashed{\epsilon}_2 \slashed{k}_2 \slashed{\epsilon}_1  \big)  + \\
& - Tr\big( \slashed{k}_1 \slashed{\epsilon}_3 \slashed{k}_2 \slashed{\epsilon}_2  \bar{\slashed{p}} \slashed{\epsilon}_1  \big) + Tr\big( \slashed{k}_2 \slashed{\epsilon}_3 \bar{\slashed{p}} \slashed{\epsilon}_2 \slashed{k}_1 \slashed{\epsilon}_1  \big) +
+ Tr\big( \slashed{k}_2 \slashed{\epsilon}_3 \slashed{k}_1 \slashed{\epsilon}_2 \bar{\slashed{p}} \slashed{\epsilon}_1  \big) + 2 Tr\big( \slashed{k}_2 \slashed{\epsilon}_3 \slashed{k}_2 \slashed{\epsilon}_2 \bar{\slashed{p}} \slashed{\epsilon}_1  \big) \Big) + \\
& \frac{1}{2} \mathcal{F}\big(2,2,0;x, \xi, \{ \beta_i \}\big) \Big( Tr\big( \slashed{k}_2 \slashed{\epsilon}_3 \slashed{k}_1 \slashed{\epsilon}_2 \slashed{k}_2 \slashed{\epsilon}_1  \big) - Tr\big( \slashed{k}_2 \slashed{\epsilon}_3 \slashed{k}_2 \slashed{\epsilon}_2 \slashed{k}_1 \slashed{\epsilon}_1  \big) \Big) + \\
& - \frac{1}{2} \mathcal{F}\big(2,2,1;x, \xi, \{ \beta_i \}\big) \Big( Tr\big( \slashed{k}_2 \slashed{\epsilon}_3 \bar{\slashed{p}} \slashed{\epsilon}_2 \slashed{k}_2 \slashed{\epsilon}_1  \big) + Tr\big( \slashed{k}_2 \slashed{\epsilon}_3 \slashed{k}_2 \slashed{\epsilon}_2 \bar{\slashed{p}} \slashed{\epsilon}_1  \big) \Big) +\\
&4s\mathcal{F}\big(1,0,0;x, \xi, \{ \beta_i \}\big) \Big( 3(k_1 \epsilon_2)(\epsilon_2 \epsilon_3) - 3 (k_1 \epsilon_3) (\epsilon_1 \epsilon_2) - 2(k_2 \epsilon_1)(\epsilon_2 \epsilon_3) \Big) + \\
&-4s\mathcal{F}\big(1,1,0;x, \xi, \{ \beta_i \}\big) \Big(  (k_2 \epsilon_1)(\epsilon_2 \epsilon_3) + (k_2 \epsilon_3)(\epsilon_1 \epsilon_2)  \Big)+\\
& \mathcal{G}(x, \xi, \{ \beta_i \}) Tr\big( \bar{\slashed{p}} \slashed{\epsilon}_3 \slashed{k}_3 \slashed{\epsilon}_2 \slashed{k}_1 \slashed{\epsilon}_1 \big) + \\
& \mathcal{A} \frac{s^2 (x-\xi)}{(\underline{p}k_3-i0)(\bar{p}k_1+i0)} \log \Big{(} \frac{1}{2\xi}\big{(}\xi - x + i0\cdot sgn(pk_3)\big{)} \Big{)}\bigg\}.
\end{aligned}
\end{equation}
Derivation of this formula is presented in Apppendix \ref{Appdx-derivation-4}. 
The expression for $Tr\big[ i\mathcal{M}^{5.R}_{1,2,3}~\slashed{p}\big]$ is obtained by changing $\xi\rightarrow -\xi$, $k_1 \rightarrow -k_3$, $k_2 \rightarrow -k_2$, $k_3 \rightarrow -k_1$. Simplification of traces leads to the formula \eqref{eq-finite-5L} presented in Appendix \ref{Appdx-full-amplitude}.
\section{Factorization theorem at the 1-loop order}\label{section-factorization}
The divergent part of the amplitude corresponding to a given permutation of photons entering the hard part is equal to
\begin{equation}\label{eq-factorization-Mcoll}
\begin{aligned}
	&Tr\big[ i\mathcal{M}_{1,2,3}~\slashed{p}\big]_{\mathrm{div.}} = \frac{2}{\varepsilon} \cdot i\frac{\alpha_S}{4\pi}C_F e_q^3\mathcal{A}_{1,2,3} \frac{4 s^{-2}}{\beta_1 \beta_3} \frac{1}{x-\xi-i0\cdot sgn(\beta_3)}\frac{1}{x+\xi+i0\cdot sgn(\beta_1)} \times \\
& \bigg( 3+ \frac{x+\xi}{\xi}\log \Big{(} \frac{1}{2\xi}\big{(}\xi - x + i0\cdot sgn(\beta_3)\big{)} \Big{)} - \frac{x-\xi}{\xi}\log \Big{(} \frac{1}{2\xi}\big{(}\xi + x + i0\cdot sgn(\beta_1)\big{)} \Big{)} \bigg).
\end{aligned}
\end{equation}
It can also be written as
\begin{equation}
\begin{aligned}
	&Tr\big[ i\mathcal{M}_{1,2,3}~\slashed{p}\big]_{\mathrm{div.}} = -\frac{2}{\varepsilon} \cdot \frac{\alpha_S}{4\pi}C_F \cdot Tr\big[ i\mathcal{M}^0_{1,2,3}~\slashed{p}\big]  \times  \\
& \bigg( 3+ \frac{x+\xi}{\xi}\log \Big{(} \frac{1}{2\xi}\big{(}\xi - x + i0\cdot sgn(\beta_3)\big{)} \Big{)} - \frac{x-\xi}{\xi}\log \Big{(} \frac{1}{2\xi}\big{(}\xi + x + i0\cdot sgn(\beta_1)\big{)} \Big{)} \bigg).
\end{aligned}
\end{equation}
The term $\mathcal{C}^q_{coll.}$ is defined by
\begin{equation}
 \sum_{\mathrm{permutations}} Tr\big[ \mathcal{M}_{1,2,3}~\slashed{p}\big]_{\mathrm{div.}} = -\frac{2}{\varepsilon}\: \mathcal{C}^q_{coll.}.
\end{equation}
Let us observe, that Formula \eqref{eq-factorization-Mcoll} bears an important similarity to that of Eq. \eqref{eq-amp-lo-1} -- it is proportional to the same trace, and the only result of interchange of photons $k_1 \leftrightarrow 3$ is the change of poles' positions if $sgn(\beta_1) \neq sgn(\beta_3)$.  It follows, that
\begin{equation}\label{eq-Cqcoll-fin}
\begin{aligned}
&\mathcal{C}^q_{coll.} = 4\frac{ie_q^3}{ s \alpha \bar{\alpha} } \frac{\alpha_S}{4\pi}C_F \bigg[ (\alpha - \bar{\alpha} )\big( \vec{\epsilon^*}_t(\mathbf{q}_1) \vec{\epsilon^*}_t(\mathbf{q}_2) \big) \big( \vec{p}_t \vec{\epsilon}_t (\mathbf{q}) \big) + \\& - \big(  \vec{p}_t \vec{\epsilon^*}_t(\mathbf{q}_1) \big) \big(  \vec{\epsilon}_t (\mathbf{q}) \vec{\epsilon^*}_t(\mathbf{q}_2) \big) + \big(  \vec{p}_t \vec{\epsilon^*}_t(\mathbf{q}_2) \big) \big(  \vec{\epsilon}_t(\mathbf{q}) \vec{\epsilon^*}_t(\mathbf{q}_1) \big) \bigg]\times \\
& \: \mathrm{Im} \bigg[ \frac{1}{x-\xi+i0}\frac{1}{x+\xi-i0} \bigg( 3+ \frac{x+\xi}{\xi}\log \Big{(} \frac{1}{2\xi}\big{(}\xi - x - i0 \big{)} \Big{)} - \frac{x-\xi}{\xi}\log \Big{(} \frac{1}{2\xi}\big{(}\xi + x - i0\big{)} \Big{)} \bigg) \bigg].
\end{aligned}
\end{equation}
The exact form of the imaginary part in the last line of the Formula \eqref{eq-Cqcoll-fin} is rather lengthy, and it is more convenient to write it as an imaginary part of a simpler expression in order to verify the factorization at the NLO order. Let us recall the form of $\mathcal{C}^q_0$:
\begin{equation}\label{eq-Cq0-recap}
\begin{aligned}
	&\mathcal{C}^q_0 (x, \xi, \dots) = 4\frac{ie_q^3}{ s \alpha \bar{\alpha} } \: \mathrm{Im} \Big( \frac{1}{x+\xi - i0} \frac{1}{x-\xi +i0} \Big)\times \\&  \bigg[(\alpha - \bar{\alpha} )\big( \vec{\epsilon^*}_t(\mathbf{q}_1) \vec{\epsilon^*}_t(\mathbf{q}_2) \big) \big( \vec{p}_t \vec{\epsilon}_t (\mathbf{q}) \big)  - \big(  \vec{p}_t \vec{\epsilon^*}_t(\mathbf{q}_1) \big) \big(  \vec{\epsilon}_t (\mathbf{q}) \vec{\epsilon^*}_t(\mathbf{q}_2) \big) + \big(  \vec{p}_t \vec{\epsilon^*}_t(\mathbf{q}_2) \big) \big(  \vec{\epsilon}_t(\mathbf{q}) \vec{\epsilon^*}_t(\mathbf{q}_1) \big) \bigg]  .
\end{aligned}
\end{equation}
The non-singlet quark kernel reads (see Eqs. (99-101) in \cite{Diehl}):
\begin{equation}
	K^{qq}(x,x') = \frac{1}{|\xi|} \frac{\alpha_S}{4\pi} C_F \Bigg[ \rho\Big(\frac{x}{\xi}, \frac{x'}{\xi} \Big) \bigg\{ \frac{x+\xi}{x'+\xi}\Big(1+\frac{2\xi}{x'-x} \Big) \bigg\} + \rho\Big(-\frac{x}{\xi}, -\frac{x'}{\xi} \Big) \bigg\{ \frac{-x+\xi}{-x'+\xi}\Big(1+\frac{2\xi}{x-x'} \Big) \bigg\} \Bigg]_+,
\end{equation}
where
\begin{equation}
	\rho\Big(\frac{x}{\xi}, \frac{x'}{\xi} \Big) = \theta\Big(\frac{x}{\xi} - \frac{x'}{\xi} \Big)\theta\Big(\frac{x'}{\xi} + 1 \Big) - \theta\Big( \frac{x'}{\xi}- \frac{x}{\xi}\Big)\theta\Big( -1 -\frac{x'}{\xi}\Big),
\end{equation}
where $\theta$ is the step function, and 
\begin{equation}
	\int dx \Big[ f(x,x') \Big]_+ g(x) = \int dx f(x,x') \big( g(x) - g(x') \big).
\end{equation}
In the considered kinematics one has $\xi \geq 0$. To prove the factorization, according to the Formula \eqref{eq-cancelation-of-div}, and using Eqs. \eqref{eq-Cq0-recap} and \eqref{eq-Cqcoll-fin} we need to verify if
\begin{equation}\label{eq-to-verify}
\begin{aligned}
&\int_{-1}^{1} dy\: K^{qq}(y,x) \mathrm{Im} \Big( \frac{1}{y+\xi - i0} \frac{1}{y-\xi +i0} \Big) \: \overset{?}{=} \\
&\frac{\alpha_S}{4\pi} C_F \: \mathrm{Im} \: \frac{1}{x-\xi+i0}\frac{1}{x+\xi-i0} \bigg( 3+ \frac{x+\xi}{\xi}\log \Big{(} \frac{1}{2\xi}\big{(}\xi - x - i0 \big{)} \Big{)} - \frac{x-\xi}{\xi}\log \Big{(} \frac{1}{2\xi}\big{(}\xi + x - i0\big{)} \Big{)} \bigg).
\end{aligned}
\end{equation}
Since the evolution kernel is real-valued, we can first compute the integral and then take the imaginary part. Moreover, to avoid complications associated with singularities in the integrand, we will substitute $i0 \rightarrow i\delta$, for some positive constant $\delta$, and then check, if the resulting integral has the appropriate limit $\delta \rightarrow 0$.
Let us denote 
\begin{equation}
g_\delta (x) = \frac{1}{x+\xi-i\delta}\frac{1}{x-\xi+i\delta}.
\end{equation}
The resulting integrals are (we omit $\frac{1}{\xi}\frac{\alpha_S}{4\pi} C_F$):
\begin{equation}\label{eq-kernel-integrated-1}
\begin{aligned}
	& \int_{-\xi}^x dy \: \frac{\xi+y}{\xi+x} \Big( 1 + \frac{2\xi}{x-y} \Big) \Big( g_\delta(y) - g_\delta(x) \Big) = \\
& = \frac{1}{x-\xi + i\delta} \log \Big(\frac{x-\xi+i\delta}{-2\xi} \Big) + \frac{1}{x+\xi}\frac{1}{x+\xi-i\delta}\frac{1}{x-\xi+i\delta} \Big( 2\xi^2 - \frac{1}{2}(x-\xi)^2 \Big) + \mathcal{O}(\delta), 
\end{aligned}
\end{equation}
\begin{equation}\label{eq-kernel-integrated-2}
\begin{aligned}
	&\int_{x}^\xi dy \: \frac{\xi-y}{\xi-x} \Big( 1 + \frac{2\xi}{y-x} \Big) \Big( g_\delta(y) - g_\delta(x) \Big) = \\
& = -\frac{1}{x+\xi - i\delta} \log \Big(\frac{x+\xi-i\delta}{2\xi} \Big) + \frac{1}{x-\xi}\frac{1}{x+\xi-i\delta}\frac{1}{x-\xi+i\delta} \Big( -2\xi^2 + \frac{1}{2}(x+\xi)^2 \Big) + \mathcal{O}(\delta). 
\end{aligned}
\end{equation}
Adding these terms we obtain
\begin{equation}
\begin{aligned}
	&\Big(\frac{1}{\xi} \frac{\alpha_S}{4\pi} C_F \Big)^{-1} \int_{-1}^{1} dy\: K^{qq}(y,x) g_\delta(y) = \\
	&\frac{1}{x-\xi + i\delta} \log \Big(\frac{x-\xi+i\delta}{-2\xi} \Big) -\frac{1}{x+\xi - i\delta} \log \Big(\frac{x+\xi-i\delta}{2\xi} \Big) + \frac{3\xi}{(x+\xi-i\delta)(x-\xi+i\delta)} + \mathcal{O}(\delta).
\end{aligned}
\end{equation}
Hence, in the limit $\delta \rightarrow 0$ we indeed obtain the relation \eqref{eq-to-verify}, so that the factorization indeed holds at the one-loop level. What is interesting, in the considered limit both imaginary and real parts of the resulting expression are equal to those in $\mathcal{C}^q_{coll.}$.

\chapter{Conclusions}
In Chapter 1 we briefly reviewed the history of studies on the nuclei structure, the relevant theoretical concepts, and, most importantly, definitions of the Generalized Parton Distributions.

In Chapter 2 we presented, how one relates the definition of GPDs to experimentally measured differential cross sections, and
what approximations one need to take, to do so. We discussed, how the perturbative corrections affect the aforementioned definitions,
in particular, how the gauge link (also known as the Wilson line) appears in calculations after performing a resummation of a relevant class of diagrams.
We described the problem with the collinear divergences, which are present because of the assumption that all partons forming hadrons are massless, and how those divergences vanish at the level of the amplitude after taking into account corrections to parton distributions. We showed, how one extracts the divergent part of GPD by using so-called renormalized string operators, presented an example
of calculation of the relevant counter-terms, and explained how one obtains the renormalization group equations of the renormalized generalized parton distributions.
It allows to write the final form of the amplitude of a given exclusive process in terms  of finite quantities, given that the collinear divergences in fact do cancel.

Results obtained in Chapter 3 show that the collinear factorization holds in the case of the photoproduction of photon pairs $\gamma p \rightarrow \gamma' \gamma" p$ at the one-loop QCD order.
That process is recently of particular interest, since, due to the charge conjugation symmetry, it sensitive to $\mathcal{C}$-odd combinations of Generalized Parton Distributions, and hence provides
a complementary (with respect to e.g. DVCS or TCS) source of data that can be used in extraction of GPDs from experiments, which will be further supported by the improved accuracy 
of the theoretical predictions, due to knowledge of the full amplitude at the order $\alpha_S$ in strong coupling constant.
Moreover, this is the simplest exclusive process, in which the hard sub-process on the parton is a $2\rightarrow 3$ reaction -- hopefully these results will be helpful in studies of this new class of processes.
This work is also a ground for the implementation of observables related to $\gamma p \rightarrow \gamma' \gamma" p$ in the PARTONS software framework \cite{partons}.\\[0.3cm]
\begin{center}
{\large \textbf{Acknowledgements}}\\
\end{center}
This work was supported by the budget for science in 2020-2021, as a research project under the "Diamond Grant" program.
\appendix

\chapter{Derivation of the finite part of 4- and 5- point loop diagrams.}\label{Appdx-derivation-4}
In the Appendix I often omit Feynman slashed inside of traces, if that does not lead to ambiguity.\\
The full form of the amplitude $4.L$:
\begin{equation}\
\begin{aligned}
&Tr \big[ i \mathcal{M}^{4.L}_{1,2,3}~\slashed{p}\big] =-C_F (ie_q)^3(ig)^2 \frac{i}{-2\underline{p}k_3 + i0} \mu^{4-d} \: 6\int\frac{d^d k'}{(2\pi)^d}\times \\
& \int_0^1 y(1-y) dr dy dz \: Tr(k) \bigg{(} k'^2 + 2y(1-y)\Big{(} z\bar{p}k_1 + zr \bar{p}k_2 + r k_1 k_2 \Big{)} + i0  \bigg{)}^{-4} \times
\\
&2 Tr\Big{\{} k_3 \epsilon_3 p \big{(} yk_1 + (1-r(1-y))k_2 +k' \big{)} \epsilon_2 \big{(} yz \bar{p} + y k_1 - r(1-y)k_2 +k' \big{)}\times \\ & \epsilon_1 \big{(} yz\bar{p} -(1-y)k_1 - r(1-y)k_2 +k' \big{)} \Big{\}}. \label{eq-M4-full-1}
\end{aligned}
\end{equation}
Let me define the following functions:
\begin{equation}
\begin{aligned}
&\int_0^1 dr dz \: r^{a}z^{b}\Big{(} -2\big{(} z\bar{p}k_1 + zr \bar{p}k_2 + r k_1 k_2 \big{)} - i0  \Big{)}^{-n} =\\
&= (-1)^n s^{-n} \int_0^1 dr dz \: r^{a}z^{b}\Big{(} \big{(} z(x+\xi)\beta_1 + zr (x+\xi)\beta_2 + r \kappa_3 \big{)} + i0  \Big{)}^{-n} = \\
&= (-1)^n s^{-n} \mathcal{F}\big(n,a,b; x, \xi, x, \xi, \{ \beta_i \} \big).
\end{aligned}
\end{equation}
$n$ depends on the power of $k'^2$ in the nominator of the loop momentum integral. We have $n=0$ and $n=1$ terms; all of $n=1$ terms are finite; there are two different $n=0$-terms with a collinearly divergent part. Note that they have also finite parts which we neglected in the previous part -- they are analyzed separately in here. Let me start from the finite $n=0$-terms -- to do so, let us take a look at the following trace:
\begin{equation}
 Tr\Big{\{} k_3 \epsilon_3 p \big{(} yk_1 + (1-r(1-y))k_2 \big{)} \epsilon_2 \big{(} yz \bar{p} + y k_1 - r(1-y)k_2 \big{)} \epsilon_1 \big{(} yz\bar{p} -(1-y)k_1 - r(1-y)k_2 \big{)} \Big{\}}.
\end{equation}
We have the general expression
\begin{equation}
\int_0^1 dr dz dy\: y^{-1}(1-y)^{-1} s^{-2} \Big{(} \big{(} z(x+\xi)\beta_1 + zr (x+\xi)\beta_2 + r \kappa_3 \big{)} + i0  \Big{)}^{-2} Tr(\dots).
\end{equation}
\begin{figure}[H]
    \centering
    \includegraphics[width = 0.9 \textwidth]{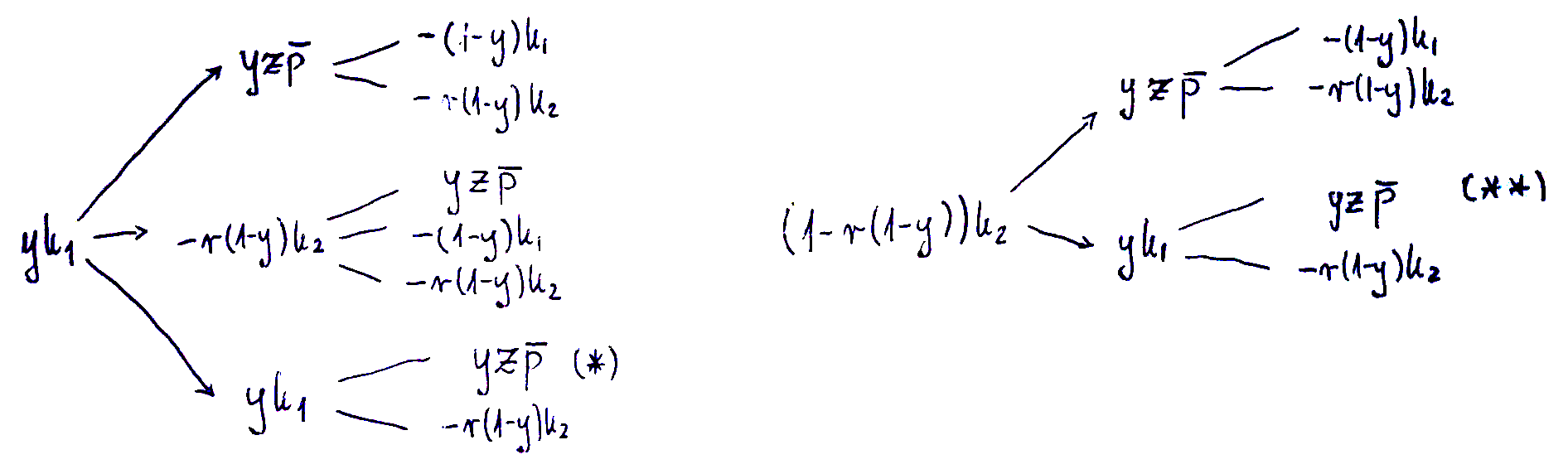}
    \caption{Graphical representation of all nonvanishing $n=0$ terms. * and ** denote terms with divergences - those are treated separately.}
    \label{graphs}
\end{figure}
All the resulting traces and integrands are
\begin{equation}
\begin{aligned}
&-y^2(1-y) z Tr\big( k_3 \epsilon_3 p k_1 \epsilon_2 \bar{p} \epsilon_1 k_1 \big) \rightarrow -\frac12 Tr\big( k_3 \epsilon_3 p k_1 \epsilon_2 \bar{p} \epsilon_1 k_1 \big) \mathcal{F}(2, 0, 1; x, \xi, \{ \beta_i \})\\
&-y^2(1-y) zr Tr\big( k_3 \epsilon_3 p k_1 \epsilon_2 \bar{p} \epsilon_1 k_2 \big) \rightarrow -\frac12 Tr\big( k_3 \epsilon_3 p k_1 \epsilon_2 \bar{p} \epsilon_1 k_2 \big) \mathcal{F}(2, 1, 1; x, \xi, \{ \beta_i \})\\
& -y^2(1-y) zr Tr\big( k_3 \epsilon_3 p k_1 \epsilon_2 k_2 \epsilon_1 \bar{p} \big) \rightarrow -\frac12 Tr\big( k_3 \epsilon_3 p k_1 \epsilon_2 k_2 \epsilon_1 \bar{p} \big) \mathcal{F}(2, 1, 1; x, \xi, \{ \beta_i \})\\
& y(1-y)^2 r Tr\big( k_3 \epsilon_3 p k_1 \epsilon_2 k_2 \epsilon_1 k_1 \big) \rightarrow \frac12 Tr\big( k_3 \epsilon_3 p k_1 \epsilon_2 k_2 \epsilon_1 k_1 \big) \mathcal{F}(2, 1, 0; x, \xi, \{ \beta_i \})\\
& y(1-y)^2 r^2 Tr\big( k_3 \epsilon_3 p k_1 \epsilon_2 k_2 \epsilon_1 k_2 \big) \rightarrow \frac12 Tr\big( k_3 \epsilon_3 p k_1 \epsilon_2 k_2 \epsilon_1 k_2 \big) \mathcal{F}(2, 2, 0; x, \xi, \{ \beta_i \})\\
& -y^2(1-y) r Tr\big( k_3 \epsilon_3 p k_1 \epsilon_2 k_1 \epsilon_1 k_2 \big) \rightarrow -\frac12 Tr\big( k_3 \epsilon_3 p k_1 \epsilon_2 k_1 \epsilon_1 k_2 \big) \mathcal{F}(2, 1, 0; x, \xi, \{ \beta_i \})\\
&-y(1-y)z Tr\big( k_3 \epsilon_3 p k_2 \epsilon_2 \bar{p} \epsilon_1 k_1 \big) \rightarrow -Tr\big( k_3 \epsilon_3 p k_2 \epsilon_2 \bar{p} \epsilon_1 k_1 \big) \mathcal{F}(2,0,1;x, \xi, \{ \beta_i \})\\
&y(1-y)^2 zr Tr\big( k_3 \epsilon_3 p k_2 \epsilon_2 \bar{p} \epsilon_1 k_1 \big) \rightarrow \frac12 Tr\big( k_3 \epsilon_3 p k_2 \epsilon_2 \bar{p} \epsilon_1 k_1 \big) \mathcal{F}(2,1,1;x, \xi, \{ \beta_i \})\\
&-y(1-y)zr Tr\big( k_3 \epsilon_3 p k_2 \epsilon_2 \bar{p} \epsilon_1 k_2 \big) \rightarrow -Tr\big( k_3 \epsilon_3 p k_2 \epsilon_2 \bar{p} \epsilon_1 k_2 \big) \mathcal{F}(2,1,1;x, \xi, \{ \beta_i \})\\
&y(1-y)^2 zr^2 Tr\big( k_3 \epsilon_3 p k_2 \epsilon_2 \bar{p} \epsilon_1 k_2 \big) \rightarrow \frac12 Tr\big( k_3 \epsilon_3 p k_2 \epsilon_2 \bar{p} \epsilon_1 k_2 \big) \mathcal{F}(2,2,1;x, \xi, \{ \beta_i \})\\
&-y^2(1-y)zr Tr\big( k_3 \epsilon_3 p k_2 \epsilon_2 k_1 \epsilon_1 \bar{p} \big) \rightarrow -\frac12 Tr\big( k_3 \epsilon_3 p k_2 \epsilon_2 k_1 \epsilon_1 \bar{p} \big) \mathcal{F}(2,1,1;x, \xi, \{ \beta_i \})\\
&-y(1-y)r Tr\big( k_3 \epsilon_3 p k_2 \epsilon_2 k_1 \epsilon_1 k_2 \big) \rightarrow -Tr\big( k_3 \epsilon_3 p k_2 \epsilon_2 k_1 \epsilon_1 k_2 \big) \mathcal{F}(2,1,0;x, \xi, \{ \beta_i \})\\
&y(1-y)^2 r^2 Tr\big( k_3 \epsilon_3 p k_2 \epsilon_2 k_1 \epsilon_1 k_2 \big) \rightarrow \frac12 Tr\big( k_3 \epsilon_3 p k_2 \epsilon_2 k_1 \epsilon_1 k_2 \big) \mathcal{F}(2,2,0;x, \xi, \{ \beta_i \}).
\end{aligned}
\end{equation}
After performing the integration, we obtain the contribution from the finite integrals with $n=0$:
\begin{equation}
\begin{aligned}
&Tr \big[ i \mathcal{M}^{4.L}_{1,2,3}~\slashed{p}\big] 	\supset -C_F e_q^3 g^2 \frac{i}{(4\pi)^2} \frac{1}{2\underline{p}k_3 - i0} s^{-2} \bigg( \\[0.3cm]
&\mathcal{F} \big(2,1,0;x, \xi, \{ \beta_i \}\big) \Big( Tr\big( k_3 \epsilon_3 p k_1 \epsilon_2 k_2 \epsilon_1 k_1 \big) - Tr\big( k_3 \epsilon_3 p k_1 \epsilon_2 k_1 \epsilon_1 k_2 \big) -2Tr\big( k_3 \epsilon_3 p k_2 \epsilon_2 k_1 \epsilon_1 k_2 \big) \Big)+\\[0.3cm]
&+\mathcal{F} \big(2,0,1;x, \xi, \{ \beta_i \}\big) \Big( - Tr\big( k_3 \epsilon_3 p k_1 \epsilon_2 \bar{p} \epsilon_1 k_1 \big) -2Tr\big( k_3 \epsilon_3 p k_2 \epsilon_2 \bar{p} \epsilon_1 k_1 \big) \Big)+\\[0.3cm]
&+\mathcal{F} \big(2,1,1;x, \xi, \{ \beta_i \}\big) \Big( -Tr\big( k_3 \epsilon_3 p k_1 \epsilon_2 \bar{p} \epsilon_1 k_2 \big) - Tr\big( k_3 \epsilon_3 p k_1 \epsilon_2 k_2 \epsilon_1 \bar{p} \big) +\\
&+ Tr\big( k_3 \epsilon_3 p k_2 \epsilon_2 \bar{p} \epsilon_1 k_1 \big) -2 Tr\big( k_3 \epsilon_3 p k_2 \epsilon_2 \bar{p} \epsilon_1 k_2 \big) -  Tr\big( k_3 \epsilon_3 p k_2 \epsilon_2 k_1 \epsilon_1 \bar{p} \big) \Big) +\\[0.3cm]
&+\mathcal{F} \big(2,2,0;x, \xi, \{ \beta_i \}\big) \Big( Tr\big( k_3 \epsilon_3 p k_1 \epsilon_2 k_2 \epsilon_1 k_2 \big) + Tr\big( k_3 \epsilon_3 p k_2 \epsilon_2 k_1 \epsilon_1 k_2 \big) \Big) + \\[0.3cm]
& + \mathcal{F} \big(2,2,1;x, \xi, \{ \beta_i \}\big) Tr\big( k_3 \epsilon_3 p k_2 \epsilon_2 \bar{p} \epsilon_1 k_2 \big) \bigg).
\end{aligned}
\end{equation}
Now let us look on the terms with $k'^2$ in the nominator (the $n=1$ terms):
\begin{equation}
\begin{aligned}
&Tr \big(k_3 \epsilon_3 p k' \epsilon_2 k' \epsilon_1 \big( yz\bar{p} - (1-y)k_1 - r(1-y)k_2 \big) \big) +\\
+&Tr \big(k_3 \epsilon_3 p k' \epsilon_2 \big( yz\bar{p} + yk_1 - r(1-y)k_2 \big) \epsilon_1 k' \big)+\\
+&Tr \big(k_3 \epsilon_3 p \big( yk_1 + \big( 1 - r(1-y) \big) k_2 \big) \epsilon_2 k'  \epsilon_1 k' \big).
\end{aligned}
\end{equation}
Use the following integration formula:
\begin{equation}
6\int \frac{d^4 k}{(2\pi)^4} \frac{k^\mu k^\nu}{\big( k^2 + \Delta + i0 \big)^4} = -\frac{ig^{\mu\nu}}{2(4\pi)^2}\Big( -\Delta - i0 \Big)^{-1}.
\end{equation}
The previous traces become
\begin{equation}
\begin{aligned}
&2Tr \big(k_3 \epsilon_3 p \epsilon_2 \epsilon_1 \big( (1-y)k_1 + r(1-y)k_2 \big) \big) +\\
-&2Tr \big(p \epsilon_3 k_3 \epsilon_2 \big( yk_1 - r(1-y)k_2 \big) \epsilon_1 \big)+\\
-&2Tr \big(k_3 \epsilon_3 p \big( yk_1 + \big( 1 - r(1-y) \big) k_2 \big) \epsilon_2 \epsilon_1\big).
\end{aligned}
\end{equation}
We have the following integral
\begin{equation}
\int_0^1 dr dz dy \:  s^{-1} \Big{(} \big{(} z(x+\xi)\beta_1 + zr (x+\xi)\beta_2 + r \kappa_3 \big{)} + i0  \Big{)}^{-1} 2Tr(\dots).
\end{equation}
Note that there was a minus sign in the expression in the biggest parenthesis -- later on I multiply the result by an additional $-1$. Recalling the previous definitions of $\mathcal{F}$ we obtain
\begin{equation}
\begin{aligned}
	&\mathcal{F}(1,0,0; x, \xi, \{ \beta_i \}) Tr\big(k_3 \epsilon_3 p \epsilon_2 \epsilon_1 k_1 \big) + \mathcal{F}(1,1,0; x, \xi, \{ \beta_i \}) Tr\big(k_3 \epsilon_3 p \epsilon_2 \epsilon_1 k_2 \big)+\\[0.3cm]
	&-\mathcal{F}(1,0,0; x, \xi, \{ \beta_i \}) Tr\big(p  \epsilon_3 k_3 \epsilon_2 k_1 \epsilon_1  \big) + \mathcal{F}(1,1,0; x, \xi, \{ \beta_i \}) Tr\big(p  \epsilon_3 k_3 \epsilon_2 k_2 \epsilon_1  \big)+\\[0.3cm]
&- \mathcal{F}(1,0,0; x, \xi, \{ \beta_i \}) Tr\big(k_3 \epsilon_3 p k_1 \epsilon_2 \epsilon_1 \big) - 2\mathcal{F}(1,0,0; x, \xi, \{ \beta_i \}) Tr\big(k_3 \epsilon_3 p k_2 \epsilon_2 \epsilon_1 \big) +\\&+ \mathcal{F}(1,1,0; x, \xi, \{ \beta_i \}) Tr\big(k_3 \epsilon_3 p k_2 \epsilon_2 \epsilon_1 \big).
\end{aligned}
\end{equation}
Now, we multiply it by $-C_F e_q^3 g^2 \frac{1}{2\underline{p}k_3 - i0} (-1) \frac{-i}{(4\pi)^2}s^{-1}$ and obtain the $k'^2$-part of the amplitude:
\begin{equation}
\begin{aligned}
	&Tr \big[ i \mathcal{M}^{4.L}_{1,2,3}~\slashed{p}\big] 	\supset -C_F e_q^3 g^2 \frac{i}{(4\pi)^2} \frac{1}{2\underline{p}k_3 - i0} s^{-1} \bigg(\\& \mathcal{F}(1,1,0; x, \xi, \{ \beta_i \}) \Big( Tr\big(k_3 \epsilon_3 p \epsilon_2 \epsilon_1 k_2 \big) + Tr\big(p  \epsilon_3 k_3 \epsilon_2 k_2 \epsilon_1  \big)+ Tr\big(k_3 \epsilon_3 p k_2 \epsilon_2 \epsilon_1 \big) \Big)+\\
	& +\mathcal{F}(1,0,0; x, \xi, \{ \beta_i \}) \Big( Tr\big(k_3 \epsilon_3 p \epsilon_2 \epsilon_1 k_1 \big) -  Tr\big(p  \epsilon_3 k_3 \epsilon_2 k_1 \epsilon_1  \big) - Tr\big(k_3 \epsilon_3 p k_1 \epsilon_2 \epsilon_1 \big) -2Tr\big(k_3 \epsilon_3 p k_2 \epsilon_2 \epsilon_1 \big) \Big)  \bigg).
\end{aligned}
\end{equation}
Finally, we extract the finite part of divergent terms. Recall that they were $-C_Fe_q^3 g^2 (2\underline{p}k_3-i0)^{-1}$ times
\begin{equation}
\begin{aligned}
&\frac{2i}{(4\pi)^{d/2}}\Gamma(4-d/2)Tr\Big{\{} k_3 \epsilon_3 p k_2 \epsilon_2 k_1 \epsilon_1 \bar{p} \Big{\}}\times\\&\times \int_0^1 dy\: y^{d/2-1}(1-y)^{d/2-3} \int_0^1 dr dz\: z\Big{(} -2\big{(} z\bar{p}k_1 + zr \bar{p}k_2 + r k_1 k_2 \big{)} - i0  \Big{)}^{d/2 - 4} ,\\
&\frac{2i}{(4\pi)^{d/2}}\Gamma(4-d/2)Tr\Big{\{} k_3 \epsilon_3 p k_1 \epsilon_2 k_1 \epsilon_1 \bar{p} \Big{\}}\times \\ &\times \int_0^1 dy\: y^{d/2}(1-y)^{d/2-3} \int_0^1 dr dz \: z\Big{(} -2\big{(} z\bar{p}k_1 + zr \bar{p}k_2 + r k_1 k_2 \big{)} - i0  \Big{)}^{d/2 - 4}.
\end{aligned}
\end{equation}
We can rewrite these terms as
\begin{equation}
\begin{aligned}
&\frac{2i}{(4\pi)^2} \Big( \frac{s e^\gamma}{4\pi \mu^2} \Big)^{-\frac{\varepsilon}{2}} (1+\gamma\varepsilon/2) \Gamma(2+\varepsilon/2) \frac{\Gamma(2-\varepsilon/2)\Gamma(-\varepsilon/2)}{\Gamma(2-\varepsilon)} Tr\Big{\{} k_3 \epsilon_3 p k_2 \epsilon_2 k_1 \epsilon_1 \bar{p} \Big{\}}s^{-2}  \times \\
& \int_0^1 dr dz\: z\Big{(} \big{(} z(x+\xi)\beta_1 + zr(x+\xi)\beta_2 + r\kappa_3 + i0  \Big{)}^{-2}\Big( 1 - \frac{\varepsilon}{2}\ln \big(- z(x+\xi)\beta_1 - zr(x+\xi)\beta_2 - r\kappa_3 - i0 \big) \Big),\\
&\frac{2i}{(4\pi)^2} \Big( \frac{s e^\gamma}{4\pi \mu^2} \Big)^{-\frac{\varepsilon}{2}} (1+\gamma\varepsilon/2) \Gamma(2+\varepsilon/2) \frac{\Gamma(3-\varepsilon/2)\Gamma(-\varepsilon/2)}{\Gamma(3-\varepsilon)} Tr\Big{\{} k_3 \epsilon_3 p k_1 \epsilon_2 k_1 \epsilon_1 \bar{p} \Big{\}}s^{-2} \times\\
& \int_0^1 dr dz\: z\Big{(} \big{(} z(x+\xi)\beta_1 + zr(x+\xi)\beta_2 + r\kappa_3 + i0  \Big{)}^{-2}\Big( 1 - \frac{\varepsilon}{2}\ln \big(- z(x+\xi)\beta_1 - zr(x+\xi)\beta_2 - r\kappa_3 - i0 \big) \Big).
\end{aligned}
\end{equation}
Expanding with respect to $\varepsilon$ we obtain
\begin{equation}
\begin{aligned}
&(1+\gamma\varepsilon/2) \Gamma(2+\varepsilon/2) \frac{\Gamma(2-\varepsilon/2)\Gamma(-\varepsilon/2)}{\Gamma(2-\varepsilon)} = -\frac{2}{\varepsilon} - 2,\\
& (1+\gamma\varepsilon/2) \Gamma(2+\varepsilon/2) \frac{\Gamma(3-\varepsilon/2)\Gamma(-\varepsilon/2)}{\Gamma(3-\varepsilon)} = -\frac{2}{\varepsilon} - \frac{5}{2}.
\end{aligned}
\end{equation}
Defining the following function:
\begin{equation}
	\mathcal{G}(x, \xi, \{ \beta_i \}) = \int_0^1 dr dz\: z\Big{(} \big{(} z(x+\xi)\beta_1 + zr(x+\xi)\beta_2 + r\kappa_3 + i0  \Big{)}^{-2}\ln \big(- z(x+\xi)\beta_1 - zr(x+\xi)\beta_2 - r\kappa_3 - i0 \big),
\end{equation}
we obtain
\begin{equation}
\begin{aligned}
	&Tr \big[ i \mathcal{M}^{4.L}_{1,2,3}~\slashed{p}\big] 	\supset -\frac{i}{(4\pi)^2} s^{-2} \bigg( \mathcal{F}(2,0,1,x, \xi, \{ \beta_i \})\Big( 4 Tr\big( k_3 \epsilon_3 p k_2 \epsilon_2 k_1 \epsilon_1 \bar{p} \big) + 5 Tr\big( k_3 \epsilon_3 p k_1 \epsilon_2 k_1 \epsilon_1 \bar{p} \big) \Big) + \\
&- 2 \mathcal{G}(x, \xi, \{ \beta_i \}) Tr\big( k_3 \epsilon_3 p k_3 \epsilon_2 k_1 \epsilon_1 \bar{p} \big) \Big) \bigg).
\end{aligned}
\end{equation}
In the last line I used the trick with interchanging $\slashed{p} (\slashed{k}_1+\slashed{k}_2) \rightarrow -\slashed{p}\slashed{k}_3$ inside the trace.\\
Hence, we get the finite part:
\begin{equation}
\begin{aligned}
&Tr\big[ i \mathcal{M}^{4.L}_{1,2,3}~\slashed{p}\big]_{\mathrm{fin.}} = -C_F e_q^3 g^2 \frac{i}{(4\pi)^2} \frac{1}{2\underline{p}k_3 - i0} s^{-2} \bigg( \\[0.3cm]
&\mathcal{F} \big(2,1,0;x, \xi, \{ \beta_i \}\big) \Big( Tr\big( k_3 \epsilon_3 p k_1 \epsilon_2 k_2 \epsilon_1 k_1 \big) - Tr\big( k_3 \epsilon_3 p k_1 \epsilon_2 k_1 \epsilon_1 k_2 \big) -2Tr\big( k_3 \epsilon_3 p k_2 \epsilon_2 k_1 \epsilon_1 k_2 \big) \Big)+\\[0.3cm]
&+\mathcal{F} \big(2,0,1;x, \xi, \{ \beta_i \}\big) \Big( - Tr\big( k_3 \epsilon_3 p k_1 \epsilon_2 \bar{p} \epsilon_1 k_1 \big) -2Tr\big( k_3 \epsilon_3 p k_2 \epsilon_2 \bar{p} \epsilon_1 k_1 \big) +\\
& - 4 Tr\big( k_3 \epsilon_3 p k_2 \epsilon_2 k_1 \epsilon_1 \bar{p} \big) - 5 Tr\big( k_3 \epsilon_3 p k_1 \epsilon_2 k_1 \epsilon_1 \bar{p} \big) \Big)\\[0.3cm]
&+\mathcal{F} \big(2,1,1;x, \xi, \{ \beta_i \}\big) \Big( -Tr\big( k_3 \epsilon_3 p k_1 \epsilon_2 \bar{p} \epsilon_1 k_2 \big) - Tr\big( k_3 \epsilon_3 p k_1 \epsilon_2 k_2 \epsilon_1 \bar{p} \big) +\\
&+ Tr\big( k_3 \epsilon_3 p k_2 \epsilon_2 \bar{p} \epsilon_1 k_1 \big) -2 Tr\big( k_3 \epsilon_3 p k_2 \epsilon_2 \bar{p} \epsilon_1 k_2 \big) -  Tr\big( k_3 \epsilon_3 p k_2 \epsilon_2 k_1 \epsilon_1 \bar{p} \big) \Big) +\\[0.3cm]
&+\mathcal{F} \big(2,2,0;x, \xi, \{ \beta_i \}\big) \Big( Tr\big( k_3 \epsilon_3 p k_1 \epsilon_2 k_2 \epsilon_1 k_2 \big) + Tr\big( k_3 \epsilon_3 p k_2 \epsilon_2 k_1 \epsilon_1 k_2 \big) \Big) + \\[0.3cm]
& + \mathcal{F} \big(2,2,1;x, \xi, \{ \beta_i \}\big) Tr\big( k_3 \epsilon_3 p k_2 \epsilon_2 \bar{p} \epsilon_1 k_2 \big)+\\[0.3cm]
&+s \mathcal{F}(1,1,0; x, \xi, \{ \beta_i \}) \Big( Tr\big(k_3 \epsilon_3 p \epsilon_2 \epsilon_1 k_2 \big) + Tr\big(p  \epsilon_3 k_3 \epsilon_2 k_2 \epsilon_1  \big)+ Tr\big(k_3 \epsilon_3 p k_2 \epsilon_2 \epsilon_1 \big) \Big)+\\
& + s \mathcal{F}(1,0,0; x, \xi, \{ \beta_i \}) \Big( Tr\big(k_3 \epsilon_3 p \epsilon_2 \epsilon_1 k_1 \big) -  Tr\big(p  \epsilon_3 k_3 \epsilon_2 k_1 \epsilon_1  \big) - Tr\big(k_3 \epsilon_3 p k_1 \epsilon_2 \epsilon_1 \big) -2Tr\big(k_3 \epsilon_3 p k_2 \epsilon_2 \epsilon_1 \big) \Big) + \\[0.3cm]
& - 2 \mathcal{G}(x, \xi, \{ \beta_i \}) Tr\big( k_3 \epsilon_3 p k_3 \epsilon_2 k_1 \epsilon_1 \bar{p} \big) \Big) \bigg).
\end{aligned}
\end{equation}
The expression for $Tr\big[ i \mathcal{M}^{4.R}_{1,2,3}~\slashed{p}\big]$ is obtained by changing $\xi\rightarrow -\xi$, $k_1 \rightarrow -k_3$, $k_2 \rightarrow -k_2$, $k_3 \rightarrow -k_1$.

Now we find the finite part of the 5.L part of the amplitude $Tr \big[ i \mathcal{M}^{5}_{1,2,3}~\slashed{p}\big]$. Let us remind that we consider the following:
\begin{equation}
\begin{aligned}
&Tr \big[ i \mathcal{M}^{5.L}_{1,2,3}~\slashed{p}\big] = \\
&-\: \frac{-2+\epsilon}{2\xi}C_F e_q^3 g^2 \mu^{4-d} 6\int \frac{d^d k'}{(2\pi)^d} \int_0^1 y(1-y) dr dy dz \Big{(} k'^2 + 2y(1-y)\Big{(} z\bar{p}k_1 + zr \bar{p}k_2 + r k_1 k_2 \Big{)} + i0  \Big{)}^{-4}\times
\\
 & Tr\Big{\{} (yz\bar{p} -(1-y)k_1 -r(1-y)k_2+k')\epsilon_3 \times \\ &(yz\bar{p} + yk_1 +\big{(}1 -r(1-y)\big{)}k_2+k')\epsilon_2 (yz\bar{p} + yk_1 -r(1-y)k_2+k')\epsilon_1 \Big{\}} . \label{eq-fin-5-0}
\end{aligned}
\end{equation}
As in the analysis in the 4-point loop case, we start from $n=0$ terms, which are presented in the following graph:
\begin{figure}[H]
    \centering
    \includegraphics[width = 0.9 \textwidth]{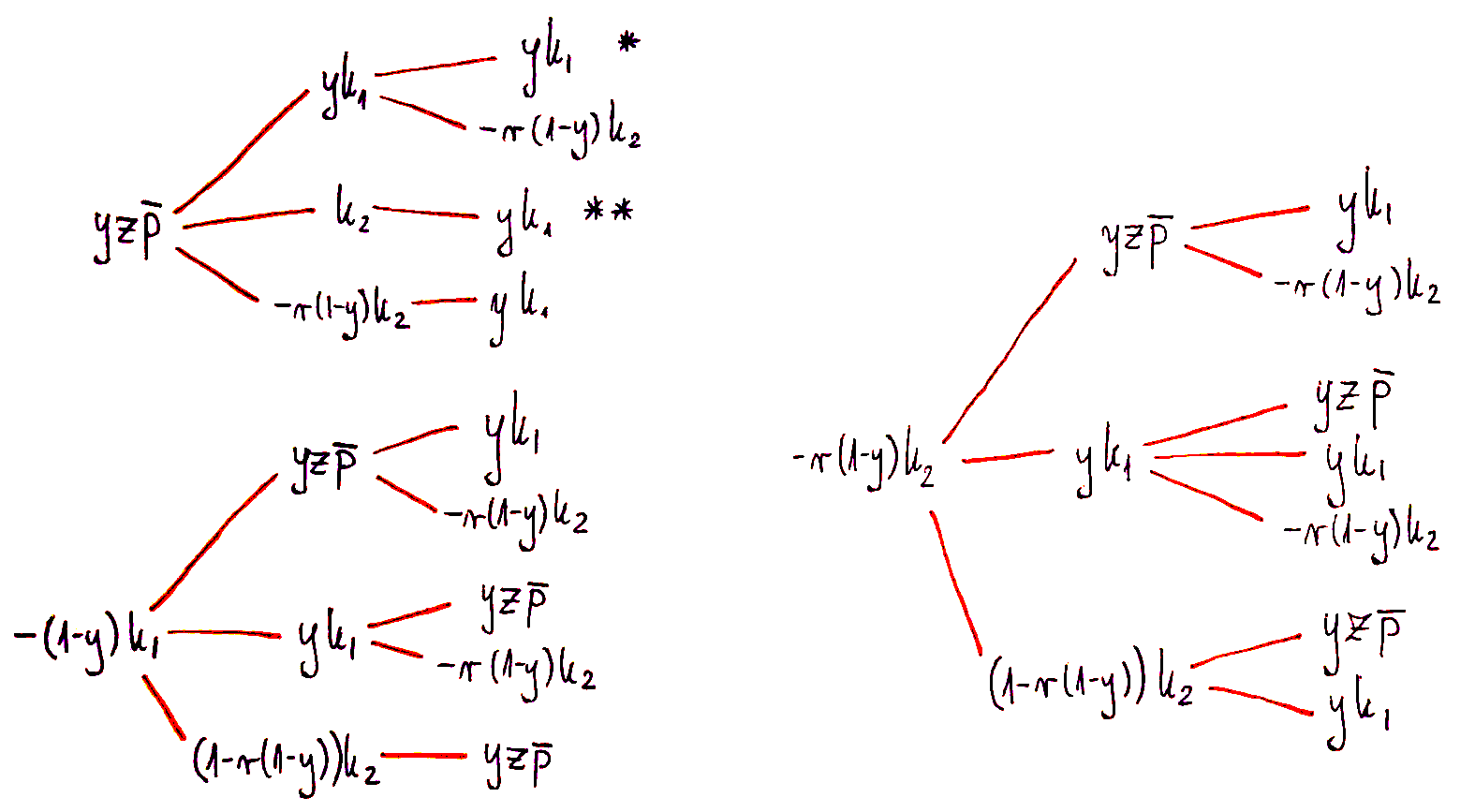}
    \caption{Graphical representation of all nonvanishing $n=0$ terms. * and ** denote terms with divergences - those are treated separately.}
    \label{graph-5}
\end{figure}
\noindent We expand the trace inside of the integral
$$6\int \frac{d^4 k }{(2\pi)^4} \int_0^1 drdydz\: y(1-y) \big( y(1-y)\Delta + i0 \big)^{-4} Tr(...) = \frac{i}{(4\pi)^2}\int_0^1 drdydz\: y^{-1}(1-y)^{-1}\big( \Delta + i0\big)^{-2} Tr(...) .$$
After computing the integrals it yields
\begin{equation}
\begin{aligned}
	& -Tr\big( \bar{p} \epsilon_3 k_1 \epsilon_2 k_1 \epsilon_1  \big) \mathcal{F}(1,1; x, \xi, \{ \beta_i \}),\\
	& -Tr\big( \bar{p} \epsilon_3 k_2 \epsilon_2 k_1 \epsilon_1  \big) \mathcal{F}(1,1; x, \xi, \{ \beta_i \}),\\
	& -Tr\big( k_1 \epsilon_3 \bar{p} \epsilon_2 k_1 \epsilon_1  \big) \mathcal{F}(0,1; x, \xi, \{ \beta_i \}),\\
	& Tr\big( k_1 \epsilon_3 \bar{p} \epsilon_2 k_2 \epsilon_1  \big) \mathcal{F}(1,1; x, \xi, \{ \beta_i \}),\\
	& -Tr\big( k_1 \epsilon_3 k_1 \epsilon_2  \bar{p}\epsilon_1  \big) \mathcal{F}(0,1; x, \xi, \{ \beta_i \}),\\
	& Tr\big( k_1 \epsilon_3 k_1 \epsilon_2  k_2\epsilon_1  \big) \mathcal{F}(1,0; x, \xi, \{ \beta_i \}),\\
	& -2 Tr\big( k_1 \epsilon_3 k_2 \epsilon_2  \bar{p} \epsilon_1  \big) \mathcal{F}(0,1; x, \xi, \{ \beta_i \}),\\
	& Tr\big( k_1 \epsilon_3 k_2 \epsilon_2  \bar{p} \epsilon_1  \big) \mathcal{F}(1,1; x, \xi, \{ \beta_i \}),\\
	& -Tr\big( k_2 \epsilon_3 \bar{p} \epsilon_2 k_1 \epsilon_1  \big) \mathcal{F}(1,1; x, \xi, \{ \beta_i \}),\\
	& Tr\big( k_2 \epsilon_3 \bar{p} \epsilon_2 k_2 \epsilon_1  \big) \mathcal{F}(2,1; x, \xi, \{ \beta_i \}),\\
	& -Tr\big( k_2 \epsilon_3 k_1 \epsilon_2 \bar{p} \epsilon_1  \big) \mathcal{F}(1,1; x, \xi, \{ \beta_i \}),\\
	& -Tr\big( k_2 \epsilon_3 k_1 \epsilon_2 k_1 \epsilon_1  \big) \mathcal{F}(1,0; x, \xi, \{ \beta_i \}),\\
	& -Tr\big( k_2 \epsilon_3 k_1 \epsilon_2 k_2 \epsilon_1  \big) \mathcal{F}(2,0; x, \xi, \{ \beta_i \}),\\
	& -2Tr\big( k_2 \epsilon_3 k_2 \epsilon_2 \bar{p} \epsilon_1  \big) \mathcal{F}(1,1; x, \xi, \{ \beta_i \}),\\
	& Tr\big( k_2 \epsilon_3 k_2 \epsilon_2 \bar{p} \epsilon_1  \big) \mathcal{F}(2,1; x, \xi, \{ \beta_i \}),\\
	& -2Tr\big( k_2 \epsilon_3 k_2 \epsilon_2 k_1 \epsilon_1  \big) \mathcal{F}(1,0; x, \xi, \{ \beta_i \}),\\
	& Tr\big( k_2 \epsilon_3 k_2 \epsilon_2 k_1 \epsilon_1  \big) \mathcal{F}(2,0; x, \xi, \{ \beta_i \}).
\end{aligned}
\end{equation}
Now, let us recall the divergent terms:
\begin{equation}
\begin{aligned}
&\frac{i}{(4\pi)^2} \Big( \frac{s e^\gamma}{4\pi \mu^2} \Big)^{-\frac{\varepsilon}{2}} (1+\gamma\varepsilon/2) \Gamma(2+\varepsilon/2) \frac{\Gamma(3-\varepsilon/2)\Gamma(-\varepsilon/2)}{\Gamma(3-\varepsilon)} Tr\big( \bar{p} \epsilon_3 k_1 \epsilon_2 k_1 \epsilon_1 \big) s^{-2}  \times \\
& \int_0^1 dr dz \:z\Big{(} z(x+\xi)\beta_1 + zr(x+\xi)\beta_2 + r\kappa_3 + i0  \Big{)}^{-2}\Big( 1 - \frac{\varepsilon}{2}\ln \big(- z(x+\xi)\beta_1 - zr(x+\xi)\beta_2 - r\kappa_3 - i0 \big) \Big),\\
&\frac{i}{(4\pi)^2} \Big( \frac{s e^\gamma}{4\pi \mu^2} \Big)^{-\frac{\varepsilon}{2}} (1+\gamma\varepsilon/2) \Gamma(2+\varepsilon/2) \frac{\Gamma(2-\varepsilon/2)\Gamma(-\varepsilon/2)}{\Gamma(2-\varepsilon)} Tr\big( \bar{p} \epsilon_3 k_2 \epsilon_2 k_1 \epsilon_1 \big) s^{-2} \times\\
& \int_0^1 dr dz\: z\Big{(} z(x+\xi)\beta_1 + zr(x+\xi)\beta_2 + r\kappa_3 + i0  \Big{)}^{-2}\Big( 1 - \frac{\varepsilon}{2}\ln \big(- z(x+\xi)\beta_1 - zr(x+\xi)\beta_2 - r\kappa_3 - i0 \big) \Big).
\end{aligned}
\end{equation}
Expansion with respect to $\varepsilon$ yields the following:
\begin{equation}
\begin{aligned}
&(1+\gamma\varepsilon/2) \Gamma(2+\varepsilon/2) \frac{\Gamma(2-\varepsilon/2)\Gamma(-\varepsilon/2)}{\Gamma(2-\varepsilon)} = -\frac{2}{\varepsilon} - 2,\\
& (1+\gamma\varepsilon/2) \Gamma(2+\varepsilon/2) \frac{\Gamma(3-\varepsilon/2)\Gamma(-\varepsilon/2)}{\Gamma(3-\varepsilon)} = -\frac{2}{\varepsilon} - \frac{5}{2}.
\end{aligned}
\end{equation}
In this way, we obtain
\begin{equation}
\begin{aligned}
	&Tr \big[ i \mathcal{M}^{5.L}_{1,2,3}~\slashed{p}\big] \supset -\frac{i}{(4\pi)^2} \frac{5}{2} Tr\big( \bar{p} \epsilon_3 k_1 \epsilon_2 k_1 \epsilon_1 \big) s^{-2} \mathcal{F}(2,0,1;x, \xi, \{ \beta_i \}) \\&+ \frac{i}{(4\pi)^2} Tr\big( \bar{p} \epsilon_3 k_1 \epsilon_2 k_1 \epsilon_1 \big) s^{-2} \mathcal{G}(x, \xi, \{ \beta_i \}) -\frac{i}{(4\pi)^2} 2 Tr\big( \bar{p} \epsilon_3 k_2 \epsilon_2 k_1 \epsilon_1 \big) s^{-2} \mathcal{F}(2,0,1;x, \xi, \{ \beta_i \}) \\&+ \frac{i}{(4\pi)^2} Tr\big( \bar{p} \epsilon_3 k_2 \epsilon_2 k_1 \epsilon_1 \big) s^{-2} \mathcal{G}(x, \xi, \{ \beta_i \}).
\end{aligned}
\end{equation}
Mind that there was an overall factor $\frac{-2+\varepsilon}{2\xi}$ multiplying everything in \eqref{eq-fin-5-0}. To take that $\varepsilon$ into account, we will just take the divergent part \eqref{eq-M5-L-div} of the expression and multiply it by $-\frac{\varepsilon}{2}$, which results in an additional term
\begin{equation}
Tr \big[ i \mathcal{M}^{5.L}_{1,2,3}~\slashed{p}\big] \supset	-\frac{i}{(4\pi)^2} C_F e_q^3 g^2 \mathcal{A}_{1,2,3} \frac{1}{(\underline{p}k_3-i0)(\bar{p}k_1+i0)} \frac{x-\xi}{\xi}\log \Big{(} \frac{1}{2\xi}\big{(}\xi - x + i0\cdot sgn(pk_3)\big{)} \Big{)}.
\end{equation}
Now we focus on the $n=1$ terms. Non-vanishing traces (see \eqref{eq-M5-L-div}) are
\begin{equation}
\begin{aligned}
	& y Tr\big( k'\epsilon_3 k' \epsilon_2 k_1 \epsilon_1 \big), \quad -r(1-y) Tr\big( k'\epsilon_3 k' \epsilon_2 k_2 \epsilon_1 \big),\\
	& y Tr\big( \epsilon_3 k_1 \epsilon_2 k' \epsilon_1 k' \big), \quad \big( 1 - r(1-y) \big) Tr\big( \epsilon_3 k_2 \epsilon_2 k' \epsilon_1 k' \big),\\
	& -(1-y) Tr\big( \epsilon_3 k' \epsilon_2 k' \epsilon_1 k_1 \big), \quad -r(1-y) Tr\big( \epsilon_3 k' \epsilon_2 k' \epsilon_1 k_2 \big).
\end{aligned}
\end{equation}
We use again the well-known formula
\begin{equation}
6\int \frac{d^4 k}{(2\pi)^4} \frac{k^\mu k^\nu}{\big( k^2 + \Delta + i0 \big)^4} = -\frac{ig^{\mu\nu}}{2(4\pi)^2}\Big( -\Delta - i0 \Big)^{-1},
\end{equation}
and $\gamma_\mu \slashed{A} \gamma^\mu = -2\slashed{A}$, so that we obtain
\begin{equation}\label{eq-App-traces-5L}
\begin{aligned}
	&Tr \big[ i \mathcal{M}^{5.L}_{1,2,3}~\slashed{p}\big] \supset\\& -\frac{i}{(4\pi)^2} s^{-1} \bigg( \mathcal{F}(1,0,0;x, \xi, \{ \beta_i \}) \Big( Tr\big( \epsilon_3 \epsilon_2 k_1 \epsilon_1 \big) + Tr\big( \epsilon_3 k_1 \epsilon_2  \epsilon_1 \big) - Tr\big( \epsilon_3 \epsilon_2  \epsilon_1 k_1 \big) + 2 Tr\big( \epsilon_3 k_2 \epsilon_2  \epsilon_1 \big) \Big) + \\
	-&\mathcal{F}(1,1,0;x, \xi, \{ \beta_i \}) \Big(  Tr\big( \epsilon_3  \epsilon_2 k_2 \epsilon_1 \big) + Tr\big( \epsilon_3 k_2 \epsilon_2  \epsilon_1 \big) + Tr\big( \epsilon_3  \epsilon_2  \epsilon_1 k_2\big)  \Big) \bigg).
\end{aligned}
\end{equation}
After the trace algebra:
\begin{equation}
\begin{aligned}
	\eqref{eq-App-traces-5L} = \:-\frac{i}{(4\pi)^2} s^{-1} 4\bigg( &\mathcal{F}(1,0,0;x, \xi, \{ \beta_i \}) \Big( 3(k_1 \epsilon_2)(\epsilon_2 \epsilon_3) - 3 (k_1 \epsilon_3) (\epsilon_1 \epsilon_2) - 2(k_2 \epsilon_1)(\epsilon_2 \epsilon_3) \Big) + \\
	-&\mathcal{F}(1,1,0;x, \xi, \{ \beta_i \}) \Big(  (k_2 \epsilon_1)(\epsilon_2 \epsilon_3) + (k_2 \epsilon_3)(\epsilon_1 \epsilon_2)  \Big) \bigg).
\end{aligned}
\end{equation}

\chapter{The full form of the amplitude}\label{Appdx-full-amplitude}
In this part we summarize all results concerning the amplitude at the higher order.

Let us first make some simplifications with polarization vectors. In the calculation of the hard part it is assumed that $p_1-p_2 \approx 2\xi p$, so that $2\xi p = q_1 + q_2 - q$. Hence, for each polarization vector $\epsilon_i$:
\begin{equation}
\epsilon_i (k_1 + k_2 + k_3 ) = 0,
\end{equation}
moreover $k_i \epsilon_i = 0$ (there is no summation over index $i$). \\
In the considered frame, only the transverse parts of polarizations vectors contribute to their products with momenta. We have:
\begin{equation}
\begin{aligned}
&q_1 \epsilon(q) = -q_2\epsilon(q) = -\vec{p}_t \vec{\epsilon}_t(q),
\\
&q \epsilon(q_1) = q_2 \epsilon(q_1) = \frac{1}{\alpha} \vec{p}_t \vec{\epsilon}_t(q_1), \qquad q \epsilon(q_2) = q_1 \epsilon(q_2) =  -\frac{1}{\bar{\alpha}} \vec{p}_t \vec{\epsilon}_t(q_2).
\end{aligned}
\end{equation}
$\vec{v}\vec{u}$ denotes the Euclidean product.
Choosing the x-axis as the direction of $p_\perp$:
\begin{equation}
\vec{p}_t = |p_t | \vec{e}_x,
\end{equation}
and introduce the basis in which the perpendicular parts of polarization vectors are either in $x$ or $y$ direction:
\begin{equation}
\begin{aligned}
&\epsilon^\mu(q_1) = \epsilon_i^\mu + \frac{2 |p_\perp| \delta_{i,x}}{\alpha s}p^\mu,
\\
&\epsilon^\mu(q_2) = \epsilon_i^\mu -
\frac{2 |p_\perp| \delta_{i,x} }{\bar{\alpha} s}p^\mu,
\\
&\epsilon^\mu(q) = \epsilon_i^\mu,
\end{aligned}
\end{equation}
$i=x,y$, and $\delta$ denotes the Kronecker delta. Let me denote by $\sigma_i$ the polarization of the photon $i$. In this real-valued basis $\epsilon = \epsilon^*$. We obtain:
\begin{equation}
\begin{aligned}
& q_1 \epsilon(q) = - |p_\perp| \delta_{\sigma_q, x}, \qquad q_2 \epsilon(q) = |p_\perp| \delta_{\sigma_q, x},
\\
& q \epsilon(q_1) = \frac{1}{\alpha} |p_\perp| \delta_{\sigma_1, x}, \qquad q \epsilon(q_2) = -\frac{1}{\bar{\alpha}} |p_\perp| \delta_{\sigma_2, x}.
\end{aligned}
\end{equation}
Product of two polarization vectors are $\epsilon_i \epsilon_j = -\delta_{\sigma_i, \sigma_j}$.\\

 At NLO, the summation over photons' permutations does not yield such simplifications as it did at the leading order, so that it will be performed numerically. To make the formulae more readable and easier in numerical implementation, we will denote products of polarization vectors and considered momenta as
\begin{equation}
\epsilon_1 k_2 = -\epsilon_1 k_3 = ek1, \quad \epsilon_2 k_1 = -\epsilon_2 k_3 = ek2, \quad \epsilon_3 k_1 = -\epsilon_3 k_2 = ek3.
\end{equation}
Product of two polarization vectors $i$ and $j$ will be denoted as $eij$. 

We have shown, that the sum over permutations of $Tr\big[ i\big( \mathcal{M}^{2.L}_{1,2,3}+\mathcal{M}^{2.R}_{1,2,3}+\mathcal{M}^{3.L}_{1,2,3}+\mathcal{M}^{3.R}_{1,2,3} \big) \slashed{p} \big]$ is proportional the LO amplitude, so that one can easily sum that combination over all photons' permutations to obtain
\begin{equation}
\begin{aligned}
&\sum_{\mathrm{perm.}}Tr\big[ i\big( \mathcal{M}^{2.L}_{1,2,3}+\mathcal{M}^{2.R}_{1,2,3}+\mathcal{M}^{3.L}_{1,2,3}+\mathcal{M}^{3.R}_{1,2,3} \big) \slashed{p} \big]=\\& i C_F e_q^3 \frac{\alpha_S}{4\pi} s^{-2} \frac{1}{\alpha \bar{\alpha} }  \bigg[ (\alpha - \bar{\alpha} )\big( \vec{\epsilon^*}_\perp(q_1) \vec{\epsilon^*}_\perp(q_2) \big) \big( \vec{p}_\perp \vec{\epsilon}(q) \big) - \big(  \vec{p}_\perp \vec{\epsilon^*}_\perp(q_1) \big) \big(  \vec{\epsilon}(q) \vec{\epsilon^*}_\perp(q_2) \big) - \big(  \vec{p}_\perp \vec{\epsilon^*}_\perp(q_2) \big) \big(  \vec{\epsilon}(q) \vec{\epsilon^*}_\perp(q_1) \big) \bigg] \\
&\times \bigg[ -\frac{i\pi}{\xi}\Big( -10 + \log\big( \alpha \bar{\alpha} \big) \Big)\Big( \delta(x-\xi) + \delta(x+\xi) \Big) \\
&+ i \: \mathrm{Im} \frac{1}{x+\xi -i0}\frac{1}{x-\xi + i0} \log\big( -(x+\xi -i0)(x-\xi+i0) \big) \bigg].
\end{aligned}
\end{equation}

$Tr\big[ i\mathcal{M}^{3.M}_{1,2,3} ~\slashed{p} \big]$ does no longer admit the previously used symmetry in interchange of photons $1$ and $3$, and it cannot be summed in such a simply manner. For a given permutation of photons one has
\begin{equation}
\begin{aligned}
&Tr\big[ i\mathcal{M}^{3.M}_{1,2,3} ~\slashed{p} \big]_{\mathrm{fin.}} = -iC_F e_q^3 \frac{\alpha_S}{4\pi}s^{-2} \frac{1}{\beta_1 \beta_3} \frac{1}{x+\xi+ i0_1} \frac{1}{x-\xi -i0_3} \times
\\
 &2s\Bigg{(} \frac{(x+\xi)\beta_1}{\big( (x+\xi)\beta_1 + (x-\xi)\beta_3 \big)^2} \big( \beta_1 e13 \cdot ek2 - \beta_1 e23 \cdot ek1 + \beta_3 e12 \cdot ek3 + \beta_3 e13 \cdot ek2 \big) \times \\ & \Big{(} \big( 2(x+\xi)\beta_1 + (x-\xi)\beta_3 \big)\log\frac{-(x-\xi)\beta_3 + i0}{(x+\xi)\beta_1 + i0} +(x+\xi)\beta_1 + (x-\xi)\beta_3  \Big{)} +\\
&-2s \frac{(x-\xi)\beta_3}{\big( (x+\xi)\beta_1 + (x-\xi)\beta_3 \big)^2} \big( \beta_1 e13 \cdot ek2+ \beta_1 e23 \cdot ek1 - \beta_3 e12 \cdot ek3 + \beta_3 e13 \cdot ek2 \big) \times \\ & \Big{(} \big( (x+\xi)\beta_1 + 2(x-\xi)\beta_3 \big)\log\frac{(x+\xi)\beta_1 + i0}{-(x-\xi)\beta_3 + i0} +(x+\xi)\beta_1 + (x-\xi)\beta_3  \Big{)}+
\\
    &+ \mathcal{A}_{1,2,3} \Big{(} 1 - \frac{(x-\xi)\beta_3 log\big( (x-\xi)\beta_3 - i0\big) + (x+\xi)\beta_1 log\big(-(x+\xi)\beta_1 - i0\big)}{(x+\xi)\beta_1 + (x-\xi)\beta_3} \Big{)} \Bigg{)}.
\end{aligned}
\end{equation}

Graphs containing 4- and 5-point loops result in far more complicated integrals over Feynman parameters than the previously presented ones, so that these amplitudes will be written using functions $\mathcal{F}$ and $\mathcal{G}$ defined in Section \ref{Section-M4}. Their exact form is presented at the end of this Appendix.
\begin{equation}
\begin{aligned}
&Tr\big[ i\mathcal{M}^{4.L}_{1,2,3} ~\slashed{p} \big]_{\mathrm{fin.}} = -2\frac{i}{(4\pi)^2}C_F e_q^3 g^2  \frac{1}{2\underline{p}k_3 - i0}\bigg\{ \\
& -\mathcal{F} \big(2,1,0;x, \xi, \{ \beta_i \}\big) s^{-1} \Big[\beta_3 \big(\kappa_3 s (e13 \cdot ek2 - 3 e12 \cdot ek3)+4 ek1 \cdot ek2 \cdot ek3\big)+\\&\quad -s \big(e23 \cdot ek1-e13 \cdot ek2\big) \big(\beta1 (\kappa_3+2 \kappa_1)-2 \beta_2 (\kappa_3+\kappa_2)\big)\Big]+\\[0.3cm]
&+\mathcal{F} \big(2,0,1;x, \xi, \{ \beta_i \}\big) (x+\xi) \big[\beta_1^2 (e13 \cdot ek2 - e23 \cdot  ek1)+ \beta_1 (2 \beta_2 e13 \cdot ek2- 2 \beta_2 e23 \cdot ek1+  \\&\quad  +3 \beta_3 e12 \cdot ek3 - 9 \beta_3 e13 \cdot ek2+4 \beta_3 e23 \cdot ek1)+2 \beta_2 \beta_3 (e12 \cdot ek3-e13 \cdot ek2)\big]\\[0.3cm]
&-\mathcal{F} \big(2,1,1;x, \xi, \{ \beta_i \}\big) (x+\xi) \big[\beta_1 \beta_3 (e23 \cdot ek1-e12 \cdot ek3)+2 \beta_2^2 (e23 \cdot ek1 - e13 \cdot ek2)+ \\&+ \beta_2 \beta_3 (-3 e12 \cdot ek3+3 e13 \cdot ek2+2 e23 \cdot ek1)\big] +\\[0.3cm]
&+\mathcal{F} \big(2,2,0;x, \xi, \{ \beta_i \}\big) s^{-1} \big[ \beta_2 \kappa_3 s (e23 \cdot ek1-e13 \cdot ek2)+ \\&\quad  +\beta_3 (-e12 \cdot ek3 \cdot \kappa_3 s+e23 \cdot ek1 \cdot \kappa_3 s+4 ek1 \cdot ek2 \cdot ek3) \big] + \\[0.3cm]
& +\mathcal{F} \big(2,2,1;x, \xi, \{ \beta_i \}\big) (x+\xi) \beta_2 \big[-\beta_2 e13 \cdot ek2+\beta_2 e23 \cdot ek1- \beta_3 e12 \cdot ek3+\beta_3 e23 \cdot ek1\big] \\[0.3cm]
& + \mathcal{F}\big(1,0,0; x, \xi, \{ \beta_i \}\big) \big[ -\beta_1 e13 \cdot ek2+ \beta_1 e23 \cdot ek1 - 2 \beta_2 e13 \cdot ek2 + \\&\quad + 2 \beta_2 e23 \cdot ek1-3 \beta_3 e12 \cdot ek3 + 3 \beta_3 e13 \cdot ek2-2 \beta_3 e23 \cdot ek1 \big] + \\[0.3cm]
&+ \mathcal{F}\big(1,1,0; x, \xi, \{ \beta_i \}\big)  \big[3 \beta2 e13 \cdot ek2-3 \beta_2 e23 \cdot ek1+ \beta_3 e12 \cdot ek3 - \beta_3 e23 \cdot ek1\big]+\\[0.3cm]
& - 2 \mathcal{G}\big(x, \xi, \{ \beta_i \}\big) (x+\xi) \beta_3 \big[ -\beta_1 e13 \cdot ek2+\beta_1 e23 \cdot ek1- \beta_3 e12 \cdot ek3 + \beta_3 e13 \cdot ek2\big] \bigg\}.
\end{aligned}
\end{equation}

\newpage
\begin{equation}\label{eq-finite-5L}
\begin{aligned}
&Tr\big[ i\mathcal{M}^{5.L}_{1,2,3} ~\slashed{p} \big]_{\mathrm{fin.}} =-2\frac{i}{(4\pi)^2} C_F e_q^3 g^2 \frac{1}{\xi}s^{-2} \bigg\{ \\
& +\mathcal{F}\big(2,1,0;x, \xi, \{ \beta_i \}\big)\big[ek2 \cdot (2 ek1 \cdot ek3- e13 \cdot \kappa_3 s)-3 ek3 \cdot (2 ek1 \cdot ek2-e12 \cdot \kappa_3 s) \big]+ \\[0.3cm]
& s\mathcal{F}\big(2,0,1;x, \xi, \{ \beta_i \}\big) (x+\xi) \big[ 2 \beta_1 e12 \cdot ek3-5 \beta_1 e13 \cdot ek2+ 3 \beta_1 e23 \cdot ek1+\beta_2 e12 \cdot ek3-\beta_2 e13 \cdot ek2\big] + \\[0.3cm]
&+ \frac12 \mathcal{F}\big(2,1,1;x, \xi, \{ \beta_i \}\big) (x+\xi) \big[ 3 \beta_1 e12 \cdot ek3 - 2 \beta_1 e13 \cdot ek2 - \beta_1 e23 \cdot ek1 +7 \beta_2 e12 \cdot ek3 - 5 \beta_2 e13 \cdot ek2\big]  + \\[0.3cm]
& + \mathcal{F}\big(2,2,0;x, \xi, \{ \beta_i \}\big) \big[  ek3 \cdot (2 ek1 \cdot ek2-e12 \cdot \kappa_3 s)-  ek1 \cdot (e23 \cdot \kappa_3 s+2 ek2 \cdot ek3) \big] + \\[0.3cm]
& +s \mathcal{F}\big(2,2,1;x, \xi, \{ \beta_i \}\big) (x+\xi) \big[ \beta_2 e23 \cdot ek1  - \beta_2 e12 \cdot ek3 \big] +\\[0.3cm]
&-8s\mathcal{F}\big(1,0,0;x, \xi, \{ \beta_i \}\big) \big[ 3 e12 \cdot ek3 - 3 e13 \cdot ek2 + 2 e23 \cdot ek1 \big] + \\[0.3cm]
&+8s\mathcal{F}\big(1,1,0;x, \xi, \{ \beta_i \}\big) \big[ e12 ek3 - e23 ek1 \big]+\\[0.3cm]
& +s\mathcal{G}(x, \xi, \{ \beta_i \}) \big[ \beta_1 e13 \cdot ek2-\beta_1 e23 \cdot ek1+\beta_3 e12 \cdot ek3 - \beta_3 e13 \cdot ek2 \big]  + \\[0.3cm]
& + \frac12 \mathcal{A}_{1,2,3} \frac{ (x-\xi)}{((x-\xi)\beta_3 -i0)((x+\xi)\beta_1+i0)} \log \Big{(} \frac{1}{2\xi}\big{(}\xi - x + i0\cdot sgn(pk_3)\big{)} \Big{)}\bigg\}.
\end{aligned}
\end{equation}

The $.R$ counter-part is obtained by substituting $k_1 \rightarrow -k_3$, $k_2 \rightarrow -k_2$, $k_3 \rightarrow -k_1$, $\xi \rightarrow -\xi$.

Integrals used in the definition of functions $\mathcal{F}$ and $\mathcal{G}$ can be computed analytically and written in terms of logarithms and dilogarithms $\mathrm{Li}_2$. 
\begin{equation}
 \int_0^1 dr \int_0^1 dz \: r \Big( az + br + czr \Big)^{-2} = \frac{\log(a+b+c) - \log(a)}{b(b+c)}.
\end{equation}

\begin{equation}
\begin{aligned}
& \int_0^1 dr \int_0^1 dz \: rz \Big( az + br + czr \Big)^{-2} =\\
&\frac{1}{6 c^2 (a+c) (b+c)} \bigg[ -6 (a+c) (b+c) \left(-\text{Li}_2\left(-\frac{c (a+b+c)}{a b}\right)+\text{Li}_2\left(-\frac{c}{a}\right)+\text{Li}_2\left(\frac{b}{b+c}\right)\right)+\\&(a+c) (b+c) \left(-3 \log ^2\left(\frac{b+c}{b}\right)+ 6 \left(\log \left(\frac{c}{b}\right)-1\right)
   \log \left(\frac{b+c}{b}\right)+6 \log (b+c)+\pi ^2\right)+ \\&6 (b+c) \log (b) \left(-(a+c) \log \left(\frac{a+c}{a}\right)-a\right)+6 (a+c) \log (a) ((b+c) (\log (b)-\log (b+c))+c)+\\&6 \left((a+c) (b+c) \log \left(\frac{(a+c) (b+c)}{a b}\right)-c (a+b+2
   c)\right) \log (a+b+c) \bigg].
\end{aligned}
\end{equation}

\begin{equation}
\begin{aligned}
& \int_0^1 dr \int_0^1 dz \: r^2 \Big( az + br + czr \Big)^{-2} =\\
& \frac{-a c^2 \log (a+b+c)+(b+c) \left(c^2-a (b+c) \left(-\log (b+c)+\log \left(\frac{b+c}{b}\right)+\log (b)\right)\right)+a c^2 \log (a)}{bc^2 (b+c)^2}.
\end{aligned}
\end{equation}

\begin{equation}
\begin{aligned}
& \int_0^1 dr \int_0^1 dz \: r^2 z \Big( az + br + czr \Big)^{-2} =\\
& \frac{2a}{c^3} \left( -\text{Li}_2\left(-\frac{c (a+b+c)}{a b}\right)+\text{Li}_2\left(-\frac{c}{a}\right)+\text{Li}_2\left(\frac{b}{b+c}\right) \right)+ \\
& \frac{1}{c^3 (a+c) (b + c)^2} \bigg[a (a+c) (b+c)^2 \log ^2\left(\frac{b+c}{b}\right)-3 a (a+c) (b+c)^2 \log (b+c)+\\&-a (a+c) (b+c)^2
   \left(2 \log \left(\frac{c}{b}\right)-3\right) \log \left(\frac{b+c}{b}\right)+ \\& \left(c \left(a^2 (2 b+3 c)+a \left(2 b^2+6 b c+5 c^2\right)+c (b+c)^2\right)-2 a (a+c) (b+c)^2 \log \left(\frac{(a+c) (b+c)}{a b}\right)\right) \log (a+b+c) \bigg]
\\& + \frac{1}{3 c^3 (a+c) (b + c)^2} \bigg[ 3 (b+c)^2 \log (b) \left(3 a^2+a c+2 a (a+c) \log \left(\frac{a+c}{a}\right)-c^2\right) + \\& -(a+c) (b+c) \left(\pi ^2 a (b+c)+3 c^2\right)+3 a (a+c) \log (a) \left(-c (2 b+3 c)-2 (b+c)^2 (\log (b)-\log (b+c))\right) \bigg].
\end{aligned}
\end{equation}

\begin{align}
& \int_0^1 dr \int_0^1 dz \: \Big( az + br + czr \Big)^{-1} = \nonumber \\
& -\frac{1}{c} \bigg[ \log \left(\frac{b}{a}\right) \log \left(\frac{a+c}{a}\right)-\log \left(\frac{(a+c) (b+c)}{a b}\right) \log
   \left(\frac{a+b+c}{a}\right)+\nonumber \\& \text{Li}_2\left(-\frac{c}{a}\right)+\text{Li}_2\left(-\frac{c}{b}\right) -\text{Li}_2\left(-\frac{c (a+b+c)}{a b}\right) \bigg]. \\[1 cm]
& \int_0^1 dr \int_0^1 dz \: r \Big( az + br + czr \Big)^{-1} = \nonumber \\
& \frac{1}{c^2 (b+c) } \Bigg[ a (b+c) \left(-\text{Li}_2\left(-\frac{c (a+b+c)}{a b}\right)+\text{Li}_2\left(-\frac{c}{a}\right)+\text{Li}_2\left(-\frac{c}{b}\right)\right)+ \nonumber \\& a \log \left(\frac{a}{b}\right) \left((b+c) \log \left(\frac{b+c}{b}\right)-c\right)+\left(c (a+b+c)-a (b+c) \log
   \left(\frac{(a+c) (b+c)}{a b}\right)\right) \log \left(\frac{a+b+c}{b}\right) \Bigg].
\end{align}

\begin{equation}
\begin{aligned}
& \int_0^1 dr \int_0^1 dz \: \Big( az + br + czr \Big)^{-1} \log \Big(- az - br - czr \Big) =\\
& \frac{1}{6 ac (a+c) } \Bigg[ 6 (a+c) \left(-\text{Li}_2\left(-\frac{c (a+b+c)}{a b}\right)+\text{Li}_2\left(-\frac{c}{a}\right)+\text{Li}_2\left(\frac{b}{b+c}\right)\right)+\\& 3 c
   \log ^2(-a-b-c)+6 (a+c) \log (-b) \log \left(\frac{a+c}{a}\right)+6 a \log (-a) \log \left(\frac{b+c}{b}\right)+ \\&3 \log \left(\frac{b+c}{b}\right)
   \left((a+c) \left(\log \left(\frac{b+c}{b}\right)-2 \log \left(\frac{c}{b}\right)\right)+ 2 c \log (-a)\right)+ \\& -6 (a+c) \log (-a-b-c) \log
   \left(\frac{(a+c) (b+c)}{a b}\right)+6 c \log (a+b+c)+ \\& -\pi ^2 (a+c)-3 c \log ^2(-b)-6 c \log (b) \Bigg].
\end{aligned}
\end{equation}

\end{document}